\documentclass[structabstract]{aa}  
\usepackage{graphicx}
\usepackage{txfonts}
\usepackage{aalongtable}
\usepackage{natbib}
\bibpunct{(}{)}{;}{a}{}{,}
\begin{document}
   \title{Astrometric confirmation of young low-mass binaries and multiple systems in the Chamaeleon star-forming regions \thanks{Based on observations made with ESO telescopes at the Paranal Observatory under program IDs 076.C-0292(A), 078.C-0535(A), 080.C-0424(A), 082.C-0489(A), 084.C-0364(B), 086.C-0638(A) \& 086.C-0600(B), the Hubble Space Telescope under program ID GO-8716 and data obtained from the ESO/ST-ECF Science Archive Facility from the Paranal Observatory under program IDs 075.C-0042(A), 076.C-0579(A), 278.C-5070(A) and from the Hubble Space Telescope under programme IDs SNAP-7387, GO-11164.
}
}
   
   \titlerunning{Confirmation of young systems in Chamaeleon}

   \author{N. Vogt\inst{1,2}
           \and
           T. O. B. Schmidt\inst{3}
           \and
           R. Neuh\"auser\inst{3}
           \and
           A. Bedalov\inst{4}
           \and
           T. Roell\inst{3}
           \and
           A. Seifahrt\inst{5,6}
           \and
           M. Mugrauer\inst{3}
}

   \offprints{Nikolaus Vogt, e-mail:Nikolaus.Vogt@uv.cl}

   \institute{Departamento de F\'isica y Astronom\'ia, Universidad de Valpara\'iso, Avenida Gran Breta\~na 1111, Valpara\'iso, Chile\\
              \email{Nikolaus.Vogt@uv.cl}
         \and
             Instituto de Astronom\'ia, Universidad Catolica del Norte, Avda.~Angamos 0610, Antofagasta, Chile
         \and
             Astrophysikalisches Institut und Universit\"ats-Sternwarte, Universit\"at Jena, Schillerg\"a\ss chen 2-3, 07745 Jena, Germany
         \and
             Faculty of Natural Sciences, University of Split, Teslina 12. 21000 Split, Croatia
         \and
             Department of Physics, University of California, Davis, CA 95616, USA
	 \and
         \   Department of Astronomy and Astrophysics, University of Chicago, Chicago, IL 60637, USA
}            

   \date{Received 2011; accepted }

% \abstract{}{}{}{}{} 
% 5 {} token are mandatory
 
  \abstract
   {The star-forming regions in Chamaeleon (Cha) are one of the nearest (distance $\sim$\,165 pc) and youngest 
(age $\sim$\,2 Myrs) conglomerates of recently formed stars and the ideal target for population studies of star formation.}
   {We investigate a total of 16 Cha targets, which have been suggested, but not confirmed 
as binaries or multiple systems in previous literature.}
  % methods heading (mandatory) ---------- COMPLETELY REVISED ----------------------
   {We used the adaptive optics instrument Naos-Conica (NACO) at the Very Large Telescope Unit Telescope (UT) 
4\,/\,YEPUN of the Paranal 
Observatory, at 2 -- 5 different epochs, in order to obtain relative and absolute astrometric measurements, as well 
as differential photometry in the J, H, and K band. On the basis of known proper motions and these observations, we analyse the 
astrometric results in our “Proper Motion Diagram” (PMD: angular separation\,/\,position angle versus time), to 
eliminate possible (non-moving) background stars, establish co-moving binaries and multiples, and search 
for curvature as indications for orbital motion.}
  % results heading (mandatory) --------- COMPLETELY REVISED ------------------------
   {All previously suggested close components are co-moving and no background stars are found. The angular separations range 
between 0.07 and 9 arcseconds, corresponding to projected distances between the components of 6 -– 845 AU. Thirteen  stars are 
at least binaries and the remaining three (RX J0919.4-7738, RX J0952.7-7933, VW Cha) are confirmed high-order multiple systems with 
up to four components. In 13 cases, we found significant slopes in the PMD´s, which are 
compatible with orbital motion whose periods (estimated from the observed gradients in the position angles) range from 60 to 
550 years.  However, in only four cases there are indications of a curved orbit, the ultimate proof of a gravitational bond.}
  % conclusions heading ---------------- COMPLETELY REVISED --------------------------
   {A statistical study based on the 2MASS catalogue confirms the high probability of all 16 stellar systems being 
gravitationally bound. Most of the secondary components are well above the mass limit of hydrogen burning stars 
(0.08 M$_{\sun}$), and have masses twice as high as this value or more. Massive primary components appear to avoid the 
simultaneous formation of equal-mass secondary components, while extremely low-mass secondary components are hard 
to find for both high and low mass primaries owing to the much higher dynamic range and the faintness of the secondaries.}


   \maketitle
%
%________________________________________________________________
\section{Introduction} % ---------- COMPLETELY REVISED ------------------------

One of the nearest dark cloud groups to the Sun can be found in the southern constellation Chamaeleon, at a distance of 
165 $\pm$ 30 pc (based on the data of \citet{1999A&A...352..574B} and \citet{1997A&A...327.1194W}). There are strong indications 
of recent star formation within these clouds: many newborn stars have been historically detected according to their variability 
and H$\alpha$ emission. Their observed location in the Hertzsprung-Russell diagram implies, according to the evolutionary models of 
\citet{2000ApJ...542..464C} and \citet{2003A&A...402..701B}, that Cha I has a mean age of $\sim$ 2 $\cdot$ 10$^{6}$ years 
\citep{2008hsf2.book..169L}. The stellar density within this region is rather low, but contains a large number of low 
mass stars. Since the Chamaeleon association is nearby and well-isolated from other young stellar populations it is an 
excellent target for studies of low-mass star formation. For a review about this region we refer to 
\cite{2008hsf2.book..169L}.

%Binarity in Chamaeleon has been studied before by \citet{1977ApJS...35..161S},
%\citet{1986A&AS...64..105T}, \citet{1988A&AS...76..189S}, \citet{1992PhDT.......255B}, \citet{1993A&A...278...81R},
%\citet{1996A&A...307..121B}, \citet{1997ApJ...481..378G}, \citet{1997A&A...328..187C}, \citet{1997A&A...326..632A},
%\citet{1998AJ....116.2975M}, \citet{1999A&AS..136..429C},
%\citet{2001ApJ...561L.199B}, \citet{2001AJ....122.3325K}, \citet{2001A&A...376..907P}, \citet{2002AJ....124.2813R},
%\citet{2002A&A...384..999N}, \citet{2003A&A...410..269M}, \citet{2003A&A...397..675T},
%\citet{2004AJ....127.1747H}, \citet{2004ApJ...602..816L}, \citet{2004ApJ...614..398L}, \citet{2006AJ....132.2675H},
%\citet{2006A&A...459..909C}, \citet{2006ApJ...649..894L}, \citet{2006A&A...448..655J}, \citet{2006A&A...454..553D},
%\citet{2007A&A...467.1147G},
%\citet{2007ApJ...671.2074A}, \citet{2007ApJ...662..413K}, \citet{2007ApJ...670.1337D}, \citet{2008ApJ...683..844L}, 
%\citet{2008A&A...492..545J}, \citet{2008A&A...484..413S}, \citet{2010AJ....140..510T}, 
%etc., who found several binary and multiple candidates, most not yet confirmed.

Our main interests are the frequency and distribution of young binaries and multiple systems, with emphasis on  
the occurrence of components near the stellar mass limit (0.08 M$_{\odot}$), and of brown dwarfs or planets. 
For this purpose, we carried out an extensive observational campaign to search for possible companion objects of the most certain
members of the Chamaeleon association with well-known proper motions. We obtained direct imaging of a total of 51 Cha 
association member stars with the European Southern Observatory (ESO) Very Large Telescope (VLT) instrument Naos-Conica 
\citep[NACO,][]{2003SPIE.4841..944L, 2003SPIE.4839..140R}, in the near-infrared (J, H, and Ks) to carry out high 
precision astrometry and photometry of them and their nearby candidate companions. In some cases, follow-up spectroscopic observations
could be obtained, as for Cha H$\alpha$ 2  \citep{2008A&A...484..413S} and CT Cha \citep{2008A&A...491..311S}. 
The purpose of the present paper is to present our results for a total of 16 of the 51 Cha members observed with 
previously known faint stellar companion candidates, which, in most cases, had been observed during only one epoch and, therefore,  
lack any confirmation of their binary nature. Here we present NACO imaging from 1 to 5 different epochs of these 
objects. Combining our astrometric data with all available literature values, we are able to show that all cases are co-moving 
binaries or multiple systems, and in many of them there are already indications of orbital motion, partly even curved orbital
motion.

In section 2, we describe the details of our observing and reduction strategy. Section 3 summarizes our most important results.
Section 4 contains a detailed discussion of each star and section 5 describes the resulting conclusions.

\section{Observations and calibrations}

\begin{table*}
\caption{Observed objects in Chamaeleon}
\label{table:1}
%\centering
\begin{tabular}{lllcccccccc}
\hline\hline
Object$^a$            & RA           & Dec                & System           & Bin. & Spectral- & SpT  & Dist. & Dist. & 2MASS        & Backg.$^e$ \\
& [h \ m \ s]$^{a,b}$ & [$^{\circ}$ \ \ ' \ \ '']$^{a,b}$ & architecture$^c$ & Ref. & types     & Ref. & [pc]  & Ref.  & K\,[mag]$^d$ & density    \\
\hline
RX J0915.5-7609 & 09 15 29.12  & -76 08 47.2  & **        & 1     & K7+?    & 2    & 168  & 3     & \ 8.488      & 0.135 \\
RX J0919.4-7738 & 09 19 24.96  & -77 38 37.0  & SB2+**    & 4,2,1 & G7+?+K0+? & 5  &  57  & 6     & \ 6.780      & 0.137 \\ 
RX J0935.0-7804 & 09 34 56.04  & -78 04 19.4  & **        & 1     & M2+?    & 2    & 168  & 3     & \ 8.889      & 0.132 \\
RX J0952.7-7933 & 09 53 13.74  & -79 33 28.5  & *+SB2     & 2,1   & F6?+F6+F6? & 2 & 170? & 7     & \ 7.994      & 0.122 \\
RX J1014.2-7636 & 10 14 08.07  & -76 36 32.8  & **        & 1     & M3+?    & 2    & 165? &       & \ 8.874      & 0.154 \\
SZ Cha          & 10 58 16.77  & -77 17 17.1  & SB?+*+w*  & 8,9   & K0+M5+? & 9,10 & 165  & Cha I & \ 7.758      & 0.144 \\
Ced 110 IRS 2   & 11 06 15.41  & -77 21 57.0  & **+w*     & 10,11 & G5+?+?  & 12   & 165  & Cha I & \ 6.419      & 0.056 \\
Cha H$\alpha$ 2 & 11 07 42.46  & -77 33 59.4  & **        & 13,14 & M6+M6?  & 15   & 165  & Cha I & 10.675       & 0.062 \\
VW Cha          & 11 08 01.49  & -77 42 28.9  & SB?+**+w* & 16,17 & K6+K7+?+M2.5& 18 & 165  & Cha I & \ 6.962 & 0.088 \\
RX J1109.4-7627 & 11 09 17.70  & -76 27 57.8  & **        & 19,20 & K7+?    & 2    & 165  & Cha I & \ 8.701      & 0.141 \\
HD 97300        & 11 09 50.02  & -76 36 47.7  & **        & 8     & B9+?    & 21   & 179  & 6     & \ 7.149      & 0.098 \\
WX Cha          & 11 09 58.74  & -77 37 08.9  & **        & 8     & K0-M1.25+?& 12 & 165  & Cha I & \ 7.970      & 0.096 \\
WY Cha          & 11 10 07.05  & -76 29 37.7  & **+w*     & 20,10 & K2-M0+?+? & 12 & 165  & Cha I & \ 8.451      & 0.127 \\
HJM C 7-11      & 11 10 38.02  & -77 32 39.9  & SB?+*+w*  & 8,22  & K3+?+?  & 2    & 165  & Cha I & \ 8.277      & 0.107 \\
Sz 41           & 11 12 24.42  & -76 37 06.4  & SB?+*+w*  & 23    & K4+?+G8 & 2,24 & 165  & Cha I & \ 7.999      & 0.166 \\
HM Anon         & 11 12 42.69  & -77 22 23.1  & **        & 8     & G8+?    & 25   & 165  & Cha I & \ 7.880      & 0.124 \\
\hline
\end{tabular}
\begin{flushleft}
\textbf{Remarks}: (a) Taken from the SIMBAD database \citep{2007ASPC..377..197W} (b) 
International Celestial Reference System (ICRS) coordinates (epoch=J2000) (c) Previously known and supposed multiplicity
of the objects: *: star, **: binary, SB: spectroscopic binary, SB2: double-lined spectroscopic binary, w*: wide stellar 
companion candidate (d) \citet{2006AJ....131.1163S,2003tmc..book.....C} (e) expected number of background stars in the NACO 
S13 field of view (see text)\\
\textbf{References}: (1) \citet{2001AJ....122.3325K} (2) \citet{1997A&A...328..187C} (3) \citet{2003A&A...404..913S} (4) 
\citet{1986A&AS...64..105T} (5) \citet{2006A&A...460..695T} (6) \citet{2007A&A...474..653V} (7) \citet{1998A&A...338..442F}
(8) \citet{1997ApJ...481..378G} (9) \citet{2002AJ....124.2813R} (10) \citet{2007ApJ...662..413K} (11) Simultaneously imaged
by \citet{2008ApJ...683..844L} and by us (here) (12) \citet{2007ApJS..173..104L} (13) supposed by \citet{2002A&A...384..999N}
(14) confirmed by \citet{2007ApJ...671.2074A}, \citet{2008ApJ...683..844L} $\&$ \citet{2008A&A...484..413S} (15) 
\citet{2008A&A...484..413S} (16) wide companion candidate: \citet{1993A&A...278...81R}, close binary: 
\citet{1992PhDT.......255B}, close triple: \citet{2001ApJ...561L.199B} (17) Possibly SB in \citet{2003A&A...410..269M}  
$\&$ \citet{2006A&A...460..695T}, but could be the signal of the close triple (18) \citet{1997A&A...321..220B}, Sz 23 
(VW Cha D?): \citet{1999A&A...343..477C} (19) \citet{2007ApJ...670.1337D} (20) \citet{2008ApJ...683..844L} (21) 
\citet{1980AJ.....85..444R} (22) SB?: \citet{2006A&A...460..695T}, w*: 
\citet{2007ApJ...662..413K} (23) **,\,w*: \citet{1992PhDT.......255B,1993A&A...278...81R}, SB? (primary and secondary in 
spectrograph entrance window): \citet{2002AJ....124.2813R} (24) \citet{1982MNRAS.201.1095H} (25) \citet{1997A&A...320..525P}
\end{flushleft}
\end{table*}

%As already described in \citet{2008A&A...484..413S,2008A&A...491..311S} we observed using the European Southern Observatory 
%(ESO) Very Large Telescope (VLT) instrument Naos-Conica \citep[NACO,][]{2003SPIE.4841..944L, 2003SPIE.4839..140R}. 

\begin{figure*}
\resizebox{\hsize}{!}{\includegraphics{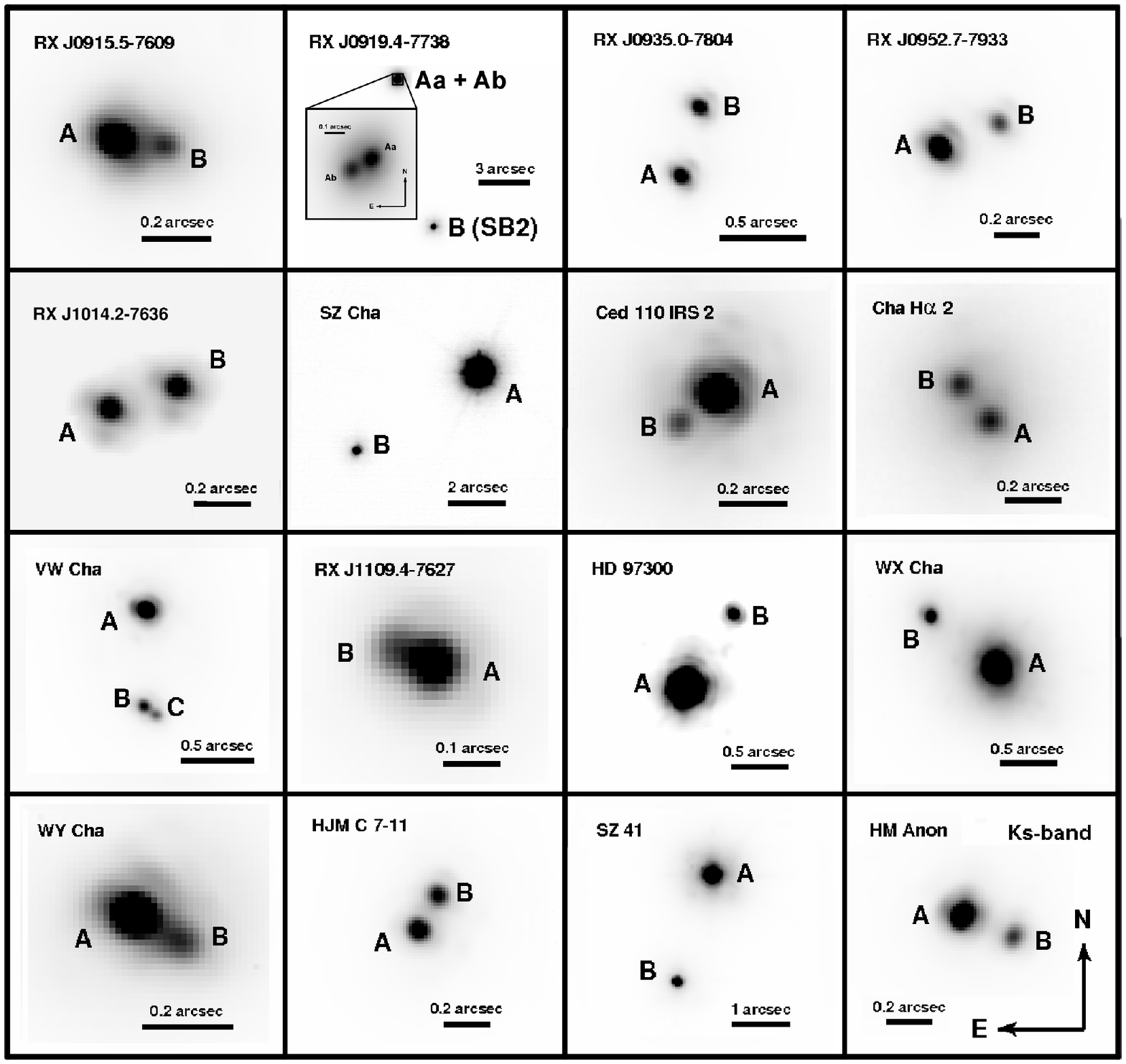}}
\caption{VLT NACO Ks band images of the binary and multiple systems. Spectroscopic binaries are only given if the binarity and
component of variation has been reliably determined. Wide stellar companion candidates, marked as w* in Table \ref{table:1}, are outside  
the FoV of our S13 observations.}
\label{Figure1}
\end{figure*}

All targeted stars with previously known companion candidates, but not yet examined to determine their possible common proper motion, 
are listed with additional information on the systems in Table \ref{table:1}. These binaries could be recovered within the campaign 
owing to their proximity to the primaries, fitting within the
NACO S13 field of view (1024\,x\,1024 pixels\,=\,13.56\,x\,13.56 arcseconds) during the search for additional fainter
companions. In addition to their positions, we list the currently
adopted system architecture and the so far known individual spectral types and distances. Finally the 2MASS 
\citep{2006AJ....131.1163S,2003tmc..book.....C} brightnesses are given and the last column of Table \ref{table:1} refers
to the expected number of fore- and background stars in each NACO field (1024\,x\,1024 pixels\,=\,13.56\,x\,13.56 arcseconds), 
according to star counts down to the 2MASS limiting magnitude (near K\,=\,16\,mag) in a cone of a radius of 
300 arcseconds around each of the targets. The mean value is 0.117 stars 
per NACO field, but the individual values vary by a factor $\sim$\,3. An overview of the current orientation and view of the 
systems is given in Fig.~\ref{Figure1}.

In Table \ref{table:2} we present observational details such as integration times (DIT), the number of combined integrations (NDIT), that is 
sometimes equivalent to the number of 
integrations within one cube (which is NDIT in the case of the cube mode), the number of individual integrations or cubes (NINT), and 
filter bands. In all cases, we used the S13 camera ($\sim$13 mas/pixel pixel scale) and the double-correlated read-out mode.

For the raw data reduction, we subtracted a mean dark from all science frames and the flatfield frames,
then divided by the normalized dark-subtracted flatfield, and subtracted the mean background using 
ESO \textit{eclipse\,/\,jitter} \citep{1997Msngr..87...19D}.

We calibrated the NACO data using the wide binary stars HIP 73357 and/or HIP 6445 
for our six epochs in 2006 -- 2011. The astrometry of these binaries were measured very accurately by the Hipparcos satellite
mission \citep{1997A&A...323L..49P}. 
However, since these measurements were performed twenty years ago, the binary has possibly undergone a large orbital motion, which now 
dominates the astrometric uncertainty (see \citet{2008A&A...484..281N} for details),
resulting in the absolute calibration given in Table~\ref{table:3}. 
Our derived pixel scale is in good agreement with earlier measurements such 
as e.g. in \citet{2005A&A...435L..13N, 2008A&A...484..281N} or \citet{2010A&A...509A..52C}.
The error bars in the pixel scale and orientation increase with time because of the increasing uncertainties in the possible orbital
motion of the calibration binaries. This systematic also dominates the error estimation in terms of the pixel scale and orientation 
as can easily be seen by the much smaller scatter among the mean values of the pixel scale, independently determined for each epoch,
than the individual uncertainties. The given orientation is relative to north in the sky and has to be added to 
calibrate any measurement done in the images.

Since in Table~\ref{table:3} a trend of increasing orientation by about 0.16 $\degr$/yr might be present, we 
checked whether this tentative trend could be due to the orbital motion of the adopted calibration binary HIP 73357. To achieve this,  
we requested all obervational data of the binary from the Washington Double Star Catalog (WDS) \citep{2001AJ....122.3466M}
from 1835 until 1998 in the version of April 4 2007 and found the binary to have an orbital motion from -0.01 to -0.04 
$\degr$/yr in position angle, neglecting two outliers, and less than -0.01 $\degr$/yr within the past 50 years. Owing to
the lack of error bars in the data, we cannot give precise numbers, although since the orbital motion seems to be 
negligible we calibrated each measurement by the orientation value given in Table~\ref{table:3}, taking the maximal 
possible orbital motion of a circular orbit according to Kepler's third law into account in the error budget.  

\begin{table*}
\caption{VLT/NACO observation log}
\label{table:2}
%\centering
\begin{tabular}{lllcccccccccc}
\hline\hline
Object          & Other name     & JD - 2453700      & Date of         & DIT       & NDIT   & Number    & Airmass & Seeing   & Filter\\
                &                & $[\mathrm{days}]$ & observation     & [s]       &        & of images &         & [arcsec] &       \\
\hline
\object{RX J0915.5-7609}
                & RX J0915.5-7608& \ \ 461.53412     & \ \ 2 Mar 2007 & 2.2        & 28     & 20        & 1.73    & 1.04   & Ks    \\
                &                & 1181.59508$^a$    & 19 Feb 2009    & 2          & 30     & 6         & 1.66    & 0.88   & Ks    \\
\object{RX J0919.4-7738}
                & HIP 45734      & \ \ 815.51911     & 19 Feb 2008    & 1.2        & 50     & 30        & 1.92    & 0.99   & Ks    \\
\object{RX J0935.0-7804}
                & 2MASS J09345604 & \ \ 461.59183    & \ \ 2 Mar 2007 & 7          & 9      & 30        & 1.71    & 1.22   & Ks    \\
                & \ \ \ \ \ \ \ \ \ \ \ \ \ \ \ \ -7804193 
                                 & \ \ 814.61953     & 18 Feb 2008    & 11         & 6      & 28        & 1.72    & 1.13   & Ks    \\
                &                & \ \ 815.58352     & 19 Feb 2008    & 30         & 2      & 38        & 1.78    & 1.05   & J     \\
                &                & 1182.58189$^a$    & 20 Feb 2009    & 3          & 20     & 30        & 1.77    & 0.73   & Ks    \\
\object{RX J0952.7-7933}
                & HD 86588       & \ \ 460.58336     & \ \ 1 Mar 2007 & 2.5/1.5    & 25/40  & 10/12     & 1.81    & 1.26   & Ks    \\
                &                & \ \ 814.64850     & 18 Feb 2008    & 3          & 20     & 13        & 1.76    & 1.01   & Ks    \\
                &                & \ \ 814.65931     & 18 Feb 2008    & 3          & 20     & 6         & 1.75    & 0.90   & J     \\
                &                & 1181.63897$^a$    & 19 Feb 2009    & 1          & 60     & 10        & 1.77    & 0.68   & Ks    \\
                &                & 1181.64617$^a$    & 19 Feb 2009    & 0.5        & 120    & 5         & 1.76    & 0.69   & H     \\
                &                & 1181.65104$^a$    & 19 Feb 2009    & 1.5        & 40     & 5         & 1.76    & 0.72   & J     \\
                &                & 1550.53614$^a$    & 23 Feb 2010    & 0.347      & 126    & 10        & 1.96    & 0.86   & Ks    \\
                &                & 1853.86365$^a$    & 23 Dec 2010    & 3          & 20     & 30        & 1.74    & 2.21   & Ks    \\
                &                & 1945.57285$^a$    & 25 Mar 2011    & 0.347      & 126    & 10        & 1.75    & 1.11   & Ks    \\
\object{RX J1014.2-7636}
                & 2MASS J10140807 & 1550.56298$^a$   & 23 Feb 2010    & 0.347      & 126    & 42        & 1.83    & 0.88   & Ks    \\
                & \ \ \ \ \ \ \ \ \ \ \ \ \ \ \ \ -7636327 
                                 & 1945.60036$^a$    & 25 Mar 2011    & 0.347      & 126    & 42        & 1.63    & 1.11   & Ks    \\
\object{SZ Cha} & Ass Cha T 2-6  & \ \ \ \ 82.64912  & 16 Feb 2006    & 1          & 50     & 20        & 1.75    & 0.73   & Ks    \\
                &                & 1181.66756$^a$    & 19 Feb 2009    & 0.5        & 120    & 22        & 1.70    & 0.65   & Ks    \\
Ced 110 IRS 2   & \object{Ass Cha T 2-21}
                                 & \ \ \ \ 82.61949  & 16 Feb 2006    & 0.3454     & 100    & 20        & 1.85    & 0.75   & Ks    \\
                &                & \ \ 815.61187     & 19 Feb 2008    & 0.3454     & 174    & 5         & 1.85    & 0.78   & Ks    \\
                &                & \ \ 815.61683     & 19 Feb 2008    & 0.5        & 120    & 5         & 1.84    & 0.80   & J     \\
Cha H$\alpha$ 2 & \object{ISO-ChaI 111}
                                 & \ \ \ \ 82.68981  & 16 Feb 2006    & 10         & 6      & 20        & 1.70    & 0.63   & Ks    \\
                &                & \ \ 460.71934     & \ \ 1 Mar 2007 & 60         & 2      & 15        & 1.66    & 1.15   & Ks    \\
                &                & \ \ 461.67171     & \ \ 2 Mar 2007 & 30         & 2      & 24        & 1.68    & 1.20   & J     \\
\object{VW Cha} & Ass Cha T 2-31 & \ \ \ \ 83.90711  & 17 Feb 2006    & 0.3454     & 100    & 20        & 1.94    & 0.49   & Ks    \\
\object{RX J1109.4-7627}
                & CHXR 37        & 1182.81186$^a$    & 20 Feb 2009    & 2          & 30     & 6         & 1.67    & 0.57   & Ks    \\
                &                & 1551.69870$^a$    & 24 Feb 2010    & 0.347      & 126    & 12        & 1.63    & 1.00   & Ks    \\
\object{HD 97300}
                & Ass Cha T 2-41 & \ \ 460.60766     & \ \ 1 Mar 2007 & 0.5        & 110    & 20        & 1.77    & 1.23   & Ks    \\
                &                & \ \ 816.71442     & 20 Feb 2008    & 0.5        & 120    & 5         & 1.63    & 0.80   & Ks    \\
                &                & \ \ 816.71936     & 20 Feb 2008    & 0.3454     & 174    & 5         & 1.63    & 0.87   & J     \\
\object{WX Cha} & Ass Cha T 2-45 & 1181.91858$^a$    & 19 Feb 2009    & 2          & 30     & 6         & 2.00    & 0.76   & Ks    \\
\object{WY Cha} & Ass Cha T 2-46 & \ \ 816.70106     & 20 Feb 2008    & 2          & 30     & 5         & 1.64    & 0.68   & Ks    \\
                &                & \ \ 816.70577     & 20 Feb 2008    & 7.5        & 8      & 5         & 1.64    & 0.73   & J     \\
\object{HJM C 7-11}
                & CHXR 47        & \ \ 461.78979     & \ \ 2 Mar 2007 & 2          & 30     & 21        & 1.71    & 1.41   & Ks    \\
                &                & \ \ 815.79485     & 19 Feb 2008    & 2          & 30     & 21        & 1.68    & 0.53   & Ks    \\
                &                & 1181.70578$^a$    & 19 Feb 2009    & 1.5        & 40     & 36        & 1.67    & 0.52   & Ks    \\
                &                & 1181.77101$^a$    & 19 Feb 2009    & 1.5        & 40     & 17        & 1.66    & 0.50   & H     \\
                &                & 1181.78711$^a$    & 19 Feb 2009    & 10         & 6      & 15        & 1.67    & 0.67   & J     \\
\object{Sz 41}  & Ass Cha T 2-51 & 1181.88546$^a$    & 19 Feb 2009    & 1.5        & 40     & 6         & 1.84    & 0.81   & Ks    \\
                &                & 1550.08623$^a$    & 23 Feb 2010    & 0.347      & 126    & 42        & 1.83    & 1.16   & Ks    \\
HM Anon         & \object{Ass Cha T 2-54}
                                 & \ \ 815.69528     & 19 Feb 2008    & 1          & 60     & 12        & 1.68    & 0.57   & Ks    \\
                &                & \ \ 815.70581     & 19 Feb 2008    & 2          & 30     & 12        & 1.67    & 0.56   & J     \\
                &                & 1182.87337$^a$    & 20 Feb 2009    & 1          & 60     & 32        & 1.84    & 0.52   & Ks    \\
\hline
\end{tabular}
\begin{flushleft}
\textbf{Remarks}: Each image consists of the number of exposures given in column 6 multiplied by the individual integration 
time given in column 5. (a) Data taken in cube mode, hence each image is a cube of the number of planes given in column 
6, each having the individual integration time given in column 5.
\end{flushleft}
\end{table*}

\begin{table}
\caption{Astrometric calibration results using the binary \object{HIP 73357}}
\label{table:3}
\centering
%\begin{tabular}{lccccccccc}
\begin{tabular}{lcccc}
\hline\hline
JD - 2453700      & Epoch        & Pixel scale & Orientation & Filter \\
$[\mathrm{days}]$ &              & [mas/Pixel] & [$\degr$]       &        \\
\hline
\ \ \ \ 88.84779  & Feb 2006     & 13.24 $\pm$ 0.18 & 0.18 $\pm$ 1.24 & Ks \\
\ \ 460.84330     & Mar 2007     & 13.24 $\pm$ 0.19 & 0.33 $\pm$ 1.32 & Ks \\
\ \ 815.91117     & Feb 2008     & 13.22 $\pm$ 0.20 & 0.73 $\pm$ 1.40 & J  \\
\ \ 815.91907     & Feb 2008     & 13.25 $\pm$ 0.20 & 0.68 $\pm$ 1.40 & Ks \\
1181.89936        & Feb 2009     & 13.25 $\pm$ 0.21 & 0.76 $\pm$ 1.48 & Ks \\
1550.82276        & Feb 2010     & 13.24 $\pm$ 0.22 & 0.72 $\pm$ 1.56 & Ks \\
1854.54334        & Dec 2010$^a$ & 13.23 $\pm$ 0.23 & 0.95 $\pm$ 1.63 & Ks \\
1945.86023        & Mar 2011     & 13.24 $\pm$ 0.24 & 1.00 $\pm$ 1.65 & Ks \\
\hline
\end{tabular}
\begin{flushleft}
\textbf{Remarks}: The values of Hipparcos \citep{1997A&A...323L..49P} were used as reference values. Measurement 
errors of Hipparcos as well as maximum possible orbital motion since the epoch of the Hipparcos observation are 
taken into account. (a) Although HIP 73357 was not observed in Dec 2010, \object{HIP 6445}, observed in Feb 2010
and Dec 2010, could be used to transfer the HIP 73357 calibration to Dec 2010.
\end{flushleft}
\end{table}

\section{Astrometric and photometric results}

\subsection{Astrometry}
\label{astrometry}

To verify the common proper motion of the tentative companions of our targets (Table \ref{table:1}), %%
we used the proper motion (PM) of the stars published in the literature (Table ~\ref{table:4}) to calculate the expected 
change in the separation and position angle of the respective components based on the hypothesis that one of the components is a 
non-moving background star.
We used either the newest values combining most of the previous data with new measurements, in this case of UCAC3
\citep{2010AJ....139.2184Z} combining data from about 140 catalogues, or the weighted mean proper motion of several
independent measurements, as in the case of the faint previously supposed brown dwarf candidate Cha H$\alpha$ 2 
\citep[see e.g.][]{2008A&A...484..413S}, whose data was in part derived by us from SuperCOSMOS Sky Survey (SSS) data 
\citep{2001MNRAS.326.1279H} and NTT/SofI and VLT/FORS measurements (see Table \ref{table:4}), to check whether
the objects have a common PM as we describe in the following sections. 
For comparison,  we give in addition the median proper motion value of the Chamaeleon I 
association derived from the UCAC2 \citep{2004AJ....127.3043Z} data by \citet{2008ApJ...675.1375L} in Table 
\ref{table:4}.

\begin{table}
\caption{Proper motions}
\label{table:4}
%\centering
\begin{center}
\begin{tabular}{llr@{\,$\pm$\,}lr@{\,$\pm$\,}l}
\hline\hline
Object & Reference             &\multicolumn{2}{c}{$\mu_{\alpha} \cos{\delta}$}    & \multicolumn{2}{c}{$\mu_{\delta}$}  \\
       &                       &\multicolumn{2}{c}{[mas/yr]}                       & \multicolumn{2}{c}{[mas/yr]}\\
\hline
RX J0915.5            \\
\ \ \ \ \ \ \ \ -7609 & UCAC 3 (1)  & -29.3     & 2.0                             & 18.9    & 1.4  \\
\hline
RX J0919.4            & Hipparcos \\
\ \ \ \ \ \ \ \ -7738 & (new) (2)   & -109.7    & 0.88                            & 72.35   & 0.8 \\
                      & Tycho-2 (3) & -82.0     & 11.3                            & 73.10   & 7.7 \\
                      & PPMX (4)    & -108.19   & 1.5                             & 68.28   & 1.6 \\
\hline
                      & used:  \\
                      & UCAC 3 (1)  & -105.3    & 1.2                             & 72.1    & 1.2 \\
\hline
RX J0935.0            \\
\ \ \ \ \ \ \ \ -7804 & UCAC 3 (1)  & -28.6     & 1.3                             & 19.2    & 1.3 \\
\hline
RX J0952.7            \\
\ \ \ \ \ \ \ \ -7933 & UCAC 3 (1)  & -13.5     & 0.9                             & 2.0     & 0.9 \\
\hline
RX J1014.2            \\
\ \ \ \ \ \ \ \ -7636 & UCAC 3 (1)  & -47.2     & 1.7                             & 30.6    & 3.6 \\
\hline
SZ Cha                & UCAC 3 (1)  & -22.5     & 6.9                             & 1.7     & 2.7 \\
\hline
Ced 110               \\
\ \ \ \ \ \ \ \ IRS 2 & UCAC 3 (1)  & -19.2     & 1.0                             & 5.0     & 1.6 \\
\hline
Cha H$\alpha$ 2 & PSSPMC (5)        & -23    & 17                                 & 1       & 17   \\
                & SSS-FORS1 (6)     & -19.4  & 14.9                               & -0.3    & 14.9 \\
                & SSS-SofI (6)      & -23.0  & 14.1                               & 7.8     & 14.1 \\
\hline
       & weighted mean                              & -21.75  & 8.77  & 3.18  & 8.77  \\
\hline
VW Cha                & UCAC 3 (1)  & -19.7  & 1.3                                & -0.8    & 3.4 \\
\hline
RX J1109.4            \\
\ \ \ \ \ \ \ \ -7627 & UCAC 3 (1)  & -22.3  & 1.8                                & -0.8    & 9.9 \\
\hline
HD 97300        & Hipparcos \\
                & (new) (2)            & -21.63  & 0.94                           & -0.72    & 0.78 \\
                & Tycho-2 (3)          & -19.1   & 1.4                            & -0.1     & 1.4  \\
                & PPMX (4)             & -21.01  & 1.4                            & -0.48    & 1.5  \\
\hline
                & used:  \\
                & UCAC 3 (1)           & -17.7   & 1.0                            & -2.7     & 1.1  \\
\hline
WX Cha          & UCAC 3 (1)           & -20.9   & 1.7                            & -0.6     & 2.3  \\
\hline
WY Cha          & ICRF \\
                & extension (7)        & -7     & 13                              & 11      & 11   \\
\hline
                & used: \\
                & UCAC 3 (1)           & -22.2  & 1.8                             & -0.5    & 1.8   \\
\hline
HJM C7-11       & UCAC 3 (1)           & -20.3  & 1.8                             & 1.7     & 1.8  \\
\hline
Sz 41           & ICRF \\
                & extension (7)        & -15    & 5                               & 9       & 4    \\
\hline
                & used: \\
                & UCAC 3 (1)           & -29.8  & 4.3                             & 4.7     & 1.7   \\
\hline
HM Anon         & Tycho-2$^a$ (3)      & -17.5  & 5.7                             & -72.3   & 5.5  \\
\hline
                & used: \\
                & UCAC 3 (1)           & -21.6  & 1.0                             & 4.8     & 1.0   \\
\hline
Median Cha I    & Luhman$^b$ (8)       & -21    & $\sim$1                         & 2     & $\sim$1 \\
\hline
\end{tabular}
\end{center}
\begin{flushleft}
\textbf{Remarks}: Only independent sources with individual error bars for the targets were taken into account. (a) As 
given in \citet{2009yCat....102023S}, the Tycho-2 coordinates are wrong and HM Anon equals \mbox{TYC 9414-1250-1} 
(b) Based on UCAC2 proper motions from (9). \\
\textbf{References}:
(1) \citet{2010AJ....139.2184Z} (2) \citet{2007A&A...474..653V} (3) \citet{2000A&A...355L..27H}
 (4) \citet{2008A&A...488..401R} (5) \citet{2005A&A...438..769D} (6) \citet{2008A&A...484..413S}; 
\citet{2001MNRAS.326.1279H} (7) \citet{2003A&A...409..361C} (8) \citet{2008ApJ...675.1375L} (9) 
\citet{2004AJ....127.3043Z}
\end{flushleft}
\end{table}

To determine the positions of both components, we constructed a reference point spread function (PSF) from both objects. 
Thus, we obtained an 
appropriate reference PSF for each single image. Using IDL/starfinder \citep{2000SPIE.4007..879D}, we scaled and shifted 
the reference PSF simultaneously to both components in each of our individual images by minimizing the residuals. 
Realistic error estimates in both the position and flux of each object were obtained from the mean and standard deviation in the 
positions found in all individual images of a single epoch. When  the objects are either too close or their
brightness difference is too strong, we subtracted the PSF of each component using an IDL rotation
routine \citep[described in more detail in][]{2008A&A...484..281N} before measuring the astrometry and photometry of the
other component, respectively.

As already discussed in the previous section, we used the Hipparcos binary HIP 73357 to calibrate our images. Hence, 
the uncertainties in the absolute astrometric results, given in Table \ref{table:5}, include the 
uncertainties in the Hipparcos astrometry, the maximum possible orbital motion of the calibration binary and the measurement 
errors in the position of the targets and their companions.
%Hence, the absolute astrometric results, given in Table \ref{table:5}, include the astrometric Hipparcos errors, maximum 
%possible orbital motion of the calibration binary (Table \ref{table:3}) and measurement errors of the position of the 
%targets and their companions. 
In Table \ref{table:5}, we list the separation between the components ($\rho$) with its uncertainty
($\delta_{\rho}$), the position angle ($PA$) with its uncertainty ($\delta_{PA}$), as well as the  of a 
measurement at epoch $i$ not being a background object ($\sigma_{\rho,\,\mathrm{back},\,i}$ and
$\sigma_{PA,\,\mathrm{back},\,i}$).  We calculated $\sigma_{\rho,\,\mathrm{back},\,i}$ and
$\sigma_{PA,\,\mathrm{back},\,i}$ from the measured difference in the separation and position angle of the components 
between epoch $i$ and the corresponding expected values at the Julian date of epoch $i$ in the case of the fainter
component being a non-moving background object by extrapolating the known proper motions of the primaries 
(Table~\ref{table:4}) with respect to the (latest) reference epoch with index 0, via
\begin{equation}
\sigma_{\rho,\,\mathrm{back},\,i} = \frac{\rho_i - \rho_{\,\mathrm{back}_{\,i}}(\rho_0)}{\sqrt{\delta_{\rho_i}^2 + 
\delta^2_{\rho_{\,\mathrm{back}_i}}(\delta_{\rho_0})}} \ ,
\end{equation}
\begin{equation}
\sigma_{PA,\,\mathrm{back},\,i} = \frac{PA_i - PA_{\,\mathrm{back}_{\,i}}(PA_0)}{\sqrt{\delta_{PA_i}^2 + 
\delta^2_{PA_{\,\mathrm{back}_i}}(\delta_{PA_0})}} \ ,
\end{equation}
where for instance $\rho_{\,\mathrm{back}_{\,i}}(\rho_0)$ would be the expected separation of the components if
the fainter of the two were a background object. The parameter $\delta_{\,\mathrm{back}_{\,i}}(\delta_0)$ is 
the associated uncertainty in that value, which is essentially the width of the background cone at epoch $i$ relative to the 
(latest) reference measurement 0 taking into account the proper motion uncertainty, the measurement uncertainty at the reference
epoch, and the brightness difference between the components.
The last quantity influences the measured proper motion of the combined light.

Finally, the likelihood that a measurement $i$ represents orbital motion is calculated similarly for each epoch 
as a measure of the deviations in the separation and position angle from the reference epoch with index 0, via
\begin{equation}
\sigma_{\rho,\,\mathrm{orb},\,i} = \frac{\rho_i - \rho_0}{\sqrt{\delta_{\rho_i}^2 + \delta_{\rho_0}^2}} \ ,
\end{equation}
\begin{equation}
\sigma_{PA,\,\mathrm{orb},\,i} = \frac{PA_i - PA_0}{\sqrt{\delta_{PA_i}^2 + \delta_{PA_0}^2}} \ .
\end{equation}
All values are given in Table \ref{table:5}.

%Our common proper motion analysis made moreover use of relative measurements or relative astrometry, hence, the 
%astrometric calibration used for our images took into account only the uncertainties 
%in separation and position angle, as well as in proper motion and orbital motion between the epochs (2006 -- 2011) of 
%interest.
%Possible maximal orbital motion in the calibration binary was calculated only for 2006 -- 2011 (or less), so that the
%errors in relative astrometry are smaller, but no absolute reference for separation and PA can be given.

In addition, we performed relative astrometric measurements between the components for our common proper motion 
analysis, which allows a more precise determination of the common proper motion and orbital motion for measurements taken 
with the same instrument.
Relative astrometric measurements deal only with the changes in the separation and position angles between the observed 
epochs, hence relaxes the constraints on the astrometric calibration. 
The calibration takes into account only the uncertainties in separation, position angle, and proper motion of the target, as well as 
the uncertainties of the separation, position angle, and possible orbital motion of the calibration binary between the NACO 
epochs (here 2006 -- 2011). The possible maximal orbital motion of the calibration
binary was calculated only for 2006 -- 2011 (or less), hence the final errors in the relative astrometric measurements for 
our targets are smaller and changes in these values can be recovered with higher significance, although no absolute reference for 
separation and PA can be given. We refer to \citet{2008A&A...484..281N} for a further discussion
of this concept.
This allows precise measurement of the relative motions of the targets and the companion candidates given in 
Table~\ref{table:6}. Analogues to the formulas given in the last paragraph, the significances for not being a background 
object and for orbital
motion from relative astrometric measurements are given in Table~\ref{table:6}, calculated using the time 
difference ($\Delta$t) between
measurements as well as the change in separation ($\Delta\,\rho$) and position angle ($\Delta\,PA$) and their uncertainties.  
The absolute astrometric measurements between the components (Table~\ref{table:5}) must instead incorporate the full 
uncertainties of the astrometric calibration (given in Table~\ref{table:3}). We emphasize that common proper motion can mostly 
be shown much more precisely based on relative 
astrometric measurements, while absolute values are given for future comparisons, also with different instruments.

\begin{table*}
\caption{Relative astrometric results}
\label{table:6}
%\centering
\begin{tabular}{llr@{\,$\pm$\,}lccr@{\,$\pm$\,}lcc}
\hline\hline
Object & Epoch differ- & \multicolumn{2}{c}{Change in sepa-} & Sign.$^{a,c}$ & Sign.$^c$ orb. & \multicolumn{2}{c}{Change in}        & Sign.$^{a,c}$  & Sign.$^c$ orb.\\
       & ence [days]   & \multicolumn{2}{c}{ration [pixel]}  & not Backg.    & motion         & \multicolumn{2}{c}{PA$^b$ [$\degr$]} & not Backg.     & motion        \\
       & \ \ \ \ \ \ \ \ $\Delta$t     
            & $\Delta\,\rho$ & $\delta_{\Delta\,\rho}$ & $\sigma_{\rho,\,\mathrm{back}}$ & $\sigma_{\rho,\,\mathrm{orb}}$  
            & $\Delta\,PA$ & $\delta_{\Delta\,PA}$ & $\sigma_{PA,\,\mathrm{back}}$ & $\sigma_{PA,\,\mathrm{orb}}$   \\
\hline
RX J0915.5-7609 AB & \ \ 720.06096 & -0.784    & 0.417 & 7.9     & 1.9  & -3.476    & 1.330 & 9.3     & 2.6 \\
RX J0935.0-7804 AB & \ \ 720.99006 & \ \ 0.564 & 0.096 & 21      & 5.9  & -1.210    & 0.229 & 17      & 5.3 \\
                   & \ \ 367.48036 & \ \ 0.317 & 0.077 & 18      & 4.1  & -0.557    & 0.156 & 16      & 3.6 \\
RXJ0952.7-7933 AB  & 1484.98949    & -0.192    & 0.128 & 13      & 1.5  & -7.690    & 0.460 & 10      & 17  \\
                   & 1130.91894    & -0.087    & 0.086 & 14      & 1.0  & -6.586    & 0.285 & 11      & 23  \\
                   & \ \ 763.93388 & \ \ 0.064 & 0.132 & 12      & 0.5  & -4.405    & 0.288 & 9.4     & 15  \\
                   & \ \ 395.03671 & \ \ 0.161 & 0.039 & 15      & 4.1  & -2.199    & 0.115 & 7.2     & 19  \\
                & \ \ \ \ 91.70920 & \ \ 0.055 & 0.039 & 8.3     & 1.4  & -0.394    & 0.080 & 3.4     & 4.9 \\
RX J1014.2-7636 AB & \ \ 395.03738 & \ \ 0.889 & 0.045 & 27      & 20   & -2.163    & 0.184 & 2.3     & 12  \\
SZ Cha AB          & 1099.01844    & \ \ 0.289 & 1.050 & 2.5     & 0.3  & -0.006    & 0.250 & 1.2     & 0.0 \\
Ced 110 IRS 2 AB   & \ \ 732.99238 & -0.404    & 0.281 & 9.8     & 1.4  & -6.244    & 1.466 & 0.4     & 4.3 \\
Cha H$\alpha$ 2 AB & \ \ 378.02953 & -0.573    & 0.188 & 1.5     & 3.1  & \ \ 0.375 & 0.875 & 2.1     & 0.4 \\
RXJ1109.4-7627  AB & 1431.18790    & -0.526    & 0.190 & 0.8     & 2.8  & -23.727   & 1.292 & 3.4     & 18  \\
                   & \ \ 368.88684 & -0.422    & 0.195 & 3.1     & 2.2  & -8.216    & 1.395 & 2,1     & 5.9 \\
HD 97300 AB        & \ \ 695.81337 & -0.271    & 0.123 & 4.2     & 2.2  & -0.505    & 0.177 & 12      & 2.9 \\
                   & \ \ 356.10676 & -0.064    & 0.078 & 4.0     & 0.8  & -0.419    & 0.096 & 12      & 4.3 \\
WX Cha AB          & 1062.02756    & \ \ 0.028 & 0.303 & 6.8     & 0.1  & \ \ 0.333 & 0.375 & 4.4     & 0.9 \\
WY Cha AB          & \ \ 697.06258 & -0.106    & 0.214 & 8.7     & 0.5  & -0.962    & 0.819 & 4.2     & 1.2 \\
HJM C 7-11 AB      & 1062.12708    & -0.208    & 0.258 & 7.0     & 0.8  & -2.925    & 0.784 & 10      & 3.7 \\
                   & \ \ 719.91599 & -0.224    & 0.275 & 5.9     & 0.8  & -1.177    & 1.229 & 7.9     & 1.0 \\
                   & \ \ 365.91093 & -0.033    & 0.149 & 5.5     & 0.2  & -0.783    & 0.676 & 8.1     & 1.2 \\
Sz 41 AB           & 1430.54160    & \ \ 0.359 & 0.529 & 4.6     & 0.7  & \ \ 0.125 & 0.328 & 5.7     & 0.4 \\
                   & \ \ 368.20077 & \ \ 0.056 & 0.154 & 4.3     & 0.4  & \ \ 0.177 & 0.089 & 6.5     & 2.0 \\
HM Anon AB         & 1062.00560    & -0.378    & 0.059 & 16      & 6.4  & 1.248     & 0.257 & 13      & 4.8 \\
                   & \ \ 648.39052 & -0.271    & 0.041 & 15      & 6.6  & 0.679     & 0.174 & 11      & 3.9 \\
                   & \ \ 367.17809 & -0.156    & 0.035 & 14      & 4.5  & 0.356     & 0.130 & 9.0     & 2.7 \\
\hline                                                   
\end{tabular}                                            
\begin{flushleft}                                        
\textbf{Remarks}: (a) Assuming the fainter component is a non-moving background star. (b) PA is measured from N over E to S. (c) Significances are given relative to the last epoch.
%(d) Results of component A relative to the center of brightness of components B and C. (e) Assuming the optically fainter component is a non-moving background star.
\end{flushleft}
\end{table*}

Finally, we performed a linear orbital movement analysis for the absolute and relative astrometric results given
in Tables \ref{table:5} \& \ref{table:6} in order to check whether the first indications of orbital motion of the systems can
already be found after a few years of observations, since the orbital periods of the projected orbital separations
(for circular orbits) is beyond
15 years in all cases, mostly beyond or significantly beyond 50 years. In Tables \ref{table:7} \& \ref{table:8} we provide 
the results of these analysis as well as the projected spatial separations, 
according to the minimal observed separation in Table \ref{table:5} and the adopted distance given in Table \ref{table:1}. 
In addition, the only comparison value found by us in the literature for
the star HM Anon is listed there, after conversion of the values from km/s to mas/yr using the distance to
Cha I of 160 pc as assumed in the source paper of \citet{2001A&A...369..249W} and originally given in 
\citet{1998MNRAS.301L..39W}.

\begin{table}
\caption{Projected orbital separations and linear orbital movement \newline fit results from absolute astrometric 
measurements}
\label{table:7}
%\centering
\begin{center}
\begin{tabular}{lcr@{\,$\pm$\,}lr@{\,$\pm$\,}l}
\hline\hline
Object & proj.  & \multicolumn{2}{c}{Change in  } & \multicolumn{2}{c}{Change     }  \\
       & sep.   & \multicolumn{2}{c}{separation} & \multicolumn{2}{c}{in PA$^a$   }  \\
       & [AU]   & \multicolumn{2}{c}{[mas/yr]  } & \multicolumn{2}{c}{[$\degr$/yr]}  \\
\hline
RX J0915.5-7609 AB    & \  \ 25   & 2.77$^b$  & 0.64 & -2.58     & 0.37 \\
RX J0919.4-7738                                                   \\
 \ \ \ \ \ \ \ \ \ \ \ \ \ \ \ \
\ \ \ \   (AaAb)B$^c$ & 511       & -2.97     & 5.64 & 0.07      & 0.06 \\
 \ \ \ \ \ \ \ \ \ \ \ \ \ \ \ \ \ \ 
\ \ \ \ \ \ \ AaAb    & \ \ \ \ 6 & -0.17     & 0.39 & -4.10     & 0.20 \\
RX J0935.0-7804 AB    & \ \ 69    & 4.61      & 0.37 & -0.65     & 0.06 \\
RX J0952.7-7933 AB    & \ \ 47    & 1.14$^b$  & 0.26 & -1.65$^b$ & 0.06 \\
RX J1014.2-7636 AB    & \ \ 41    & 10.67     & 0.53 &  1.94$^b$ & 0.45 \\
SZ Cha AB             & 845       & 0.26      & 2.44 &  0.01     & 0.07 \\
Ced 110 IRS 2 AB      & \ \ 24    & -2.18     & 2.32 & -2.95     & 1.15 \\
Cha H$\alpha$ 2 AB    & \ \ 35    & -4.09     & 2.20 &  0.57     & 0.83 \\
VW Cha A(BC)$^d$      & 120       & -0.98     & 0.57 &  0.27     & 0.07 \\
\ \ \ \ \ \ \ \ \ \ \ \
\ \ \ \ \ \ \ \ BC    & \ \ 19    & -3.59$^b$ & 0.42 & -0.57     & 0.26 \\
RX J1109.4-7627 AB    & \ \ 13    & -1.34     & 0.69 & -5.82     & 0.55 \\
HD 97300 AB           & 139       & -4.10     & 2.96 & -0.02     & 0.16 \\
WX Cha AB             & 130       & -2.48     & 2.70 & -0.19     & 0.16 \\
WY Cha AB             & \ \ 20    & -0.73     & 2.00 & -0.50     & 1.07 \\
HJM C7-11 AB          & \ \ 29    & -1.11$^e$ & 1.63 & -0.39     & 0.33 \\
Sz 41 AB              & 326       & -0.97     & 2.28 & -0.03     & 0.10 \\
HM Anon AB            & \ \ 45    & -1.77     & 1.35 &  0.75     & 0.33 \\
%\hline
%HM Anon AB$^f$ \ \ (1)& \ \ 45    & -3.49     & 9.80 &  4.08     & 3.96 \\
\hline
\end{tabular}
\end{center}
\begin{flushleft}
\textbf{Remarks}: (a) PA is measured from N over E to S (b) Possible indications present for curved orbital motion
(c) See (d) in Table \ref{table:5} (d) See (e) in Table  \ref{table:9} (e) Not using the data point by 
\citet{1997ApJ...481..378G} given in Table \ref{table:5} 
%(f) Values converted from km/s to mas/yr using the distance to
%Cha I of 160 pc as assumed in (1) and given in (2) and from km/s to $\degr$/yr assuming the same distance and a separation
%of the components of 245 mas \\
%\textbf{References}: (1) \citet{2001A&A...369..249W} (2) \citet{1998MNRAS.301L..39W}
\end{flushleft}
\end{table}

\begin{table}
\caption{Projected orbital separations and linear orbital movement \newline fit results from relative astrometric
measurements}
\label{table:8}
%\centering
\begin{center}
\begin{tabular}{lcr@{\,$\pm$\,}lr@{\,$\pm$\,}l}
\hline\hline
Object & proj. & \multicolumn{2}{c}{Change in    } & \multicolumn{2}{c}{Change     }  \\
       & sep.  & \multicolumn{2}{c}{separation  } & \multicolumn{2}{c}{in PA$^b$   }  \\
       & [AU]  & \multicolumn{2}{c}{[mas/yr]$^a$} & \multicolumn{2}{c}{[$\degr$/yr]}  \\
\hline
RX J0915.5-7609 AB & \ \ 25    & -5.26     & 2.80 & -1.76     & 0.67 \\
RX J0935.0-7804 AB & \ \ 69    &  3.88     & 0.59 & -0.59     & 0.10 \\
RX J0952.7-7933 AB & \ \ 47    &  0.02$^c$ & 0.25 & -2.04$^c$ & 0.06 \\
RX J1014.2-7636 AB & \ \ 41    & 10.88     & 0.56 & -2.00     & 0.17 \\
SZ Cha AB          & 845       &  1.27     & 4.62 &  0.00     & 0.08 \\
Ced 110 IRS 2 AB   & \ \ 24    & -2.66     & 1.85 & -3.11     & 0.73 \\
Cha H$\alpha$ 2 AB & \ \ 35    & -7.35     & 2.40 &  0.36     & 0.85 \\
RX J1109.4-7627 AB & \ \ 13    & -1.31     & 0.58 & -5.76     & 0.28 \\
HD 97300 AB        & 139       & -1.52     & 0.75 & -0.34     & 0.07 \\
WX Cha AB          & 130       &  0.13     & 1.38 &  0.11     & 0.13 \\
WY Cha AB          & \ \ 20    & -0.73     & 1.48 & -0.50     & 0.43 \\
HJM C7-11 AB       & \ \ 29    & -0.98     & 0.94 & -0.93     & 0.24 \\
Sz 41 AB           & 326       &  1.00     & 1.36 &  0.10     & 0.06 \\
HM Anon AB         & \ \ 45    & -1.86     & 0.21 &  0.40     & 0.07 \\
\hline
\end{tabular}
\end{center}
\begin{flushleft}
\textbf{Remarks}: (a) Using a nominal pixel scale of 0.01324 arcsec/pixel to convert from pixel to mas (b) PA is measured 
from N over E to S. (c) Possible indications present for curved orbital motion.
\end{flushleft}
\end{table}

\subsection{Photometry}

As described in section \ref{astrometry}, from the PSF fitting of both components we also obtained their flux ratio, which is 
given for each pair in Table \ref{table:9}. Owing to the lack of photometric conditions or photometric calibrators in different
nights, we are unable to provide the individual brightnesses of the objects. However, we calculated the mean brightness of each
individual object by assuming the combined brightness of the objects, measured by 2MASS \citep{2006AJ....131.1163S} in
one epoch, and dividing the brightness according to our measured flux ratios in each band. However, we should remember 
that all of the objects are young and thus likely variable, so that we give these values in Table \ref{table:9} with 
reservations as a preliminary orientation for the reader only and hence, without error bars.

\begin{table*}
\caption{Measured brightness differences and mean apparent magnitudes}
\label{table:9}
\centering
\begin{tabular}{llccc|cccc}
\hline\hline
Object & Epoch & J-band & H-band & Ks-band & Ob-    & J-band & H-band & Ks-band \\
       &       & [mag]  & [mag]  & [mag]   & ject   & [mag]  & [mag]  & [mag]   \\
\hline   
RX J0915.5-7609 A-B & \ \ 2 Mar 2007  &                 && 1.501 $\pm$ 0.023 & A  &              && \ \ 8.736 \\
                    & 19 Feb 2009     &                 && 1.448 $\pm$ 0.081 & B  &              && 10.212    \\ 
RX J0919.4-7738 Aa-B & 19 Feb 2008    &                 && 0.796 $\pm$ 0.130 & Aa &              && \ \ 8.236 \\
\ \ \ \ \ \ \ \ \ \ \ \ \ \ \ \ \ \ \ \ \ \ \ \ \ \ \ \ \ 
               Ab-B  &                &                 && 1.141 $\pm$ 0.114 & Ab &              && \ \ 8.581 \\
\ \ \ \ \ \ \ \ \ \ \ \ \ \ \ \ \ \ \ \ \ \ \ \ \ \ \
               Aa-Ab &                &                 && 0.344 $\pm$ 0.043 & B*$^{,d}$ &       && \ \ 7.440 \\
RX J0935.0-7804 A-B & \ \ 2 Mar 2007  &                 && 0.113 $\pm$ 0.021 & A  & 10.482       && \ \ 9.587 \\ 
                    & 18/19 Feb\,2008 & 0.133 $\pm$ 0.009 && 0.100 $\pm$ 0.118 & B & 10.615      && \ \ 9.701 \\
                    & 20 Feb 2009     &                 && 0.130 $\pm$ 0.015 &    &              &&           \\
RX J0952.7-7933 A-B & \ \ 1 Mar 2007  &                 && 1.575 $\pm$ 0.055 & A  & \ \ 8.580 & \ \ 8.307 & \ \ 8.244 \\
                    & 18 Feb 2008   & 1.541 $\pm$ 0.017 && 1.647 $\pm$ 0.148 & B  & 10.127    & \ \ 9.944 & \ \ 9.718 \\
                    & 19 Feb 2009    & 1.553 $\pm$ 0.020 & 1.637 $\pm$ 0.256 & 1.486 $\pm$ 0.059 &&           \\
                    & 23 Feb 2010     &                 && 1.411 $\pm$ 0.014 &    &              &&           \\
                    & 23 Dec 2010     &                 && 1.365 $\pm$ 0.017 &    &              &&           \\
                    & 25 Mar 2011     &                 && 1.361 $\pm$ 0.014 &    &              &&           \\
RX J1014.2-7636 A-B & 23 Feb 2010     &                 && 0.073 $\pm$ 0.014 & A  &              && \ \ 9.598 \\
                    & 25 Mar 2011     &                 && 0.045 $\pm$ 0.010 & B  &              && \ \ 9.656 \\
SZ Cha A-B          & 16 Feb 2006     &                 && 3.518 $\pm$ 0.050 & A* &              && \ \ 7.758 \\
                    & 19 Feb 2009     &                 && 3.406 $\pm$ 0.031 & B  &              && 11.220    \\
Ced 110 IRS 2 A-B   & 16 Feb 2006 &                     && 2.358 $\pm$ 0.060 & A  & \ \ 7.689    && \ \ 6.529 \\
                    & 19 Feb 2008 & 2.628 $\pm$ 0.011   && 2.507 $\pm$ 0.007 & B  & 10.318       && \ \ 8.962 \\
Cha H$\alpha$ 2 A-B & 25\,Mar\,2005$^{a,b}$ &           && 0.085 $\pm$ 0.028 & A  & 12.906       && 11.374    \\
                    & 16 Feb 2006           &           && 0.111 $\pm$ 0.032 & B  & 13.023       && 11.484    \\
                    & 1/2 Mar\,2007         & 0.116 $\pm$ 0.048 && 0.135 $\pm$ 0.065 &           &&           \\
VW Cha A-B          & 17 Feb 2006           &           && 1.663 $\pm$ 0.007 & A  &              && \ \ 7.280 \\
\ \ \ \ \ \ \ \ \ \ \ \ \ \ \ 
                A-C &                       &           && 2.268 $\pm$ 0.006 & B  &              && \ \ 8.943 \\
\ \ \ \ \ \ \ \ \ \ \ \ \ \ \ 
                B-C &                       &           && 0.605 $\pm$ 0.005 & C  &              && \ \ 9.548 \\
RX J1109.4-7627 A-B & 26\,Mar\,2006$^c$ && 1.174 $\pm$ 0.008 & 1.089 $\pm$ 0.010 & A && \ \ 9.361 & \ \ 9.033 \\ 
                    &  20 Feb 2009    &                 && 1.175 $\pm$ 0.005 & B  &   & 10.535    & 10.149    \\
                    & 24 Feb 2010     &                 && 1.084 $\pm$ 0.020 &    &              &&           \\
HD 97300 A-B        & 26\,Mar\,2006$^c$ && 3.463 $\pm$ 0.054 & 3.119 $\pm$ 0.017 & A & \ \ 7.673 & \ \ 7.391 & \ \ 7.215 \\
                    & \ \ 1 Mar 2007    &&                   & 2.946 $\pm$ 0.010 & B & 11.478    & 10.854    & 10.224    \\
                    & 20 Feb 2008       & 3.805 $\pm$ 0.016 && 2.962 $\pm$ 0.023 &   &           &&           \\
WX Cha A-B          & 25\,Mar\,2006$^c$ && 2.276 $\pm$ 0.034 & 2.678 $\pm$ 0.023 & A && \ \ 8.894 & \ \ 8.073 \\
                    & 19 Feb 2009       &               && 2.356 $\pm$ 0.020 & B  &   & 11.170    & 10.590    \\
WY Cha A-B          & 25\,Mar\,2006$^c$ && 1.554 $\pm$ 0.058 & 1.773 $\pm$ 0.018 & A & 10.205 & \ \ 9.197 & \ \ 8.678 \\
                    & 20 Feb 2008       & 1.268 $\pm$ 0.004 && 1.415 $\pm$ 0.004 & B & 11.473 & 10.751    & 10.272    \\
HJM C7-11 A-B       & 25\,Mar\,2006$^c$ && 0.490 $\pm$ 0.018 & 0.520 $\pm$ 0.029 & A & 10.280 & \ \ 9.241 & \ \ 8.813 \\
                    & \ \ 2 Mar 2007    &               && 0.596 $\pm$ 0.058     & B & 10.759 & \ \ 9.684 & \ \ 9.303 \\
                    & 19 Feb 2008       &               && 0.472 $\pm$ 0.027 &   &               &&           \\
                    & 19 Feb 2009       & 0.479 $\pm$ 0.004 & 0.397 $\pm$ 0.024 & 0.373 $\pm$ 0.018 &   && &  \\
Sz 41 A-B           & 25\,Mar\,2006$^c$ && 2.571 $\pm$ 0.116 & 2.351 $\pm$ 0.039 & A  && \ \ 8.622  & \ \ 8.137 \\
                    & 19 Feb 2009       &               && 1.999 $\pm$ 0.008     & B  && 11.193     & 10.315    \\
                    & 23 Feb 2010       &               && 2.185 $\pm$ 0.028     &    &          &&           \\
HM Anon A-B         & 26\,Mar\,2006$^c$ &               && 1.486 $\pm$ 0.012 & A & \ \ 8.853 & \ \ 8.285 & \ \ 8.137 \\
                    & 13\,May\,2007$^c$ & 1.763 $\pm$ 0.023 & 1.500 $\pm$ 0.009 & 1.443 $\pm$ 0.013 & B & 10.542 & \ \ 9.786 & \ \ 9.572 \\
                    & 19 Feb 2008       & 1.615 $\pm$ 0.019 && 1.386 $\pm$ 0.006 &    &          &&           \\
                    & 20 Feb 2009       &               && 1.426 $\pm$ 0.017     &    &          &&           \\
\hline
\end{tabular}
\begin{flushleft}
\textbf{Remarks}: Mean apparent magnitudes based on combined brightness measurements of 2MASS \citep{2006AJ....131.1163S}. In
case of resolved measurements by 2MASS the component for which the brightness was used is marked by a *. (a) 
\citet{2008A&A...484..413S}, improved \\ (b) see also \citet{2007ApJ...671.2074A} and \citet{2008ApJ...683..844L} 
%(c) Optically brighter component 
(c) Rereduced, see also \citet{2008ApJ...683..844L} (d) Aa+Ab are slightly brighter than B, being itself an unresolved
spectroscopic binary (SB2)
\end{flushleft}
\end{table*}

\section{Description of the individual targets}

\subsection{The Proper Motion Diagram (PMD)}
\begin{figure*}
\includegraphics[width=0.49\textwidth]{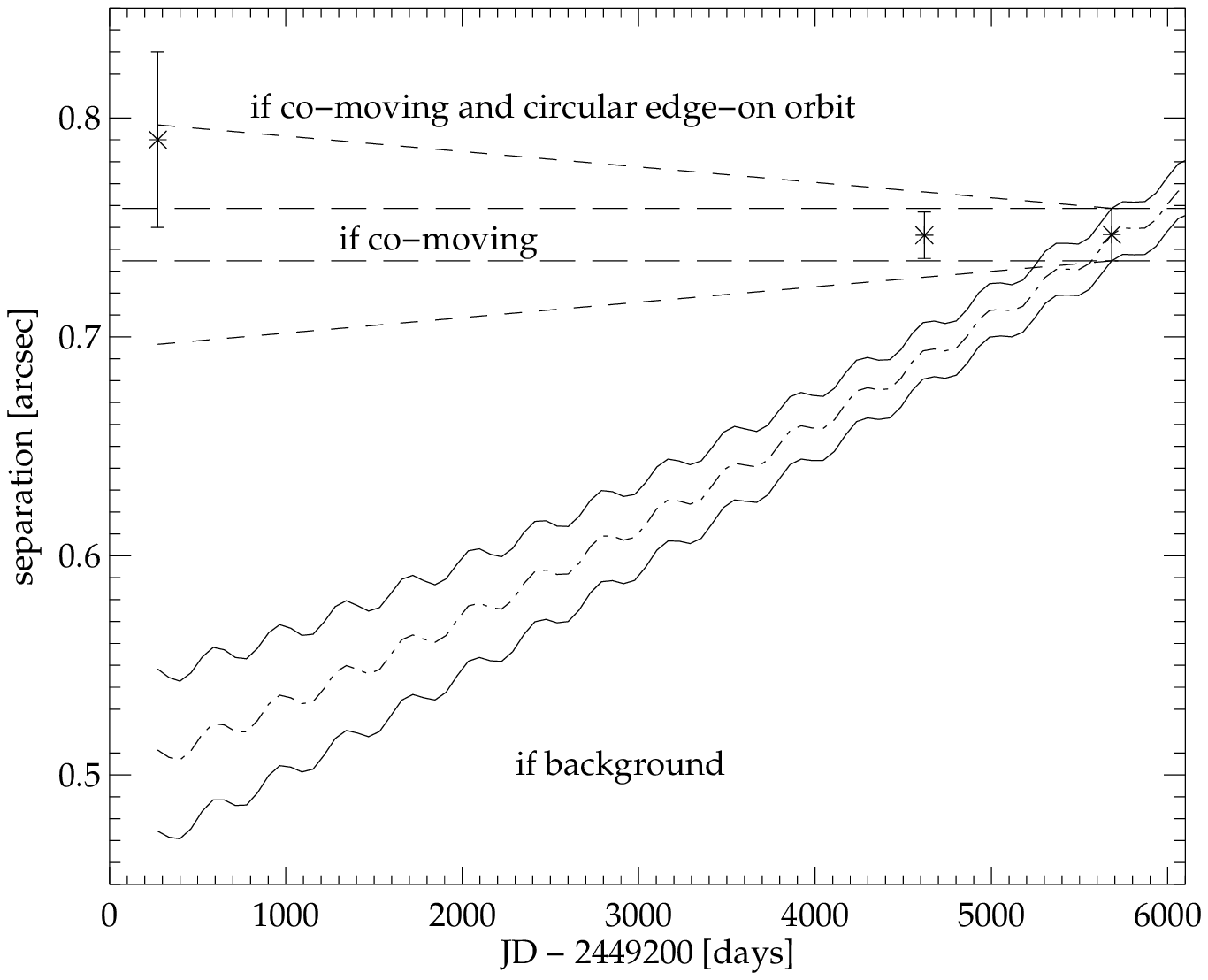}
\includegraphics[width=0.49\textwidth]{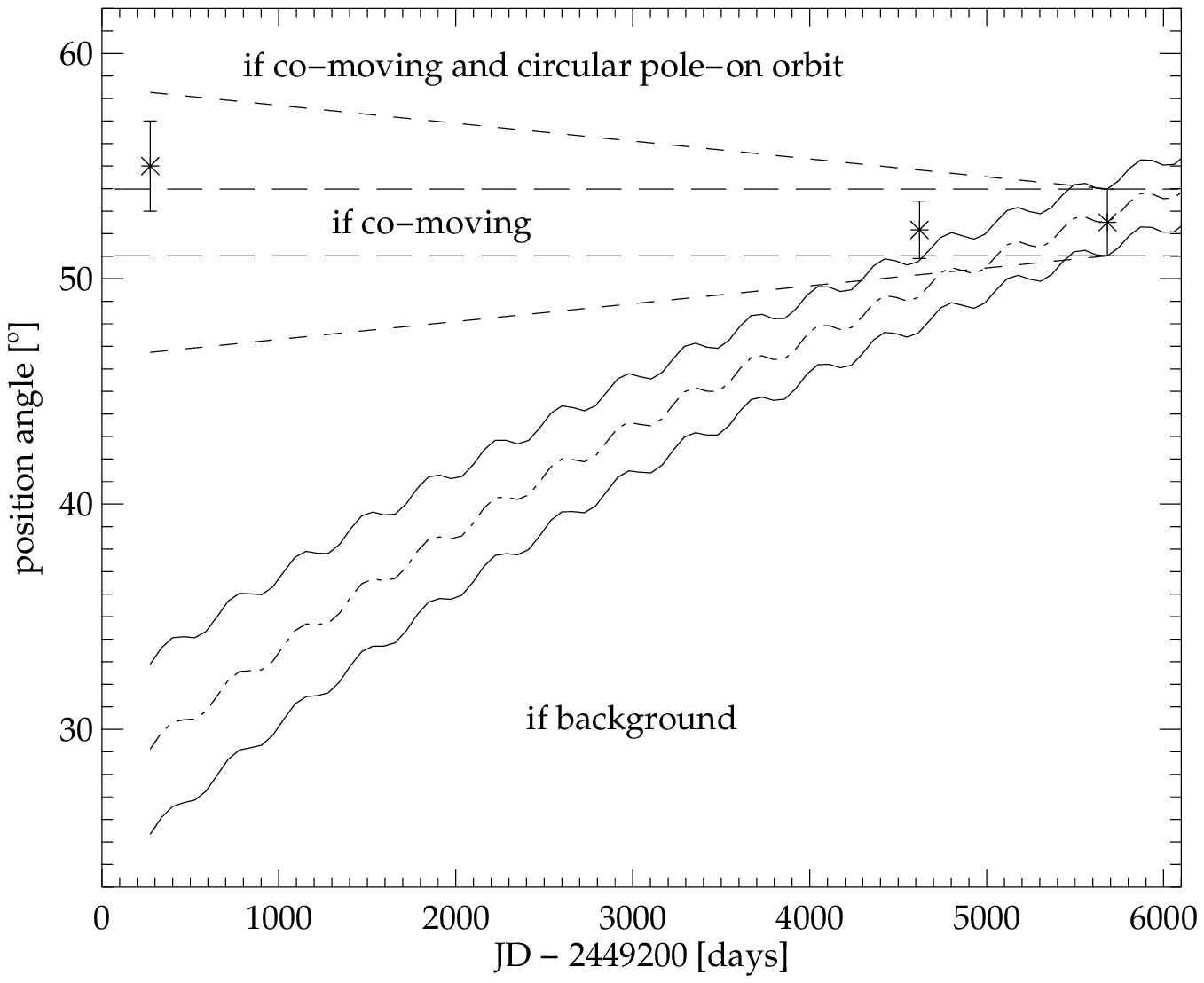}
\includegraphics[width=0.49\textwidth]{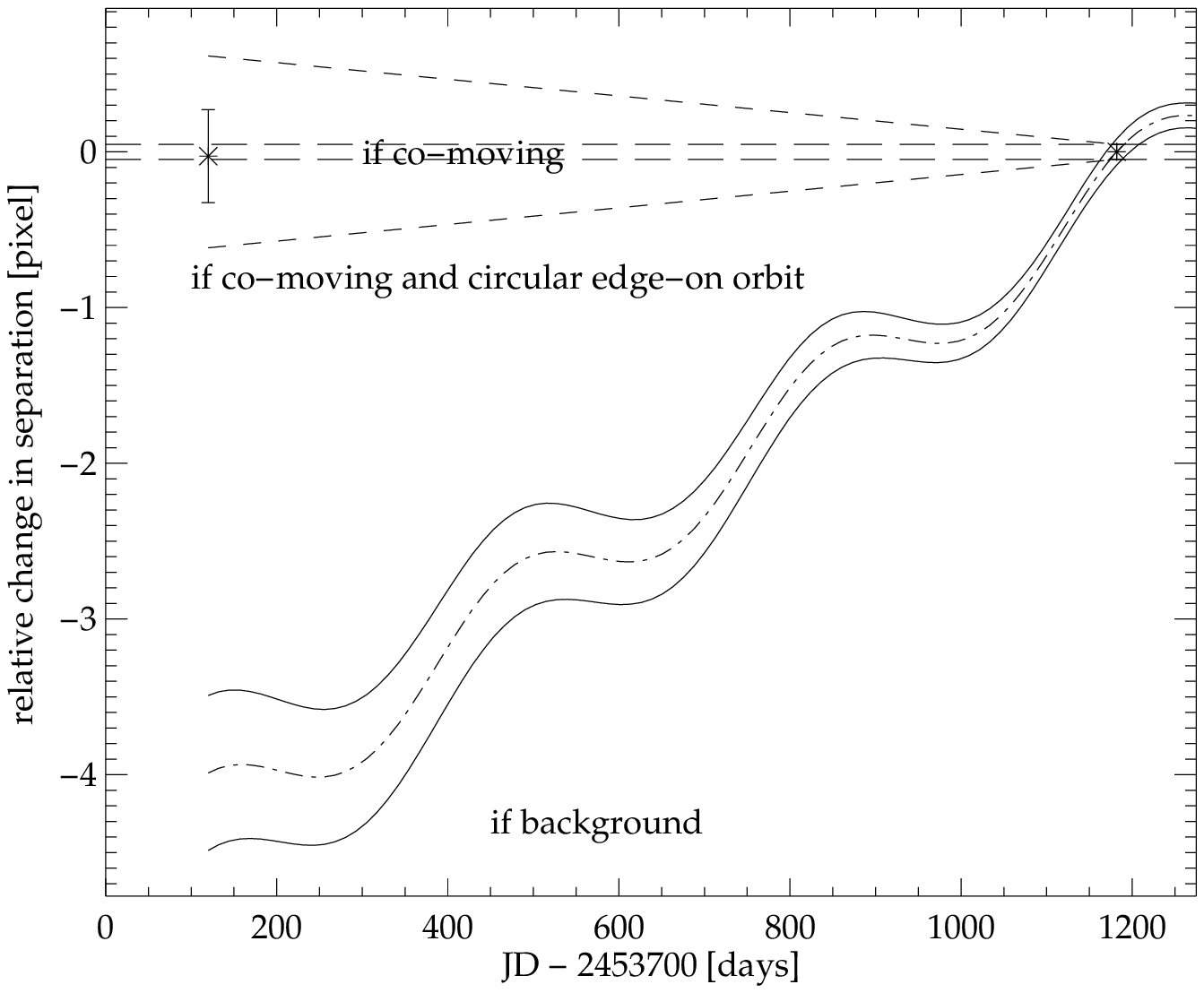}
\includegraphics[width=0.49\textwidth]{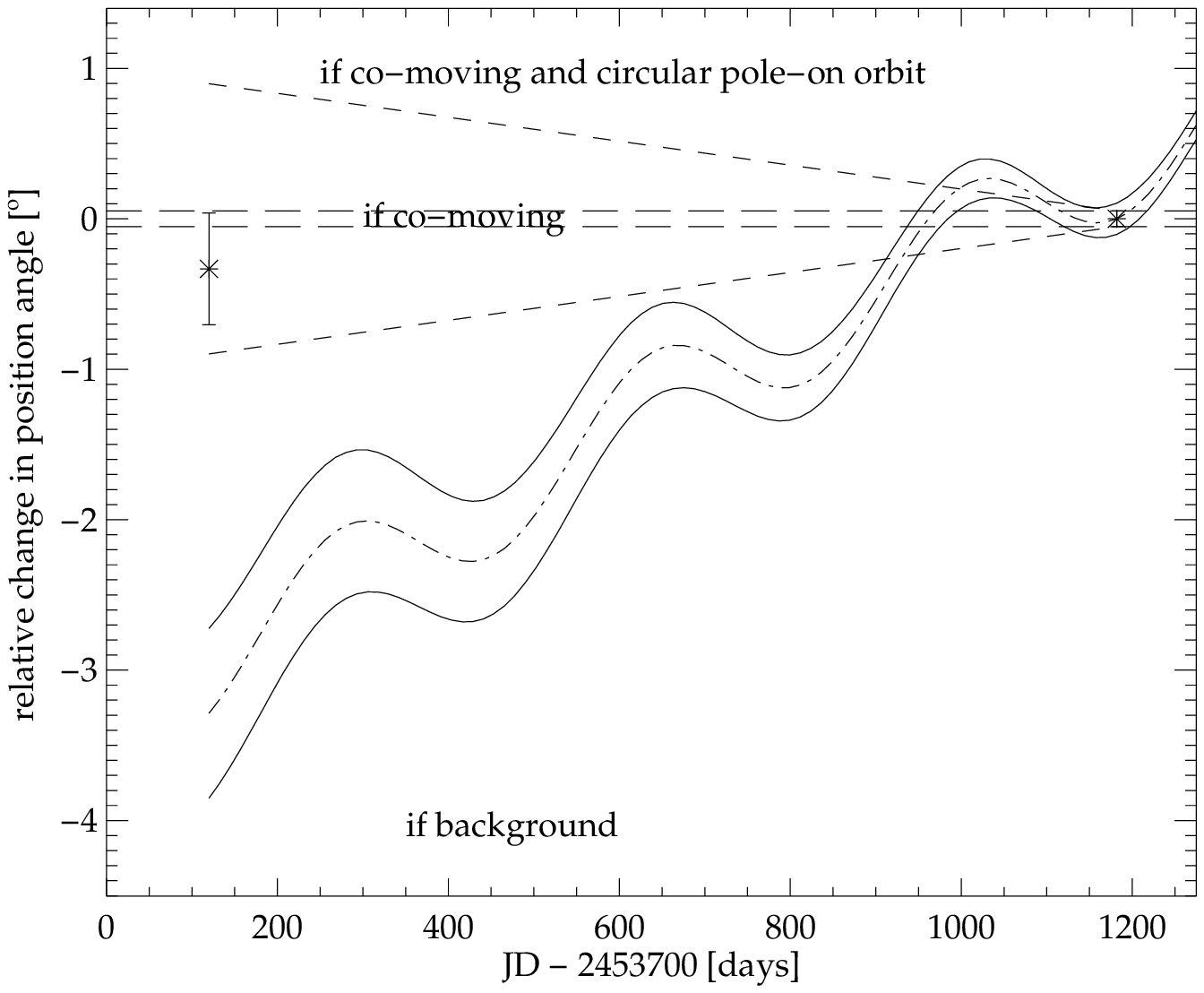}
%\resizebox{\hsize}{!}{\includegraphics{Anhangfigures/RXJ09527_AB_Poswinkel_1196_abs.eps}}
%\caption{Example proper Motion Diagram (PMD) from absolute astrometry for the position angle change in the RX J0952.7-7933 AB system. See text for more information.} 
\caption{Example Proper Motion Diagrams (PMD) ) for separation and 
position angle change (left to right) from absolute astrometric measurements (top) and from relative astrometric measurements 
(bottom)  in the WX Cha AB system. The long dashed lines enclose the area for constant separation, as expected for a co-moving object. 
The dash-dotted line is the change expected if the WX Cha companion is a non-moving background star. The opening cone 
enclosed by the continuous lines its estimated errors. The waves of this cone show the differential 
parallactic motion that has to be taken into account if the other component is a non-moving background star 
with negligible parallax. The opening short-dashed cone represents a combination of co-motion and the maximum 
possible orbital motion for a circular edge-on orbit (for separation) or circular pole-on orbit (for PA).
See text for more information.}
\label{Figure2}
\end{figure*}

To analyse and interpret the astrometric data we apply a 'proper motion diagram' (PMD), which contains the measured 
values of either the separation or the position angle with their errors versus time (Fig.~\ref{Figure2}). 
%We always use the latest or our latest measurement as a reference because normally it has smaller errors than earlier ones. 
In each case, we consider the 
hypothesis that the fainter component of the supposed binary is a non-moving background star. The proper motions in Table
\ref{table:4} are obtained by evaluating the centroid measurements of the unresolved system, or they are proper motions of unresolved
systems from literature catalogs, except for the resolved system RX J0919.4-7738, which was treated accordingly.
Hence, if the fainter component is a 
non-moving background object, the true proper motion of the brighter object must be corrected to ensure that the centroid 
reproduces the observed proper motion. For this purpose, the formulas given by the Astronomical Almanac 2005 of the
\citet{2003asal.book.....U} were applied. The location and movement of the centroid was determined according to the 
apparent Ks magnitudes in Table \ref{table:9}. 
The corrected proper motion values of the brighter components were used to calculate the separations and position angles, which
gives the dashed-dotted central line in the PMD, surrounded by solid lines according to the errors in the proper
motions. The waves in these curves are due to the differential parallactic motion of the brighter component, which depends on the
adopted distance; for the non-moving background object we assume a parallax of zero. In most cases, we used the mean value 
165 pc for Cha I (based on the data of \citet{1999A&A...352..574B} and \citet{1997A&A...327.1194W}; see 
\citet{2008A&A...491..311S} for more information); different distance values are mentioned in the description of individual 
stars in Table \ref{table:1}.

The long-dashed lines in the PMD enclose the area of constant separation (or position angle), while the opening short-dashed 
cone indicates the range of maximal amplitudes of the circular orbital motions according to Kepler's third law. Here we used our 
photometry in Table \ref{table:9} and converted the Ks magnitudes into mass estimates according to the models of 
\citet{1998A&A...337..403B}.
Assumed edge-on orbits define the maximal variations in the separation, while face-on orbits would cause maximal variations in 
the position angle.

The PMDs confirm that all 16 binaries and multiple systems are co-moving, which can also be concluded from the 
significances $\sigma_{\rho,\,\mathrm{back}}$ and $\sigma_{PA,\,\mathrm{back}}$ that the faint object is not a background 
star. The relative astrometric measurements 
(Table \ref{table:6}) give for both of them combined always $\sigma >$ 2.8 .  The slopes of the linear least square fits 
through the individual data points in the PMDs and their errors are listed in Table \ref{table:7} for the 
absolute astrometric measurements and in Table \ref{table:8} for the relative astrometric measurements. 

In principle, the a priori hypothesis that the fainter component is a background star is arbitrary. It could be that the brighter 
component is a projected distant, but more luminous non-member while the fainter one belongs to the 
Cha complex. The corresponding calculations were carried out and revealed that the resulting changes in the significances
for not being a background object are insignificant in the case of co-moving systems, as for all objects in this paper. 
Hence, these reversed PMDs are therefore not considered in greater detail. 

Moving background stars can only be rejected if curved orbital motion is detected.

%The 4 PMDs for each of our 16 program stars are published in the online material appendix. 
In only a few remarkable cases are the corresponding PMDs shown in the following sections. 
A typical example is given in Fig.~\ref{Figure2}. The remaining PMDs are in the online appendix.

\subsection{Co-moving binaries without orbit indications}

We discuss here three stars that are co-moving but for which we could not detect orbital motions. 
Their significance in relative astrometric measurements is $\sigma_{\mathrm{orb}} \leq$ 1.2 $\sigma$  (see Table \ref{table:6}). 
Consequently, their slopes in separation and position angles are also insignificant (see Tables \ref{table:7} \& 
\ref{table:8}).

\subsubsection*{SZ Cha}
With about 5 arcseconds separation, this is the widest here reported binary. The first measurement by 
\citet{1997ApJ...481..378G} has a rather large error. The absolute PMD (Fig.~\ref{SZ Cha}) is therefore compatible with a 
co-moving binary, 
but we can exclude from these values the background hypothesis of a non-moving fainter star in the background
by 2.8 $\sigma$. However, in the relative PMDs we can exclude this possibility with a 3.0 $\sigma$ significance based on a time
difference $\Delta$t of only 3 years instead of almost 15 years in the absolute comparison. 
In Table \ref{table:5}, we did not include a measurement by \citet{1997A&A...326..632A}, since the given time of observation
between July 1993 and May 1994 was rather imprecise, but more importantly the separation value of 5.2\,$\pm$\,0.025 arcsec 
was in quite good agreement with our measurements, while the position angle of 145.5\,$\pm$\,0.5 $\degr$ is more than 
20\,$\degr$ from our measurements and the measurement of \citet{1997ApJ...481..378G}, taken possibly within the same year.  
According to \citet{1997ApJ...481..378G}, there is also a wide visual companion candidate at 12.5 arcseconds separation, 
although this companion was outside the field of view (FoV) of our NACO S13 observations. Interestingly,  
\citet{2003A&A...410..269M} list SZ Cha as a spectroscopic single star, while \citet{2002AJ....124.2813R} list it as a 
good candidate for a PMS spectroscopic binary with a 5 day period. Finally, we note that our mean
brightness value of SZ Cha B of Ks = 11.22 mag is closely consistent with that of a spectral type M5 star in Cha I, as given
by \citet{2007ApJ...662..413K}.

\subsubsection*{WX Cha}
Only one epoch was observed by us. To be able to derive relative astrometric results, we re-reduced the 
data of \citet{2008ApJ...683..844L}, and show in Fig.~\ref{Figure2} \& \ref{WX Cha} all four PMDs. 
The background star hypothesis is excluded with high significance. 
From our relative photometry on 2006 March 25 (Table \ref{table:9}), we find the difference in Ks to be approximately 
0.4 mag larger than in the H band,
leading to the interesting result that WX Cha A seems to be redder than WX Cha B. This could for example be evidence of an 
edge-on disk around the secondary, actually being the primary, as found by us in the case of the very young star 
[MR81] H$\alpha$ 17 in Corona Australis \citep{2009A&A...496..777N}. Moreover, we see a strong variation in the Ks band
photometry between the data by \citet{2008ApJ...683..844L}, rereduced by us, and our data.

\subsubsection*{WY Cha}
As in the case of WX Cha, we obtained a second epoch of relative astrometric measurements by re-reducing the data of 
\citet{2008ApJ...683..844L} (Fig.~\ref{WY Cha}).
There could be marginal indications of orbital changes in the position angle (1.2 $\sigma$ level).
As in the case of WX Cha, we also see from our photometry (Table \ref{table:9}) that WY Cha A seems to be redder than WY Cha
B by quite a large amount. In this case, we even saw the effect twice between H and Ks bands on 2006 March 25 and between
J and Ks bands on 2008 February 20. Moreover, we see a strong variation in the Ks band photometry between the dates.
Near-infrared photometry with AO of both components as in the case of [MR81] H$\alpha$ 17 in Corona Australis 
\citep{2009A&A...496..777N} could either confirm or reject this assumption for WY Cha as well as WX Cha.
In addition, \citet{2007ApJ...662..413K} published an ultrawide visual companion candidate at an approximately 28 arcsecond
separation outside the FoV of our NACO S13 observations. 

\subsection{\textbf{Binaries with indications of orbital motion}}

We present here binaries for which we found significant slopes in the PMDs, probably corresponding to orbital
motion. In many cases, however, the evidence is still tentative, requiring confirmation by additional observations.

\subsubsection*{RX J0915.5-7609}
For this star, we adopt a distance of 168 pc \citep{2003A&A...404..913S} and an age of 1.3 Myrs \citep{2010ApJ...724..835W}.
Since we observed during two epochs, we can give a relative PMD (Fig.~\ref{RX J0915.5-7609}). This excludes the background 
hypothesis with high significance. The early observation of \citet{2001AJ....122.3325K} suggests that there is a curved 
orbital motion in the separation (but not in position angle).
%This star might be a runaway T Tauri star, as the relative proper motion points away from the
%Cha I star forming region into the northeastern direction.
The observed linear slope in the absolute position angle PMD is consistent with a possible orbital period of the order of 
140 years. 

\subsubsection*{RX J0935.0-7804}
Our measurement at JD 2454515.1 refers to a weighted mean of observations in two band passes, J and Ks, each one calibrated 
independently with corresponding HIP 73357 frames. The absolute, as well as the relative PMDs (Fig.~\ref{RX J0935.0-7804})
exclude the background 
hypothesis and reveal rather strong linear orbital displacements in both separation and position angle.
This system also seems to be very young with an age estimate of 1.1 Myrs \citep{2010ApJ...724..835W}.
%This star might be a runaway T Tauri star, as the relative proper motion points away from the
%Cha I star forming region into the northeastern direction.
The observed linear slope in the absolute position angle PMD is consistent with a possible orbital period of the order of 
550 years.
Since the separation changes for both absolute and relative values almost as much as expected for a maximum change in a circular
(edge-on) orbit, the PA should not change, if edge-on and circular. However, the PA also changes, hence the orbit 
might be either eccentric or inclined.

\subsubsection*{RXJ 1014.2-7636}
There are rather strong variations in the separation and position angle. The relative PMD (Fig.~\ref{RX J1014.2-7636})
in position angle indicates a 
significantly non-circular orbit, a quite inclined orbit, or a strongly curved orbit because its sign is the opposite of that 
in the early observation of \citet{2001AJ....122.3325K}.
According to \citet{1997A&A...319..184A}, this is a very young star of spectral type M2 at an age of 0.24\,$\pm$\,0.29 Myrs.
For this star, we could not find a convincing distance value in the literature. Although the higher proper motion found for
this star (Table \ref{table:4}) indicates that it is close to the Sun, as does the quite high orbital motion found
(Tables \ref{table:7} \& \ref{table:8}), the object might not be as close as the distance 14 pc given in \citet{2006AJ....132..866R}.
This is particularly true since it is unlikely to find a very young (see above) star in the solar neighbourhood. The
amount of orbital motion is still consistent with the distance of Cha I of 165 pc.
%, a second possible conclusion from the
%higher proper motion is, that the star might be a runaway T Tauri star, as the relative proper motion points away from the
%Cha I star forming region into the northeastern direction.
The slope in the (more long-term) absolute PMD of the position angle is consistent with an orbital period of the order of 
190 years. However, the relative PMD indicates that there is a significant position angle difference between the latest two measurements,
with a sign that is opposite of the long-term trend. This could be an indication of a curved orbit.  

\subsubsection*{Ced 110 IRS 2}
Data for this companion was published for the first time by \citet{2008ApJ...683..844L}, although we (Table
\ref{table:5}) also imaged this system 37 days before the measurements of \citet{2008ApJ...683..844L}. 
While it had already been found to be a visual binary, we waited for a second epoch of data
before presenting  here for the first time the high significance of a common proper motion.   
The background hypothesis can be excluded from the separation behaviour, but not from that of the position angle. On the 
other hand, there is a significant temporal variation in the position angle of absolute and relative astrometric 
measurements, but only a marginal variation in the relative separation PMD (Fig.~\ref{Ced 110 IRS 2}).
In addition, \citet{2007ApJ...662..413K} published an ultrawide visual companion candidate displaying an approximately 22 arcsecond
separation outside the FoV of our NACO S13 observations. 
The slopes in the PMDs of the position angle are consistent with an orbital period of the order of 120 years. 

\subsubsection*{Cha H$\alpha$ 2}
\begin{figure}
\includegraphics[width=0.49\textwidth]{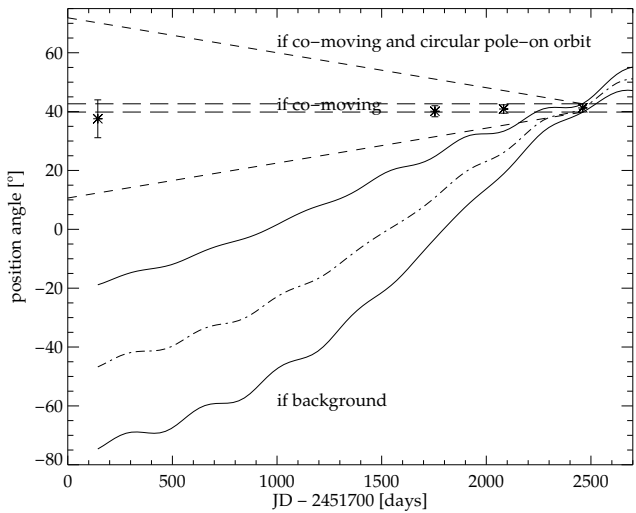}
\caption{Proper Motion Diagram (PMD) from absolute astrometric measurements for the position angle in the Cha 
H$\alpha$ 2 AB system. See text for more information.}
\label{Figure3}
\end{figure}

The binary nature of this object was suggested by \citet{2002A&A...384..999N} from HST data. Our two epochs of data are 
complemented by an archival observation previously discussed in \citet{2008A&A...484..413S}, where we confirm the 
very close binary nature of the object and derive the masses of both components of the order of 0.1\,$M_{\sun}$, 
near the lower stellar mass limit, with error bars down to 0.07\,$M_{\sun}$ for one of the components. 
The components of Cha H$\alpha$ 2 are among the faintest and least massive member stars found in the Cha star-forming region 
and one is still a candidate brown dwarf. The absolute PMDs (Fig.~\ref{Cha Ha 2}) were presented in \citet{2008A&A...484..413S}; here 
we add an older measurement from \citet{2002A&A...384..999N} to the regular PMD (see Fig.~\ref{Figure3}) after slightly improving the
astrometric reduction quality of the original presented data. In addition, we also present the relative PMDs 
that confirm the linear orbital motion in the separation found in \citet{2008A&A...484..413S}. The position angle does not 
show significant orbital variations, but it excludes the background hypothesis, confirms the co-moving nature of the 
components of Cha H$\alpha$ 2, and indicates that the orbit is edge-on.

\subsubsection*{RX J1109.4-7627}
There are small variations in the separation but a rather strong slope in the position angle, which is consistent with a
circular pole-on orbit of a period of the order of 62 years. The relative and absolute PMDs (Fig.~\ref{RX J1109.4-7627})
are almost identical in this case, 
owing to the lack of earlier observations and the small separation of the components, reducing the influence of the more 
imprecise absolute astrometric calibration. 

\subsubsection*{HD 97300}
This is the most massive binary of our sample with a spectral type B9V \citep{1980AJ.....85..444R,1999A&A...346..205G}. For 
this star, we adopt a distance of 179 pc \citep{2007A&A...474..653V} and an age of 5.6\,$\pm$\,2.0 Myrs 
\citep{2011MNRAS.410..190T}. Two epochs were observed by us. In addition, we re-reduced the data of 
\citet{2008ApJ...683..844L} in order to be able to include this epoch in our relative PMDs (Fig.~\ref{HD 97300}).
There is a quite significant 
linear orbital motion in both of the separation PMDs, as well as in the relative astrometric measurements of the 
position angle. The background hypothesis is rejected with high significance in all diagrams.

\subsubsection*{HJM C7-11}
We consider the strongly deviating value of the early separation measurement of 1.2 arcsec by \citet{1997ApJ...481..378G} 
as a misprint, as it might mean 0.2 arcsec. Disregarding this value, we obtain a co-moving pair with a rather constant 
separation, but slight indications of a position angle slope (Fig.~\ref{HJM C 7-11}). A circular pole-on orbit would have 
a period of the 
order of 500 years. While \citet{2006A&A...460..695T} give a multiplicity flag of 'SB?' (spectroscopic binary),
this possibility was not mentioned in other publications \citep{1997A&A...328..187C,2006A&A...448..655J}.
In addition, \citet{2007ApJ...662..413K} published an ultrawide visual companion candidate at an approximately 13.6 arcsecond
separation outside the FoV of our NACO S13 observations. 

\subsubsection*{Sz 41}
There is an early observation of the 4 Myr \citep{2010ApJ...724..835W} old star \citep{1997ApJ...481..378G} that is of very low 
accuracy. This and the remaining 
observations are consistent with a co-moving binary without any significant orbit evidence in separation. The position 
angle change for relative astrometric measurements might either indicate orbital motion or a 2\,$\sigma$ outlier of our 
first own measurement. The point from the re-reduced data of \citet{2008ApJ...683..844L} is consistent with no orbital 
motion at all, although it does have a larger time difference and hence, larger error bars due to the possible larger 
amount of orbital motion of the calibration binary. The background hypothesis is ruled out with high significance, especially 
according to the relative PMD (Fig.~\ref{Sz 41}),
which contains three epochs. According to \citet{1992PhDT.......255B} \& \citet{1993A&A...278...81R}, there is a second wide
visual companion candidate at 11.4 arcsec separation outside the FoV of our NACO S13 observations. \citet{2002AJ....124.2813R} 
report from their spectroscopy data, with the primary and secondary situated in the spectrograph entrance window, that 
the object is a spectroscopic binary candidate with a possible period of about 125 days. While the spectro-astrometric
displacement found by \citet{2003A&A...397..675T} is small and might be caused by the binarity also recovered 
by us at $\sim$\,2 arcsec,
\citet{1997A&A...328..187C}, \citet{2003A&A...410..269M}, \citet{2007A&A...467.1147G} and \citet{2008A&A...492..545J} 
all found no evidence of a spectroscopic binarity.  

\subsubsection*{HM Anon}
In addition to our two epochs of data we re-reduced the data of \citet{2008ApJ...683..844L} in order to be able to 
include these epochs in our relative PMD (Fig.~\ref{HM Anon}). 
All diagrams reveal that there are significant linear orbital motions in separation and position 
angle, and enable us to reject the background hypothesis with large significance. 
As given in \citet{2009yCat....102023S},u the Tycho-2 coordinates are wrong and HM Anon is identical with  \mbox{TYC 9414-1250-1}.
Since the proper motion of Tycho-2 \citep{2000A&A...355L..27H} differs greatly from the mean value for Cha I 
\citep{2008ApJ...675.1375L}, we chose to use the value of UCAC 3 \citep{2010AJ....139.2184Z}; this is because this measurement is 
consistent with the Cha I median value (see Table \ref{table:4}), although UCAC 3 found fewer than two good matches, while
including the catalogues used for Tycho-2.
HM Anon is the only object in our sample for which we could find values of the orbital motion in the literature 
\citep{2001A&A...369..249W}. If we convert their units (km/s) into those of our Table \ref{table:7} (mas/yr), using the 
distance to Cha I of 160 pc as assumed there and given in \citet{1998MNRAS.301L..39W} and from km/s to $\degr$/yr assuming 
the same distance and a separation of the components of 245 mas, we derive the trends of -3.5\,$\pm$\,9.8 mas/yr in separation 
and 4.1\,$\pm$\,4.0 $\degr$/yr in position angle observed by \citet{2001A&A...369..249W}, consistent with our new, more 
precise  values listed in Table \ref{table:7}.
As discussed for RX J0935.0-7804, the orbit of this binary might be inclined and/or eccentric.

%The values of \citet{2001A&A...369..249W} are consistent within the error bars and in both orientations of the directions 
%with our new values (see Table \ref{table:7}).  
%\hline
%HM Anon AB$^f$ \ \ (1)& \ \ 45    & -3.49     & 9.80 &  4.08     & 3.96 \\
%(f) Values converted from km/s to mas/yr using the distance to
%Cha I of 160 pc as assumed in (1) and given in (2) and from km/s to $\degr$/yr assuming the same distance and a separation
%of the components of 245 mas \\
%\textbf{References}: (1) \citet{2001A&A...369..249W} (2) \citet{1998MNRAS.301L..39W}

\subsection{Triples and quadruples}

\subsubsection*{RX J0919.4-7738}
\begin{figure}
\includegraphics[width=0.49\textwidth]{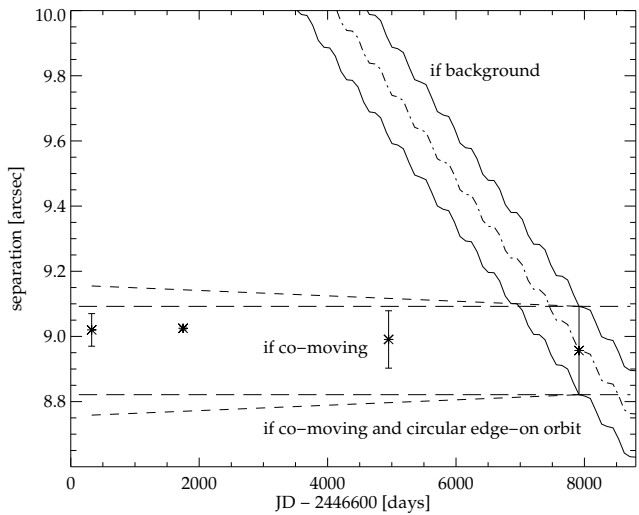}
\caption{Proper Motion Diagram (PMD) from \textbf{absolute astrometric measurements} for the separation in the RX J0919.4-7738 AaAbB 
system. See text for more information.}
\label{Figure4}
\end{figure}

For this star, we adopt a distance of 56.8$\,\pm\,$2.8 pc \citep{2007A&A...474..653V} and an age of 16.1\,$\pm$\,2.5 Myrs 
\citep{2011MNRAS.410..190T}, consistent with the earlier value of 14 Myrs given by \citet{1998A&A...330L..29N}.
This is a strong quadruple system candidate. It consists of a wide binary (separation 9 arcseconds) whose southern 
component is a double-lined spectroscopic binary \citep{1997A&A...328..187C} called B, while the northern component is 
visually double, beeing resolved for the first time by \citet{2001AJ....122.3325K} called Aa/Ab.  Our observations 
comprise only one epoch of Aa/Ab/B, therefore no relative 
PMDs can be obtained. We used the catalogues of HIPPARCOS \citep{1997A&A...323L..49P} and 2MASS 
\citep{2006AJ....131.1163S,2003tmc..book.....C} in order to derive separations and position angles between Aa/Ab and B, using 
the centroid in magnitude for Aa/Ab (which were not resolved by HIPPARCOS and 2MASS). These data do not reveal significant 
variations in either the separation or position angle of the wide binary, and their PMD is consistent with a co-moving binary 
(Fig.~\ref{Figure4}). 
The PMD of the Aa/Ab pair (Fig.~\ref{RX J0919.4-7738})
shows strong variability in the position angle, while the distance seems to be constant during the 
time interval of about 4300 days between the two observations available at present. This is consistent with a co-moving 
binary in a circular pole-on orbit, with an orbital period of the order of 90 years.
We did not include in our analysis the position values of \citet{1998AJ....116.2975M}, \citet{2010AJ....140..510T}, and the 
earlier ones as in \citet{1986A&AS...64..105T}, noting that there has been little or no change in the orbit of 
the wide binary since 1872, as no error bars are given for the separation and position angle.
For a further discussion of the system, we refer to \citet{2006A&A...454..553D} and \citet{2010AJ....140..510T}.

\subsubsection*{RX J0952.7-7933}
For this star, we adopt an age of 3 Myrs, since \citet{1998A&A...338..442F} give this value for their subgroup 1 of stars,
mentioning that RX J0952.7-7933 might be a member of this subgroup judging from its proper motion alone, while warning that 
the reflex motion of the sun is very similar to the typical proper motion of Chamaeleon member stars. 
The object is a spectroscopic triple star \citep{1997A&A...328..187C}, partly resolved by \citet{2001AJ....122.3325K} and
us here, although we are unable to judge from the currently available data which resolved star of the components is the 
additional spectroscopic binary. 
Components A and B are co-moving with only marginal variations in the separation: the six data points in the relative 
separation PMD could reveal some indication of an orbital curvature for the triple system AB. 
This curvature is strongly consistent with a first astrometric detection of the third component of the spectroscopic
triple, having in this case an orbital period of approximately five years.
On the other hand, there is a rather strong slope in both of the
position angle PMDs (Fig.~\ref{RX J0952.7-7933}), which is 
consistent with a circular pole-on obit of the wide binary and a period of the order of 200 years.

\begin{figure}
\includegraphics[width=0.49\textwidth]{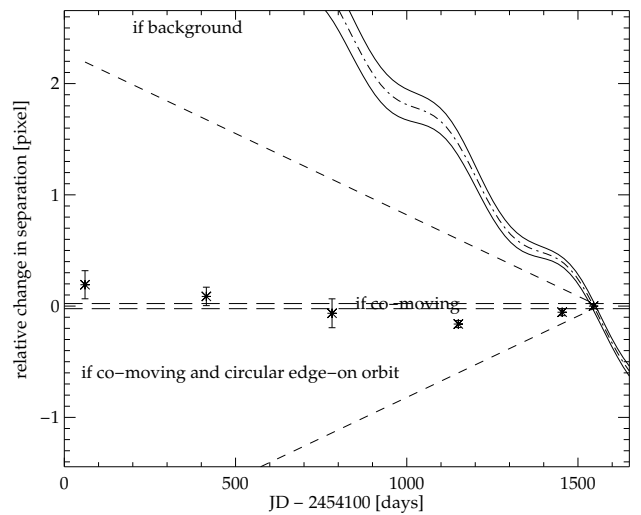}
\caption{Proper Motion Diagram (PMD) from relative astrometric measurements for the separation in the RX J0952.7-7933 AB 
system. See text for more information.}
\label{Figure5}
\end{figure}

\begin{figure*}
   \includegraphics[width=0.49\textwidth]{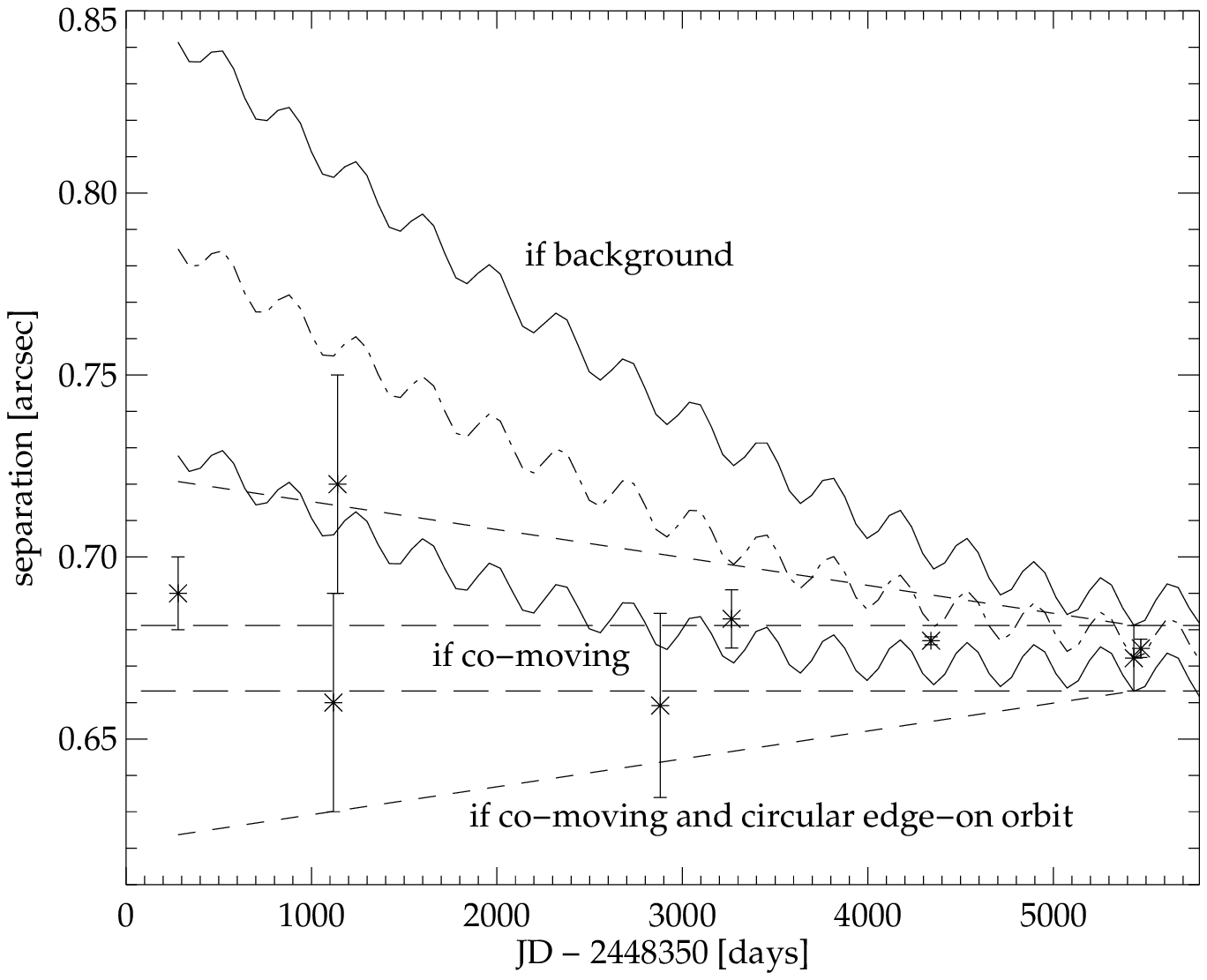}
   \includegraphics[width=0.49\textwidth]{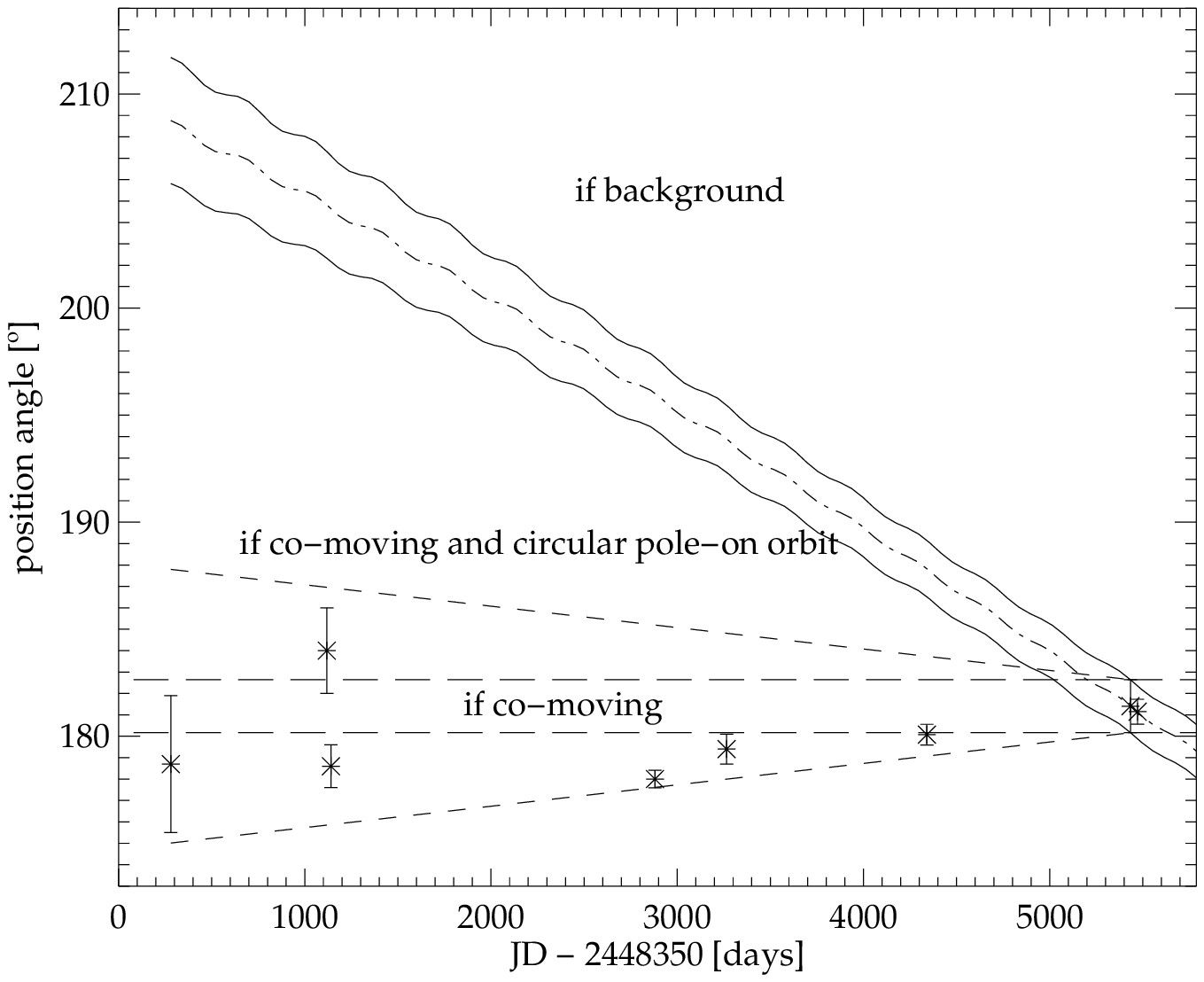}
   \includegraphics[width=0.49\textwidth]{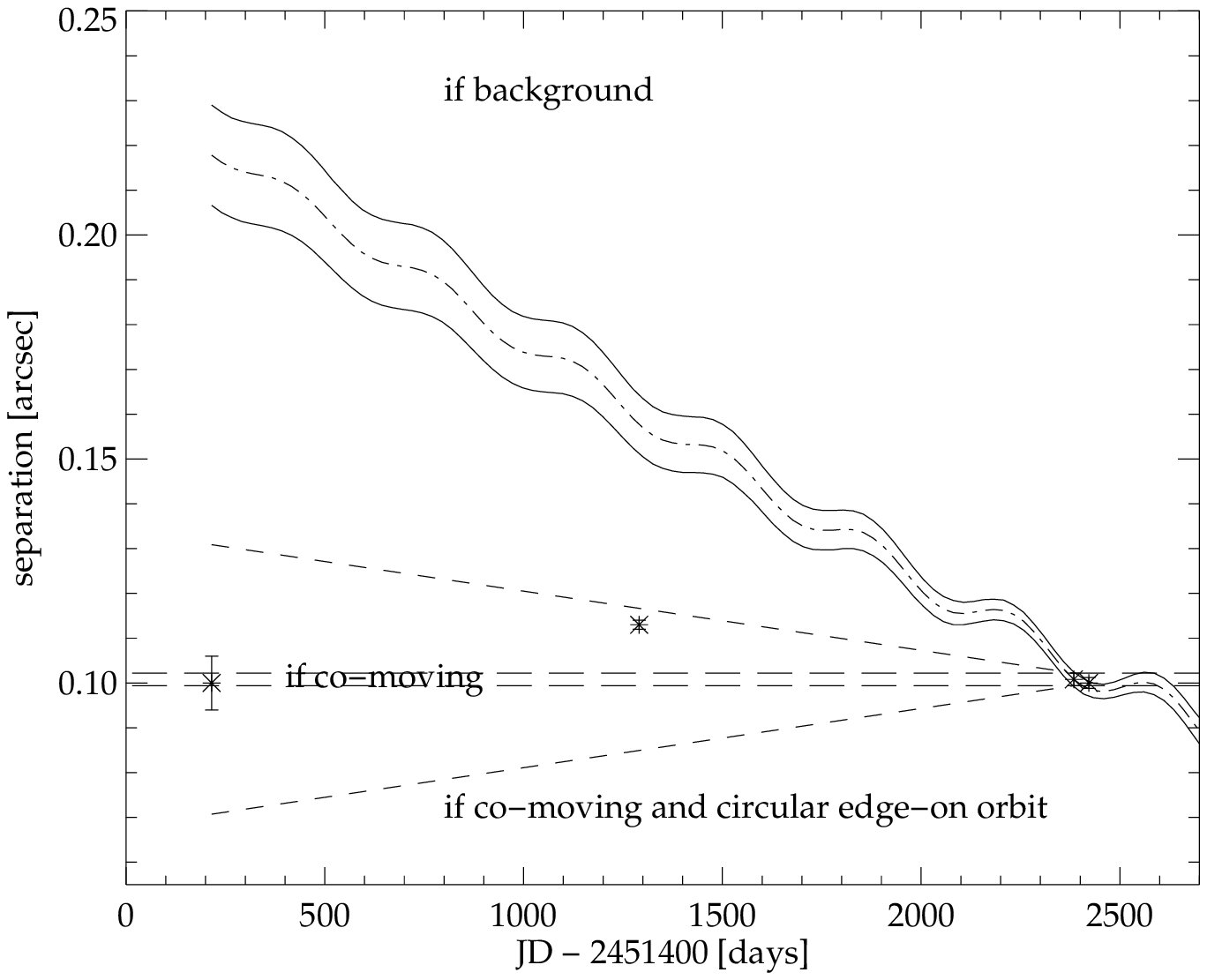}
   \includegraphics[width=0.49\textwidth]{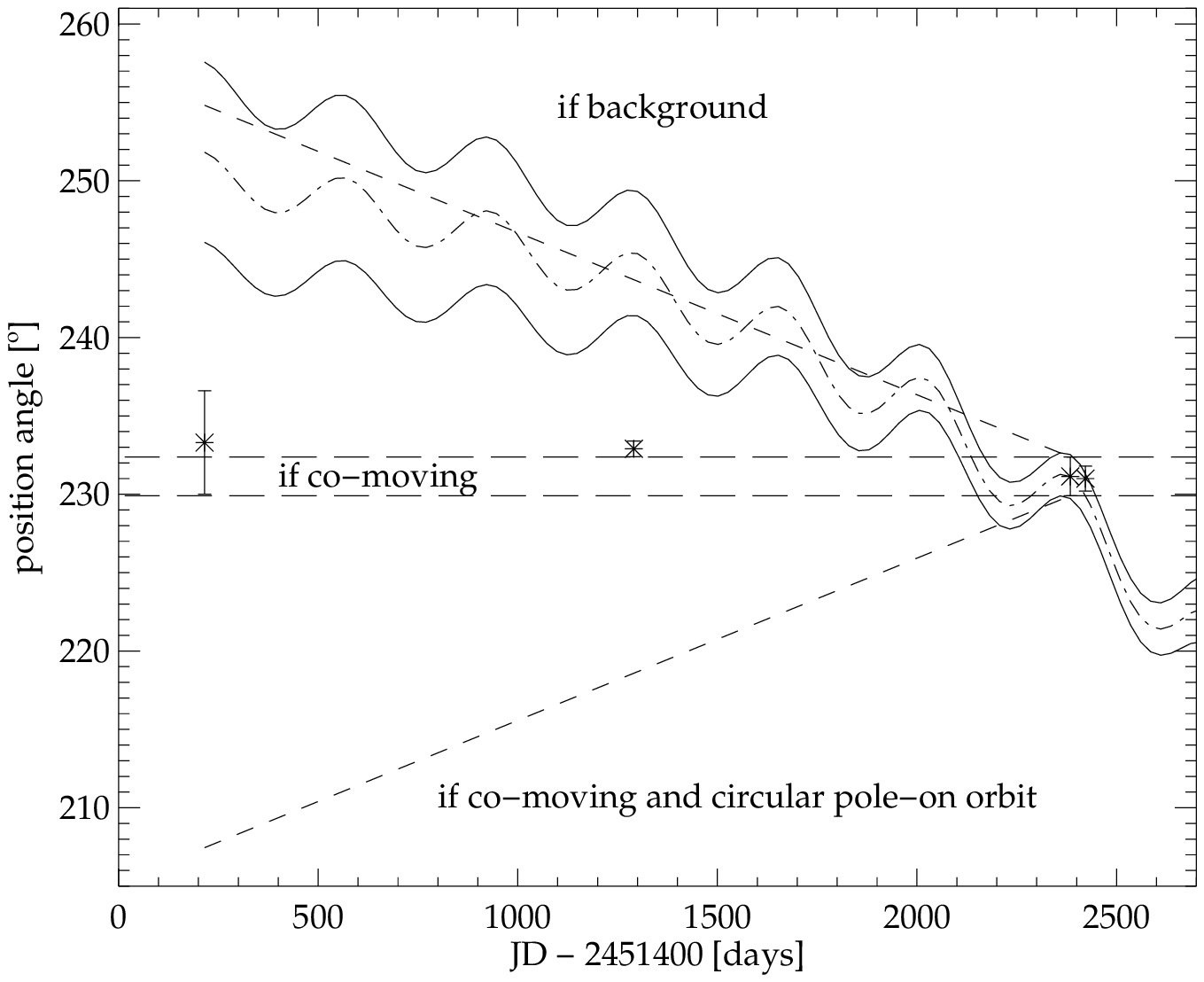}
%\resizebox{\hsize}{!}{\includegraphics{Anhangfigures/RXJ09527_AB_Poswinkel_1196_abs.eps}}
%\caption{Example Proper Motion Diagram (PMD) from absolute astrometry for the position angle change in the RX J0952.7-7933 AB system. 
%See text for more information.} 
\caption{Proper Motion Diagrams (PMD) from absolute astrometric measurements (top) for separation and position angle change (left
to right) of VW Cha A relative to the centroid of B \& C  
and from absolute astrometric measurements (bottom) for separation and position angle change (left to right) of VW Cha B relative to
VW Cha C. See text for more information.}
\label{Figure6}
\end{figure*}

\subsubsection*{VW Cha}
This object was recognized as a binary by \citet{1992PhDT.......255B}. Additional measurements were published by 
\citet{1997ApJ...481..378G} and \citet{1996A&A...307..121B} and were found in the ESO/ST-ECF science archive (see also 
\citet{2001ASPC..231..620S}, 
\citet{2001AAS...199.6009P} for more information on the observations with the HST). \citet{2001ApJ...561L.199B} resolved, 
for the first time, the southern component of the original binary into 
two stars, now called B/C. Consequently, the triple system was measured by \citet{2006A&A...459..909C} and 
\citet{2008ApJ...683..844L}, 
as well as in one epoch by us (therefore no relative PMD is available). Figs.~\ref{Figure6} and \ref{VW Cha} show the PMDs 
of A versus (vs.) B/C and B vs. C. The position angle variations in the former PMD are consistent with a circular pole-on orbit, 
excluding the background hypothesis,
while the separation values do not allow firm conclusions.  The B/C pair is bound, and the three points in the separation PMD
could indicate curved orbital motion. \citet{1993A&A...278...81R} and \citet{2006A&A...459..909C} found evidence of a possible fourth 
component in this system, outside the field of our NACO observations at 16.7 arcseconds separation.
VW Cha was found to be spectroscopically multiple by \citet{2003A&A...410..269M} and \citet{2006A&A...460..695T}, although 
according to the details in the discussion of \citet{2003A&A...410..269M} the three found components might in fact be the 
three resolved inner components by \citet{2001ApJ...561L.199B}.

\section{Conclusions}

Our astrometric results from VLT NACO observations based on new data of one to five epochs confirm the physical connection
of the secondary stars in all (previously known, but hitherto unconfirmed) 16 binary and multiple members of the star-forming
region in Chamaeleon. The angular separation between the components range from 0.07 to 9.0 arcseconds (Table \ref{table:5}), 
corresponding to
projected minimum distances of between 6 and 845 AU (Table \ref{table:7}). All secondary components are brighter than 11.5 mag 
in the Ks band (Table \ref{table:9}), despite our observations reaching limiting magnitudes that vary between
16.5 and 18 mag in Ks.
On average, we expect only 0.117 stars per field, according to 2MASS (c.f. Table \ref{table:1}). However, only about 7\,\% of 
all 2MASS stars in our statistics are as bright as K $\leq$ 11.5 mag, therefore the probability of a chance coincidence at 
this brightness level ranges between 0.4 \% and 1.2 \%, according to the different star densities in our fields. 
Statistically, we should observe about 125 different fields, in order to expect one chance coincidence of a star 
brighter than 11.5 mag with one of our targets. Therefore, it is unsurprising that background stars in our small NACO 
fields are missing.

However, from a statistical point of view, there is an additional strong argument for a true physical connection between 
most components of our targets: In only two fields are the angular separations larger than 5 arcseconds (RX J0919.4-7738 and 
SZ Cha), i.e. of the same order of magnitude as our NACO field size. There is one additional case (Sz 41) with a separation 
of 1.9 arcseconds, but in all the other 13 cases the components are closer than 1 arcsecond. Therefore, a realistic estimation
of a chance coincidence is still much lower than the above value.

On the other hand, we should emphasize that, in principle, a real proof of a gravitational bond between the components is only
given, if curvature in the orbit can be measured (which would have to also differ from a hyperbolic ejection orbit). 
Marginal indications of possible curvatures are found in four targets (RX J0915.5-7609, RX J0952.7-7933, 
RX J1014.2-7636, and VW Cha).
Future observations should confirm this and improve the orbital parameters. 
We also refer to \citet{2010A&A...516A.112N} and \citet{2010A&A...523L...1M} for further detailed discussions and sub-stellar 
examples of the first indications of orbital curvature. As an alternative explanation of apparently co-moving  
binaries, we could suppose that we have seen two seperate members of the star-forming region in Chamaeleon, aligned by chance along our 
line of sight. Our orbital motions of a few to several mas/yr (Tables \ref{table:7} \& \ref{table:8}) would be typical 
of the velocity dispersion in star-forming regions such as Cha I \citep{2005A&A...438..769D}.
However, the aforementioned statistical arguments do not support this idea. We therefore have 
observed, with very high probablilty, 16 physically bound binaries or multiple systems.
Even if one turns out to be an unbound case of two young Chamaeleon members, the age and distance (within the given 
uncertainties) would be the same for both objects, and likewise the masses of these individual Chamaeleon member objects.

The 2MASS catalogue separates the components only for the two cases with separations 
$\geq$\,5\,arcseconds mentioned above, in all other cases the positions and magnitudes in 2MASS refer to the combined light 
(c.f. Section 3). The limiting magnitude of 2MASS is near K\,=\,16\,mag, far below the limit of the 
faintest secondary components investigated in this paper (K\,=\,11.5\,mag, see Table \ref{table:9}). 
Might we be missing additional low-mass components of larger separations, outside of our NACO field of
view? To investigate this possibility, we searched in the 2MASS catalogue for additional components at angular
distances of up to 20 arcseconds, around all 16 targets presented here, as well as around CT Cha \citep{2008A&A...491..311S}. 
This search revealed a total of 17 candidate stars, i.e. on average 1.0 stars per field, with separations ranging between 
7.5 and 16.5 arcseconds. Since these search fields are 6.83 times larger than the FoV of NACO, we expect on average about 
0.117 $\cdot$ 6.83 = 0.8 stars per field, very close to the observed number. 
Assuming that we can also find objects within the additional amount we jittered the images, although we could not go as
deep as in the inner parts of the images, the search fields are only
5.2 times larger than the NACO FoV plus two of the four arcsecond jitter box width used, hence we expect on average
about 0.61 stars per field. On average, we found 0.2 -- 0.39 objects too many in the search areas per star or in total
3.4 -- 6.6 candidate stars that could be additional multiple-star components.
Moreover, there is a large excess of bright stars: a total of 4 
candidate stars (= 24 \%) are in the range K $<$ 11.99 mag, while  our 2MASS counts reveal that only 9 \% of all stars are in this 
brightness range.  A similar excess is present in the magnitude range 12.0 $<$ K $<$ 13.99 mag in which we found 6 stars 
(=35 \%), but expect only 19 \%.  Consequently, there is a deficit of the faintest stars K $>$ 14.0 mag (7 stars = 41 \% 
observed, but 72 \%  expected), of faint background stars behind the dark cloud. According to these results, a total of 5 -- 6 
candidate stars with K $<$ 14.0 mag could be additional multiple star components of our targets. 

The given magnitude limit of K\,=\,11.5 mag of our target stars has already been mentioned. It correponds at the distance of 
the Cha I region rather precisely to the mass limit of 0.08 M$_{\sun}$ of hydrogen burning stars. We should emphasize that 
this limit is only reached by one binary star (Cha H$\alpha$ 2, see also \citet{2008A&A...484..413S}), both of whose components 
are very near to this lower limit of possible stellar masses, one of them still being a brown dwarf candidate. SZ Cha B is 
the only other star with K\,$>$\,11\,mag, while all the remaining targets of our sample contain much brighter components 
with K\,$\leq$\,10.6\,mag, corresponding to a mass of 0.16 M$_{\sun}$ at the average age (2 Myrs) and the distance of Cha I
(165 pc) using the \citet{1998A&A...337..403B} models.
%Apparently, 
In contrast to many studies \citep{2000A&A...360..997T,2003A&A...397..159H}, we find that
binary or multiple systems with relatively massive primary components tend to
avoid the simultaneous formation of equal-mass secondary components
(as also found by \citet{2003ApJ...599.1344M}).
Extremely low-mass secondary components are 
hard to find for high and low mass primaries owing to the much higher dynamic range and the faintness of the secondaries.

%__________________________________________________________________

\begin{acknowledgements}
We would like to thank the ESO Paranal Team, the ESO User Support department and all other very helpful ESO 
services, the anonymous referee for helpful comments and finally the language editor Claire Halliday for corrections.

NV acknowledges support by Comit\'e Mixto ESO-Gobierno de Chile, as well as by the Gemini-CONICYT fund 32090027.
TOBS acknowledges support from the Evangelisches Studienwerk e.V. Villigst, from the German National Science 
Foundation (Deutsche Forschungsgemeinschaft, DFG) in grant NE 515/30-1 and from Friedrich Schiller University Jena / State 
of Thuringia / Germany. AS acknowledges support from NSF under grants AST-0708074 and AST-1108860.

This publication makes use of data products from the Two Micron All Sky Survey, which is a joint project of the University of
Massachusetts and the Infrared Processing and Analysis Center/California Institute of Technology, funded by the National 
Aeronautics and Space Administration and the National Science Foundation.
This research has made use of the VizieR catalog access tool and the Simbad database, both operated at the Observatoire 
Strasbourg.
This reasearch makes use of the Hipparcos Catalogue, the primary result of the Hipparcos space astrometry mission, undertaken
by the European Space Agency.
This research has made use of NASA's Astrophysics Data System Bibliographic Services.
\end{acknowledgements}

\bibliographystyle{aa}

%\bibliography{references.bib}

%end of the main text
\listofobjects

\begin{small}
\begin{flushleft}
\begin{longtable}{llcr@{\,$\pm$\,}lccr@{\,$\pm$\,}lccc}
\caption{\label{table:5} Absolute astrometric results}\\
%\centering
\hline\hline
Object        & JD-2446000        & Ref. & \multicolumn{2}{c}{Separation} & Sign.$^a$  & Sign. orb. & \multicolumn{2}{c}{PA$^b$ }   & Sign.$^a$  & Sign. orb.\\
              & $[\mathrm{days}]$ &      & \multicolumn{2}{c}{[arcsec]}   & not Backg. & motion     & \multicolumn{2}{c}{[$\degr$]} & not Backg. & motion    \\
              &         & & $\rho$ & $\delta_{\rho}$ & $\sigma_{\rho,\,\mathrm{back}}$ & $\sigma_{\rho,\,\mathrm{orb}}$  
                        & $PA$   & $\delta_{PA}$   & $\sigma_{PA,\,\mathrm{back}}$   & $\sigma_{PA,\,\mathrm{orb}}$   \\
\hline
\endfirsthead
\caption{continued.}\\
\hline\hline
Object        & JD-2446000        & Ref. & \multicolumn{2}{c}{Separation} & Sign.$^a$  & Sign. orb. & \multicolumn{2}{c}{PA$^b$ }   & Sign.$^a$  & Sign. orb.\\
              & $[\mathrm{days}]$ &      & \multicolumn{2}{c}{[arcsec]}   & not Backg. & motion     & \multicolumn{2}{c}{[$\degr$]} & not Backg. & motion    \\
              &         & & $\rho$ & $\delta_{\rho}$ & $\sigma_{\rho,\,\mathrm{back}}$ & $\sigma_{\rho,\,\mathrm{orb}}$  
                        & $PA$   & $\delta_{PA}$   & $\sigma_{PA,\,\mathrm{back}}$   & $\sigma_{PA,\,\mathrm{orb}}$   \\
\hline
\endhead
\hline
\endfoot
RX J0915.5-7609 \\
\ \ \ \ \ \ \ \
\ \ \ \ \  AB       & 4172.50000         & 1     & 0.111  & 0.007  & 16   & 3.1  & 292.5   & 4.3   & 0.5  & 7.1    \\
                    & 8161.53412         &       & 0.1489 & 0.0031 & 6.9  & 1.6  & 262.763 & 1.412 & 5.3  & 1.5    \\
                    & 8881.59508         &       & 0.1385 & 0.0055 & $^c$ & $^c$ & 259.287 & 1.920 & $^c$ & $^c$   \\
RX J0919.4-7738 \\
\ \ \ \ \ \ \ \
\ \ \ \ \ AaB       & 8515.51911         &       & 8.9727 & 0.1357 &      &      & 194.218 & 1.400 &      &        \\
\ \ \ \ \ \ \ \
\ \ \ \ \ AbB       & 8515.51911         &       & 8.9351 & 0.1352 &      &      & 194.858 & 1.400 &      &        \\
\ \ \ \ \ \ \ \
\ \ \ \ \ (AaAb)B$^d$ & \ \ 925.53114    & 2     & 9.020  & 0.050	& 12   & 0.4  & 192.350	& 0.740 & 7.1  & 1.3    \\ 
                    & 2348.81400         & 3     & 9.0250 & 0.0051 & 10   & 0.5  & 194.154 & 0.036 & 5.7  & 0.2    \\
                    & 5549.68730         & 4     & 8.9907 & 0.0881 & 4.5  & 0.2  & 194.396 & 0.797 & 2.6  & 0.1    \\
                    & 8515.51911         &       & 8.9567 & 0.1355 & $^c$ & $^c$ & 194.487 & 1.400 & $^c$ & $^c$   \\
\ \ \ \ \ \ \ \
\ \ \ \ \ AaAb      & 4172.50000         & 1     & 0.109  & 0.003  & 43   & 0.4  & 173.9   & 1.2   & 29   & 21     \\
                    & 8515.51911         &       & 0.1070 & 0.0035 & $^c$ & $^c$ & 125.175 & 2.008 & $^c$ & $^c$   \\
RX J0935.0-7804 \\
\ \ \ \ \ \ \ \
\ \ \ \ \ AB        & 4173.50000         & 1     & 0.360  & 0.003  & 21   & 8.0  & 353.9   & 0.2   & 16   & 5.5    \\
                    & 8161.59183         &       & 0.4114 & 0.0059 & 9.6  & 0.8  & 346.840 & 1.328 & 5.0  & 0.6    \\
                    & 8515.10153         &       & 0.4146 & 0.0048 & 6.2  & 0.5  & 346.215 & 1.009 & 3.3  & 0.3    \\
                    & 8882.58189         &       & 0.4189 & 0.0067 & $^c$ & $^c$ & 345.630 & 1.483 & $^c$ & $^c$   \\
RX J0952.7-7933 \\
\ \ \ \ \ \ \ \
\ \ \ \ \ AB        & 4173.50000         & 1     & 0.267  & 0.003  & 13   & 2.7  & 314.6   & 0.5   & 12   & 15     \\
                    & 8160.58336         &       & 0.2852 & 0.0043 & 7.4  & 0.4  & 296.912 & 1.356 & 4.2  & 3.6    \\
                    & 8514.65391         &       & 0.2838 & 0.0032 & 6.9  & 0.2  & 295.658 & 1.033 & 3.8  & 3.3    \\
                    & 8881.63897         &       & 0.2818 & 0.0048 & 4.8  & 0.1  & 293.627 & 1.498 & 2.3  & 2.0    \\
                    & 9250.53614         &       & 0.2805 & 0.0047 & 2.8  & 0.3  & 291.421 & 1.565 & 1.2  & 1.0    \\
                    & 9553.86365         &       & 0.2819 & 0.0050 & 0.7  & 0.1  & 289.616 & 1.633 & 0.2  & 0.2    \\
                    & 9645.57285         &       & 0.2826 & 0.0050 & $^c$ & $^c$ & 289.222 & 1.653 & $^c$ & $^c$   \\
RX J1014.2-7636 \\
\ \ \ \ \ \ \ \
\ \ \ \ \ AB        & 4174.50000         & 1     & 0.091  & 0.007  & 22   & 19   & 259.6   & 6.5   & 5.8  & 4.4    \\
                    & 9250.56298         &       & 0.2392 & 0.0041 & 14   & 1.9  & 291.248 & 1.571 & 1.0  & 0.9    \\
                    & 9645.60036         &       & 0.2510 & 0.0045 & $^c$ & $^c$ & 289.085 & 1.654 & $^c$ & $^c$   \\
SZ Cha \\
\ \ \ \ \ \ \ \
\ \ \ \ \  AB       & 3476.50000         & 5     & 5.3    & 0.2    & 2.0  & 0.8  & 121     & 2     & 1.6  & 0.9    \\
                    & 5101.55583         & 6     & 5.118  & 0.012  & 2.0  & 0.0  & 123.00  & 0.15  & 1.0  & 0.2    \\
                    & 7782.64912         &       & 5.1151 & 0.0685 & 0.5  & 0.0  & 123.264 & 1.236 & 0.2  & 0.0    \\
                    & 7821.89696         & 6     & 5.122  & 0.0146 & 0.7  & 0.0  & 122.9   & 0.6   & 0.4  & 0.2    \\
                    & 8881.66756         &       & 5.1189 & 0.0819 & $^c$ & $^c$ & 123.258 & 1.481 & $^c$ & $^c$   \\
Ced 110 IRS 2 \\
\ \ \ \ \ \ \ \
\ \ \ \ \ AB        & 7782.61949         &       & 0.1481 & 0.0027 & 8.7  & 1.1  & 128.196 & 1.428 & 0.3  & 2.6    \\
                    & 7819.92133         & 6     & 0.140  & 0.0063 & 4.7  & 0.4  & 126.1   & 2.8   & 0.6  & 1.2    \\
                    & 8515.61187         &       & 0.1427 & 0.0039 & $^c$ & $^c$ & 121.952 & 1.888 & $^c$ & $^c$   \\
Cha H$\alpha$ 2 \\
\ \ \ \ \ \ \ \
\ \ \ \ \  AB       & 5843.40756         & 7,8   & 0.21   & 0.05   & 0.1  & 1.0  & 37.6    & 6.4   & 2.9  & 0.6    \\
                    & 7454.63563         & 7,9   & 0.1652 & 0.0037 & 1.0  & 1.3  & 40.121  & 1.871 & 2.1  & 0.5    \\
                    & 7782.68981         & 7     & 0.1667 & 0.0031 & 1.5  & 1.9  & 40.875  & 1.436 & 2.0  & 0.2    \\
                    & 8160.71934         & 7     & 0.1591 & 0.0026 & $^c$ & $^c$ & 41.250  & 1.402 & $^c$ & $^c$   \\
VW Cha \\
\ \ \ \ \ \ \ \
\ \ \ \ \  AB       & 7783.90711         &       & 0.6491 & 0.0087 &      &      & 178.929 & 1.236 &      &        \\
\ \ \ \ \ \ \ \
\ \ \ \ \ AC        & 7783.90711         &       & 0.7153 & 0.0096 &      &      & 185.324 & 1.236 &      &        \\
\ \ \ \ \ \ \ \
\ \ \ \ \ A(BC)$^e$ & 2630.83699         & 10    & 0.69   & 0.01   & 1.6  & 1.3  & 178.7   & 3.2   & 6.9  & 0.8    \\
                    & 3468.50000         & 5     & 0.66   & 0.03   & 1.7  & 0.4  & 184     & 2     & 6.5  & 1.1    \\
                    & 3490.53749         & 11    & 0.72   & 0.03   & 0.6  & 1.5  & 178.6   & 1.0   & 9.8  & 1.8    \\
                    & 5230.38638         & 12,13 & 0.6592 & 0.0253 & 1.2  & 0.6  & 178.001 & 0.405 & 10   & 2.6    \\
                    & 5615.50000         & 14    & 0.683  & 0.008  & 0.5  & 0.9  & 179.4   & 0.7   & 8.4  & 1.4    \\
                    & 6690.70028         & 15    & 0.6770 & 0.0011 & 0.2  & 0.5  & 180.076 & 0.483 & 5.5  & 1.0    \\
                    & 7783.90711         &       & 0.6722 & 0.0090 & $^f$ & $^f$ & 181.404 & 1.236 & $^f$ & $^f$   \\
                    & 7821.93038         & 6     & 0.6749 & 0.0025 & 0.2  & 0.3  & 181.148 & 0.581 & 0.2  & 0.2    \\
\ \ \ \ \ \ \ \
\ \ \ \ \ BC        & 5615.50000         & 14    & 0.100  & 0.006  & 9.3  & 0.1  & 233.3   & 3.3   & 2.8  & 0.6    \\
                    & 6690.70028         & 15    & 0.113  & 0.001  & 6.7  & 7.1  & 232.9   & 0.5   & 3.1  & 1.3    \\
                    & 7783.90711         &       & 0.1008 & 0.0014 & $^f$ & $^f$ & 231.136 & 1.238 & $^f$ & $^f$   \\
                    & 7821.93038         & 6     & 0.100  & 0.0012 & 0.4  & 0.4  & 231.0   & 0.8   & 0.3  & 0.1    \\
RX J1109.4-7627 \\
\ \ \ \ \ \ \ \
\ \ \ \ \ AB        & 7820.51080         & 16    & 0.0806 & 0.0018 & 0.8  & 2.3  & 85.733  & 1.376 & 3.4  & 10.1   \\
                    & 8882.81186         &       & 0.0792 & 0.0020 & 3.1  & 1.8  & 70.222  & 1.708 & 2.1  & 3.2    \\
                    & 9251.69870         &       & 0.0736 & 0.0024 & $^c$ & $^c$ & 62.006  & 1.914 & $^c$ & $^c$   \\
\\
HD 97300 \\
\ \ \ \ \ \ \ \
\ \ \ \ \  AB       & 3468.50000         & 5     & 0.84   & 0.04   & 1.9  & 1.4  & 327     & 2     & 5.9  & 0.2    \\
                    & 7820.90105         & 16    & 0.7855 & 0.0106 & 0.8  & 0.2  & 326.994 & 1.246 & 1.5  & 0.3    \\
                    & 8160.60766         &       & 0.7827 & 0.0112 & 0.4  & 0.0  & 326.908 & 1.320 & 0.8  & 0.2    \\
                    & 8516.71442         &       & 0.7819 & 0.0118 & $^c$ & $^c$ & 326.489 & 1.250 & $^c$ & $^c$   \\
WX Cha \\
\ \ \ \ \ \ \ \
\ \ \ \ \  AB       & 3473.50000         & 5     & 0.79   & 0.04   & 5.1  & 1.0  & 55      & 2     & 6.1  & 1.0    \\
                    & 7819.89102         & 16    & 0.7464 & 0.0106 & 3.2  & 0.0  & 52.169  & 1.275 & 1.5  & 0.2    \\
                    & 8881.91858         &       & 0.7467 & 0.0120 & $^c$ & $^c$ & 52.502  & 1.482 & $^c$ & $^c$   \\
WY Cha \\
\ \ \ \ \ \ \ \
\ \ \ \ \  AB       & 7819.63848         & 16    & 0.1241 & 0.0033 & 7.5  & 0.4  & 241.737 & 1.474 & 2.8  & 0.5    \\
                    & 8516.70106         &       & 0.1227 & 0.0019 & $^c$ & $^c$ & 240.775 & 1.407 & $^c$ & $^c$   \\
HJM C 7-11 \\
\ \ \ \ \ \ \ \
\ \ \ \ \  AB       & 3476.50000         & 5     & 1.2$^g$& 0.1    & 5.2  & 10$^g$ & 336   & 5     & 7.1  & 0.8    \\
                    & 7819.57870         & 16    & 0.1772 & 0.0041 & 6.3  & 0.6  & 334.781 & 1.440 & 7.7  & 1.4    \\
                    & 8161.78979         &       & 0.1774 & 0.0044 & 4.9  & 0.6  & 333.033 & 1.788 & 5.6  & 0.5    \\
                    & 8515.79485         &       & 0.1748 & 0.0032 & 3.8  & 0.1  & 332.639 & 1.542 & 4.1  & 0.4    \\
                    & 8881.70578         &       & 0.1744 & 0.0029 & $^c$ & $^c$ & 331.856 & 1.491 & $^c$ & $^c$   \\
Sz 41 \\
\ \ \ \ \ \ \ \
\ \ \ \ \  AB       & 3476.50000         & 5     & 1.5    & 0.8    & 0.4  & 0.6  & 150     & 20    & 1.4  & 0.6    \\
                    & 6661.86278         & 15    & 1.974  & 0.002  & 2.6  & 0.1  & 162.4   & 0.5   & 3.7  & 0.3    \\
                    & 7819.54463         & 16    & 1.9720 & 0.0266 & 1.2  & 0.1  & 162.737 & 1.245 & 1.8  & 0.1    \\
                    & 8619.97904         & 12    & 1.9623 & 0.0169 & 0.1  & 0.4  & 162.200 & 0.193 & 1.4  & 0.4    \\
                    & 8881.88546         &       & 1.9760 & 0.0316 & 0.3  & 0.0  & 162.685 & 1.482 & 0.5  & 0.1    \\
                    & 9250.08623         &       & 1.9767 & 0.0334 & $^c$ & $^c$ & 162.862 & 1.564 & $^c$ & $^c$   \\
HM Anon \\
\ \ \ \ \ \ \ \
\ \ \ \ \  AB       & 3470.50000         & 5     & 0.27   & 0.03   & 9.6  & 0.9  & 236     & 5     & 6.2  & 2.3    \\
                    & 7820.86777         & 16    & 0.2469 & 0.0033 & 9.1  & 1.0  & 246.743 & 1.245 & 4.8  & 0.6    \\
                    & 8234.48285         & 16    & 0.2455 & 0.0036 & 6.0  & 0.7  & 247.312 & 1.338 & 2.5  & 0.3    \\
                    & 8515.69528         &       & 0.2440 & 0.0037 & 3.9  & 0.4  & 247.635 & 1.401 & 1.9  & 0.2    \\
                    & 8882.87337         &       & 0.2419 & 0.0039 & $^c$ & $^c$ & 247.991 & 1.483 & $^c$ & $^c$   \\
\hline
\end{longtable}
\textbf{Remarks}: (a) Assuming the fainter component is a non-moving background star (b) Position angle (PA) is measured
from N over E to S (c) Significances are given relative to the last epoch (d) Results of component A relative to the 
centre of brightness (brightness ratio $\sim$\,1.37 at Ks band) (e) Results of component A relative to the centre of 
brightness (brightness ratio $\sim$\,1.75 at Ks band as well as to the centre of mass (masses from apparent magnitudes, Table 
\ref{table:8}, and distance of Chamaeleon cloud of 165 $\pm$ 30 pc, using the models of \citet{1998A&A...337..403B}, giving 
for B 0.37 $M_{\sun}$ and for C 0.21 $M_{\sun}$ at 1\,Myr, resulting in approximately the same ratio of $\sim$\,1.76 as the 
brightnesses) of components B and C (f) Significances are given relative to the second to last epoch. (g) Included as given 
in \citet{1997ApJ...481..378G}, this might be a typo, being actually 0.2 arcsec \\
\textbf{References}: (1) \citet{2001AJ....122.3325K}; date given, assuming midnight for calculation of JD (2) 
\citet{1988A&AS...76..189S} (3) Calculated from the HIPPARCOS catalogue \citep{1997A&A...323L..49P} position and positional 
error at the catalogue epoch (J1991.25) (4) Calculated from the 2MASS catalogue \citep{2006AJ....131.1163S} position and 
positional error (5) \citet{1997ApJ...481..378G}; date given, assuming midnight for calculation of JD (6) 
\citet{2008ApJ...683..844L} (7) \citet{2008A&A...484..413S}, improved (8) \citet{2002A&A...384..999N} (9) see also 
\citet{2007ApJ...671.2074A} and \citet{2008ApJ...683..844L} (10) \citet{1992PhDT.......255B} (11) \citet{1996A&A...307..121B}
(12) HST data from ESO/ST-ECF science archive, only position measurement error in RA and Dec considered. (13) From ESO/ST-ECF 
science archive, see also \citet{2001ASPC..231..620S}, \citet{2001AAS...199.6009P} (14) \citet{2001ApJ...561L.199B}; date 
given, assuming midnight of central of three observing days for calculation of JD (15) \citet{2006A&A...459..909C} (16) Rereduced, 
see also \citet{2008ApJ...683..844L}.
\end{flushleft}
\end{small}

\appendix
\section{Proper Motion Diagrams (PMDs)}

\onlfig{1}{
\begin{figure}
   \includegraphics[width=0.49\textwidth]{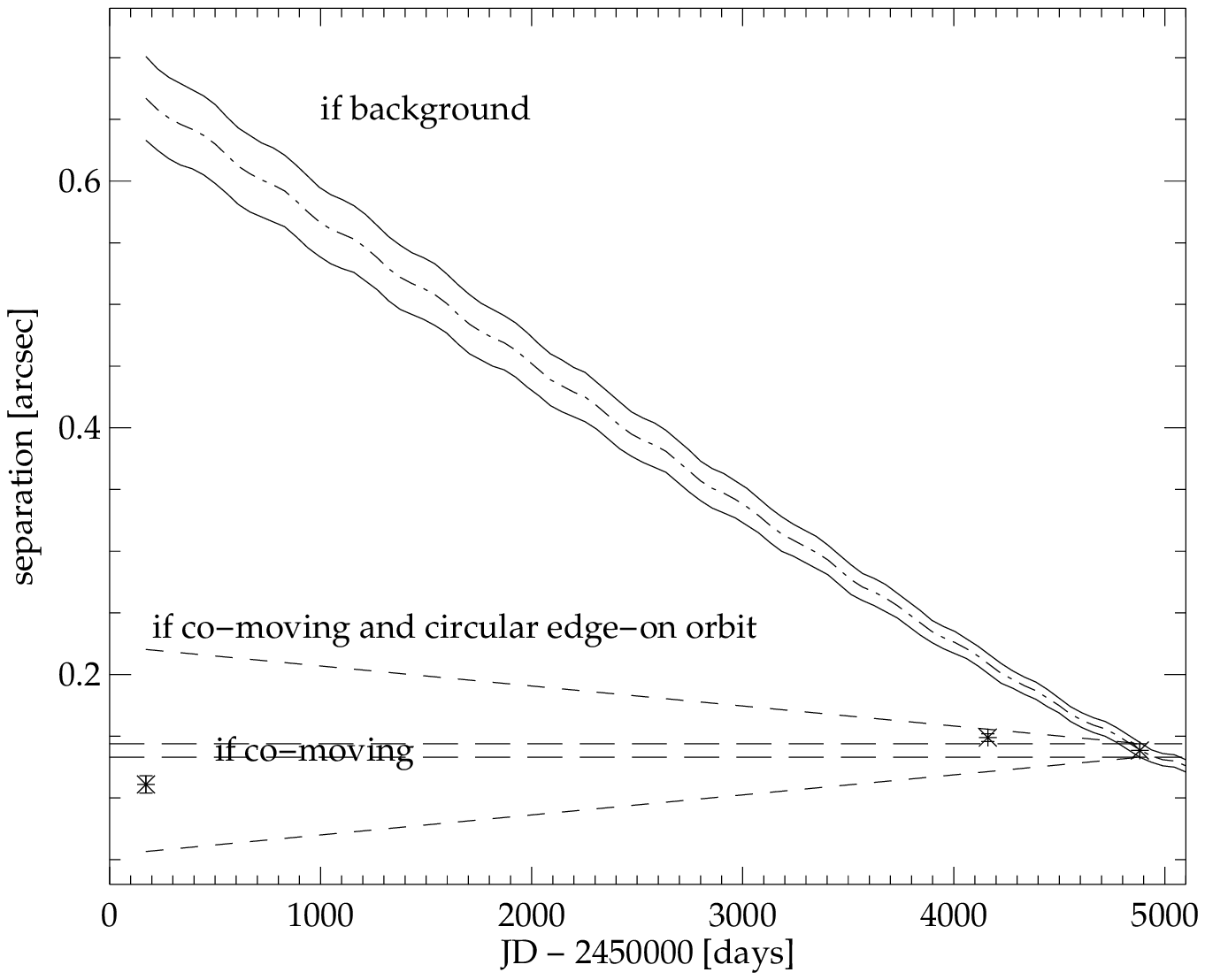}
   \includegraphics[width=0.49\textwidth]{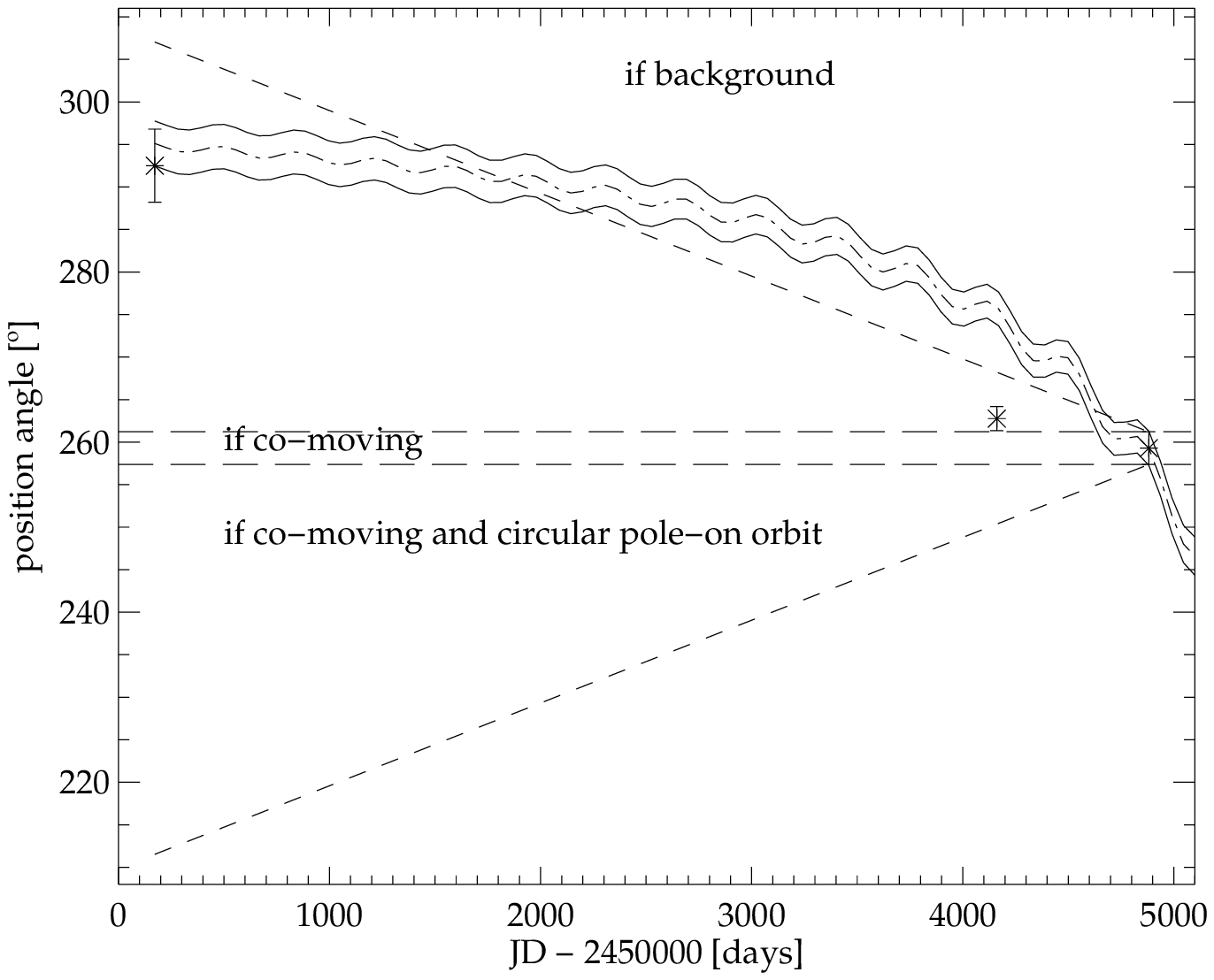}
   \includegraphics[width=0.49\textwidth]{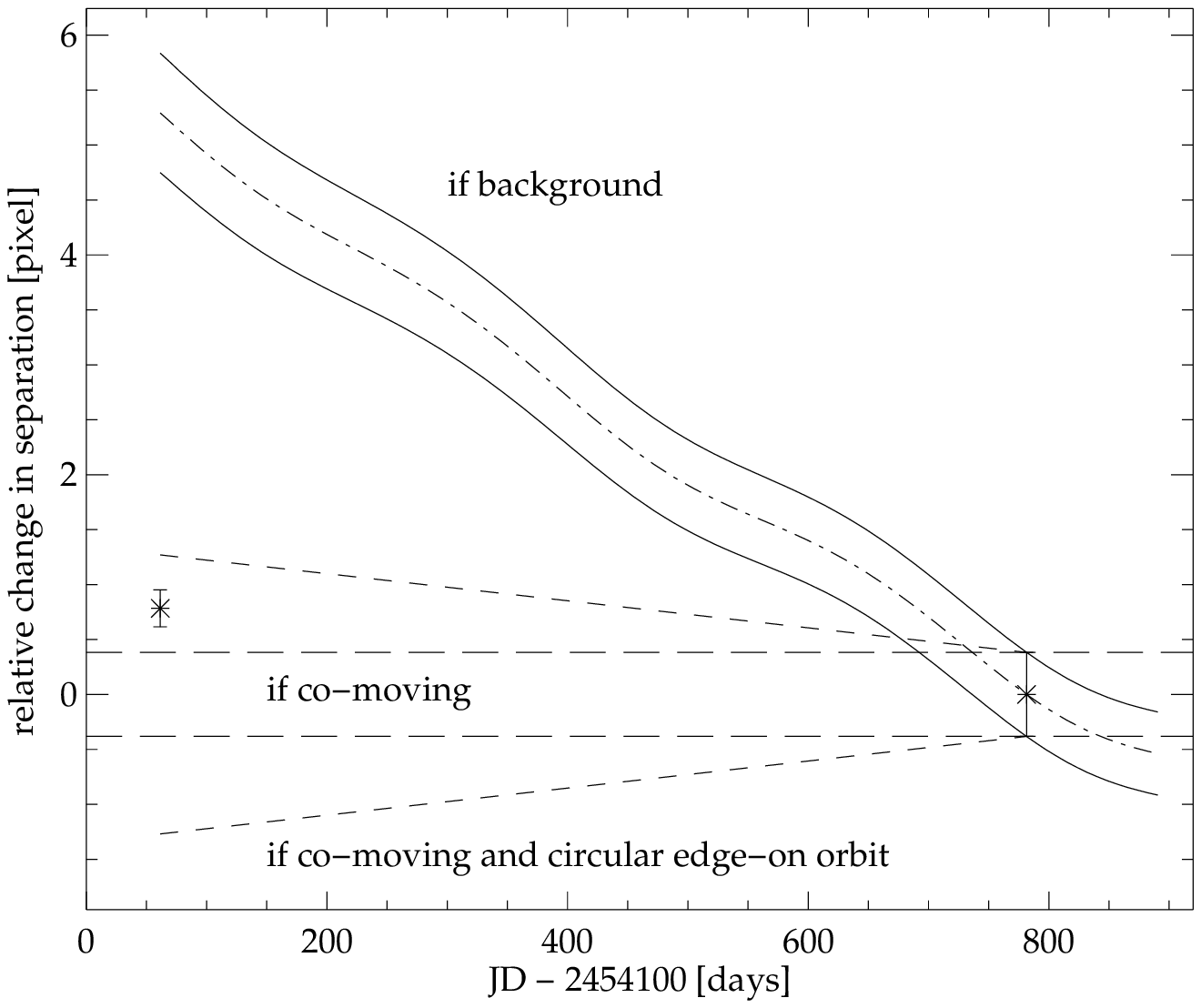}
   \includegraphics[width=0.49\textwidth]{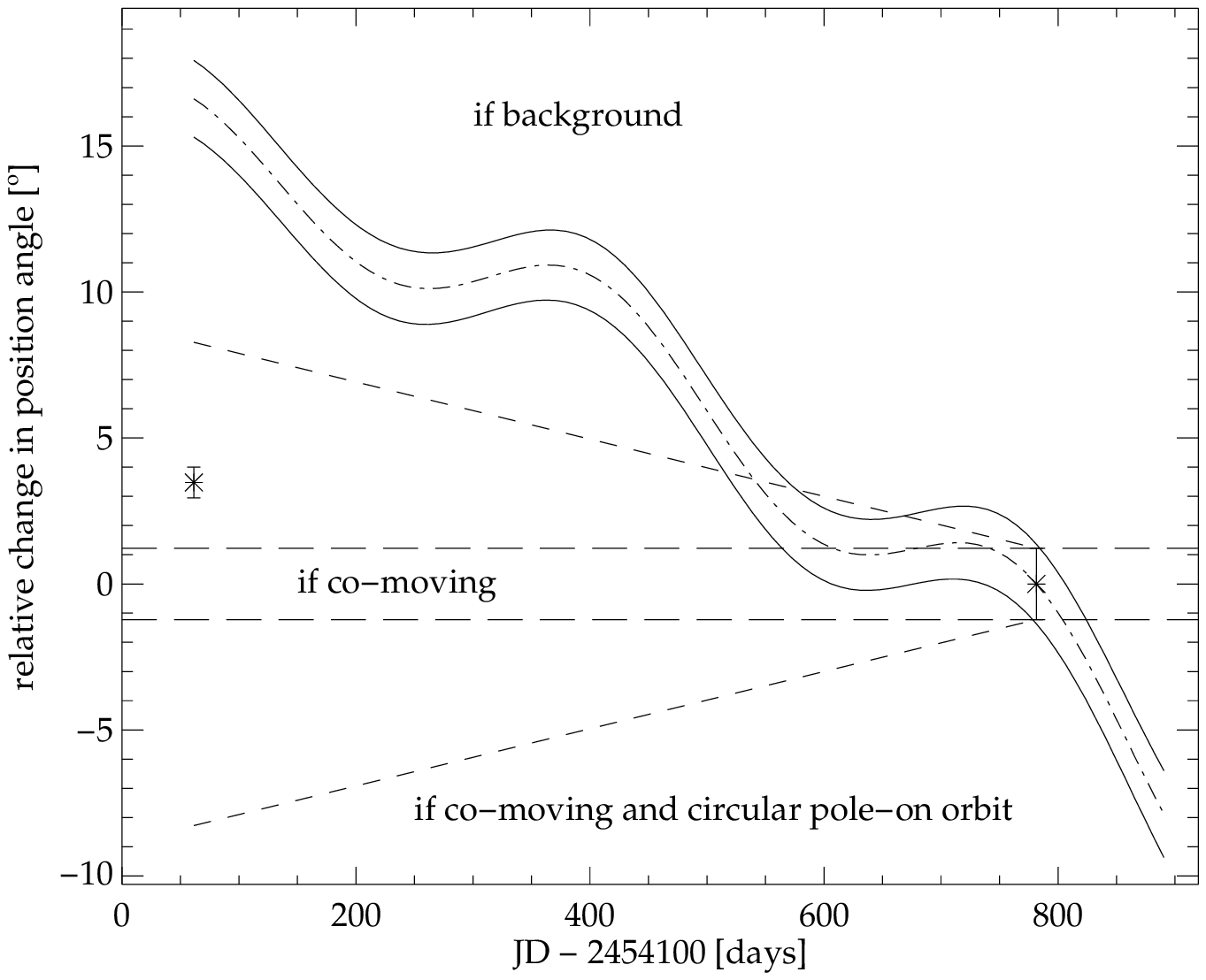}
   \caption{Proper Motion Diagrams (PMD) for separation and position angle change from absolute astrometric measurements (top, left to right) 
    and from relative astrometric measurements (bottom, left to right) in the 
   RX J0915.5-7609 AB system. See text for more information.}
   \label{RX J0915.5-7609}
   \end{figure}
}
     
\onlfig{2}{
\begin{figure}
   \includegraphics[width=0.49\textwidth]{Anhangfigures/RXJ0919.4-7738_A_BC_Separation_0887_abs.eps}
   \includegraphics[width=0.49\textwidth]{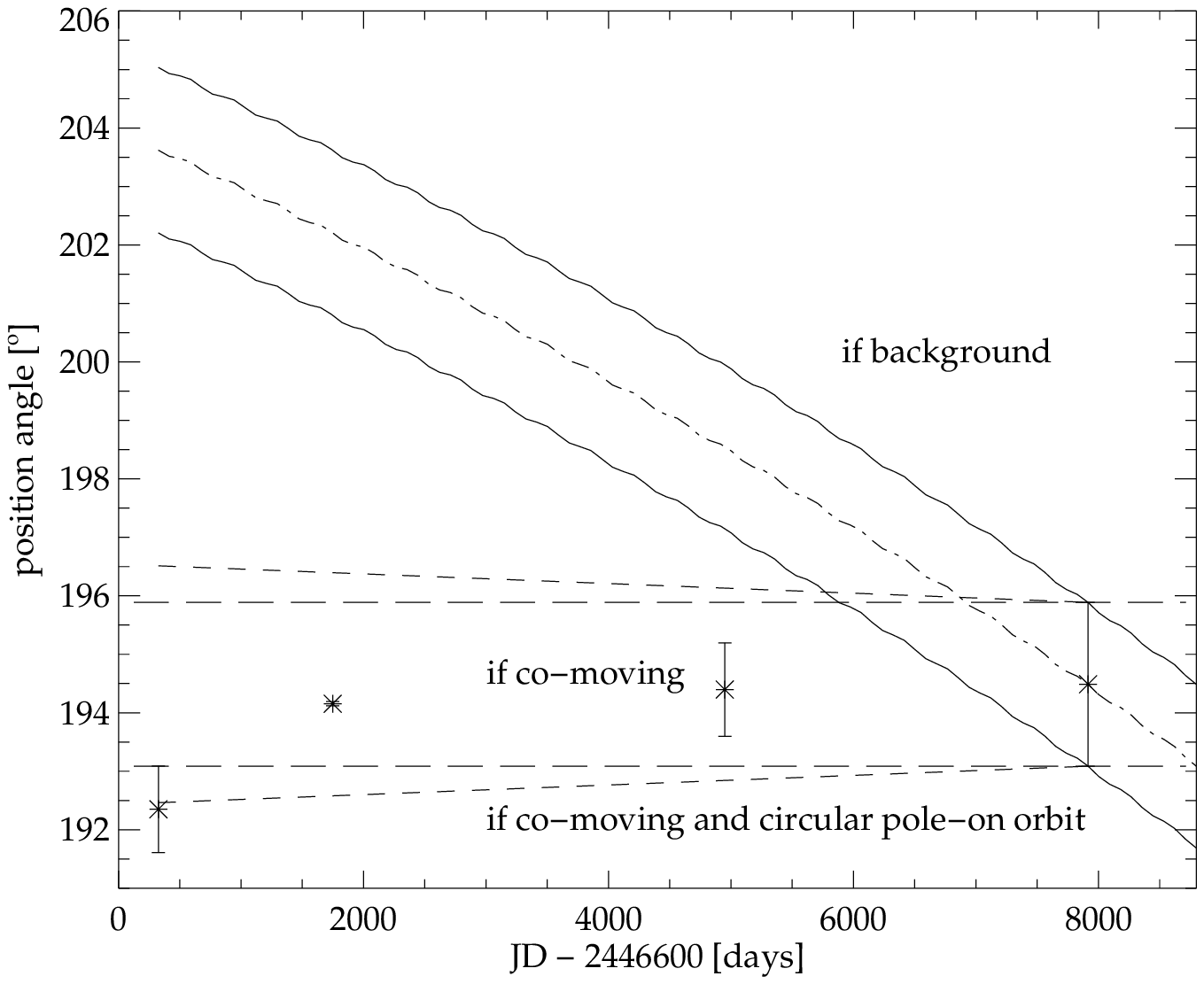}
   \includegraphics[width=0.49\textwidth]{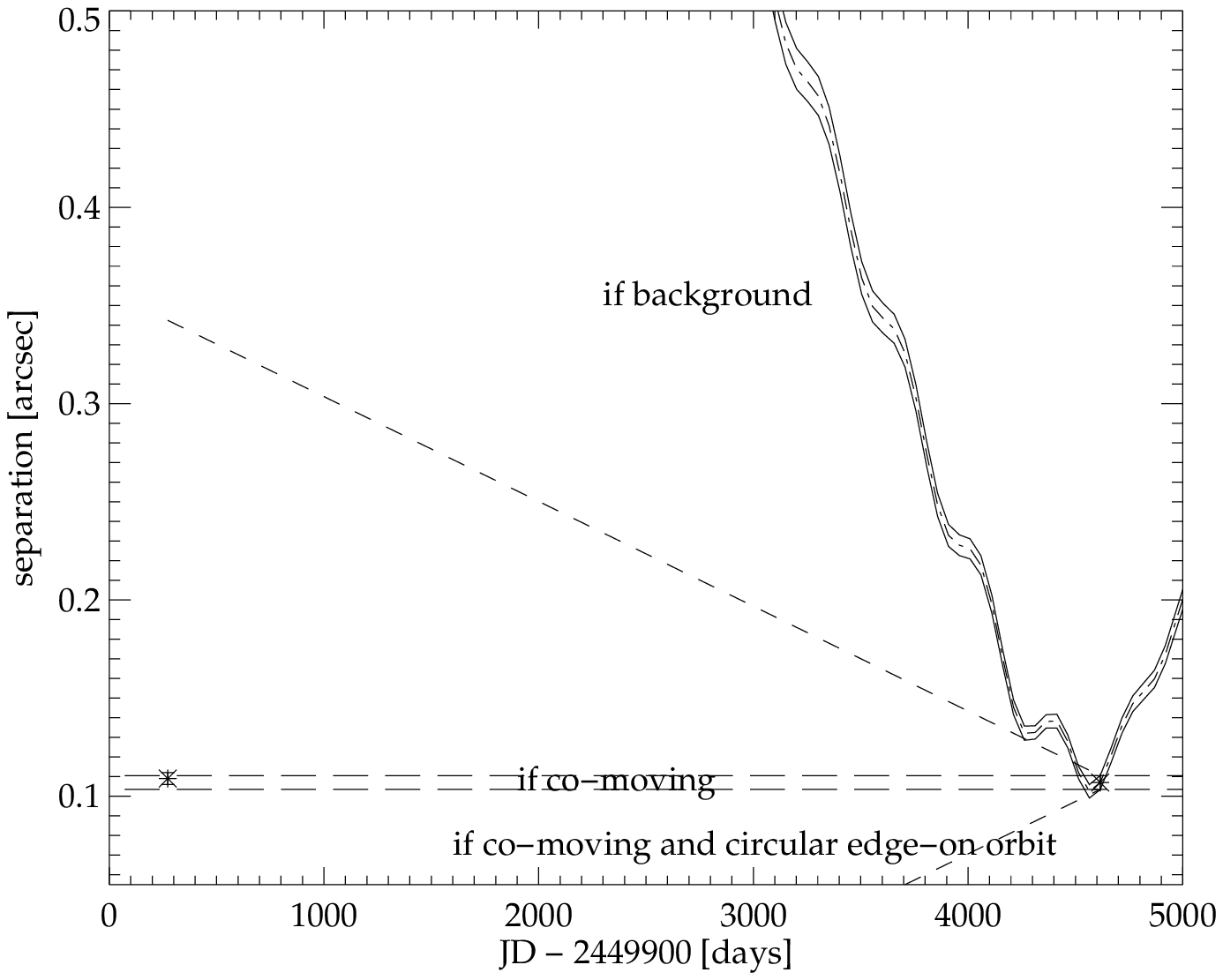}
   \includegraphics[width=0.49\textwidth]{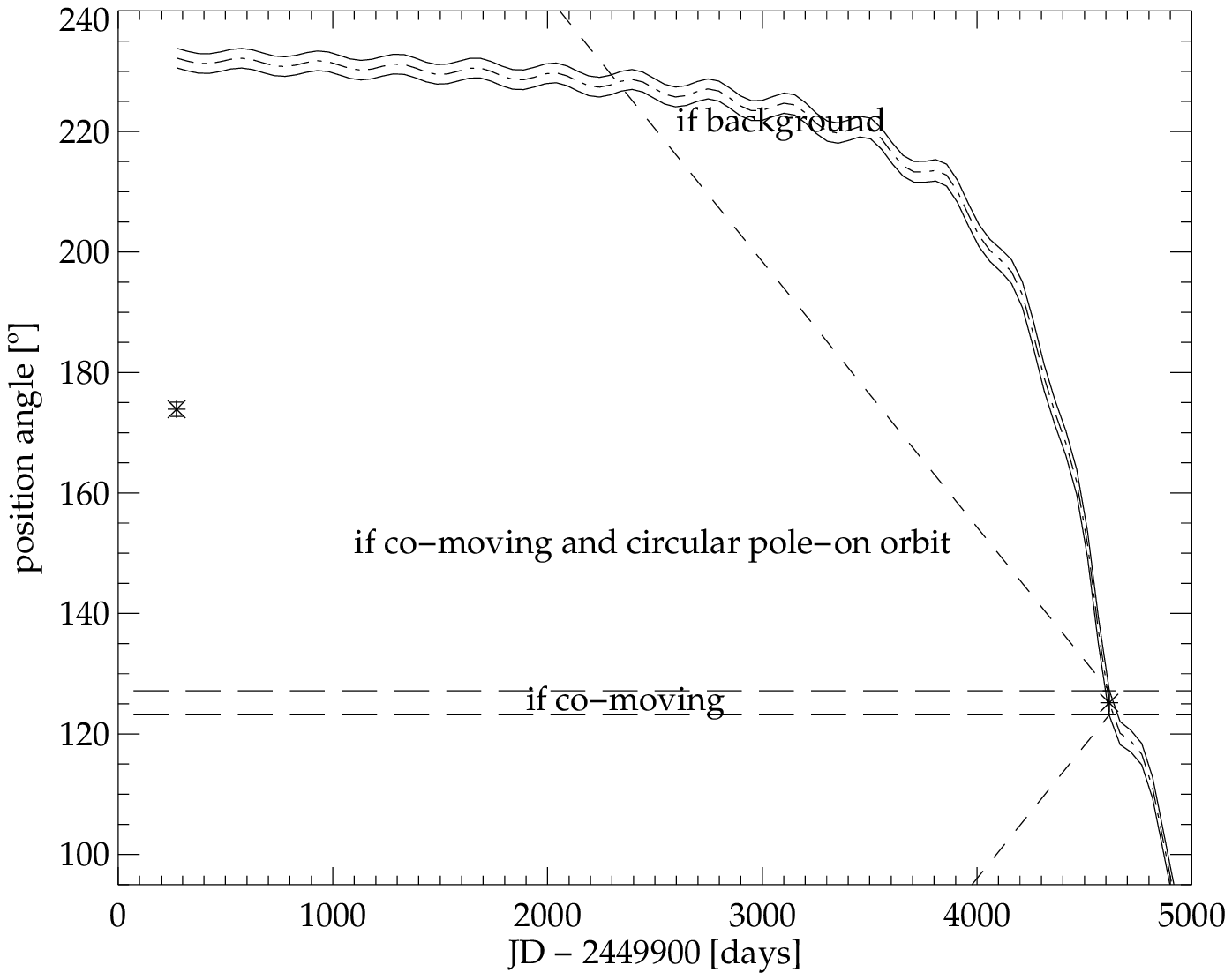}
   \caption{Proper Motion Diagrams (PMD) for separation and position angle change from absolute astrometric measurements 
   of RX J0919.4-7738 B relative to the centroid of Aa \& Ab (top, left to right) and from absolute astrometric measurements 
   of RX J0919.4-7738 Aa relative to RX J0919.4-7738 Ab (bottom, left to right). See text for more information.}
   \label{RX J0919.4-7738}
   \end{figure}
}

\onlfig{3}{
\begin{figure}
   \includegraphics[width=0.49\textwidth]{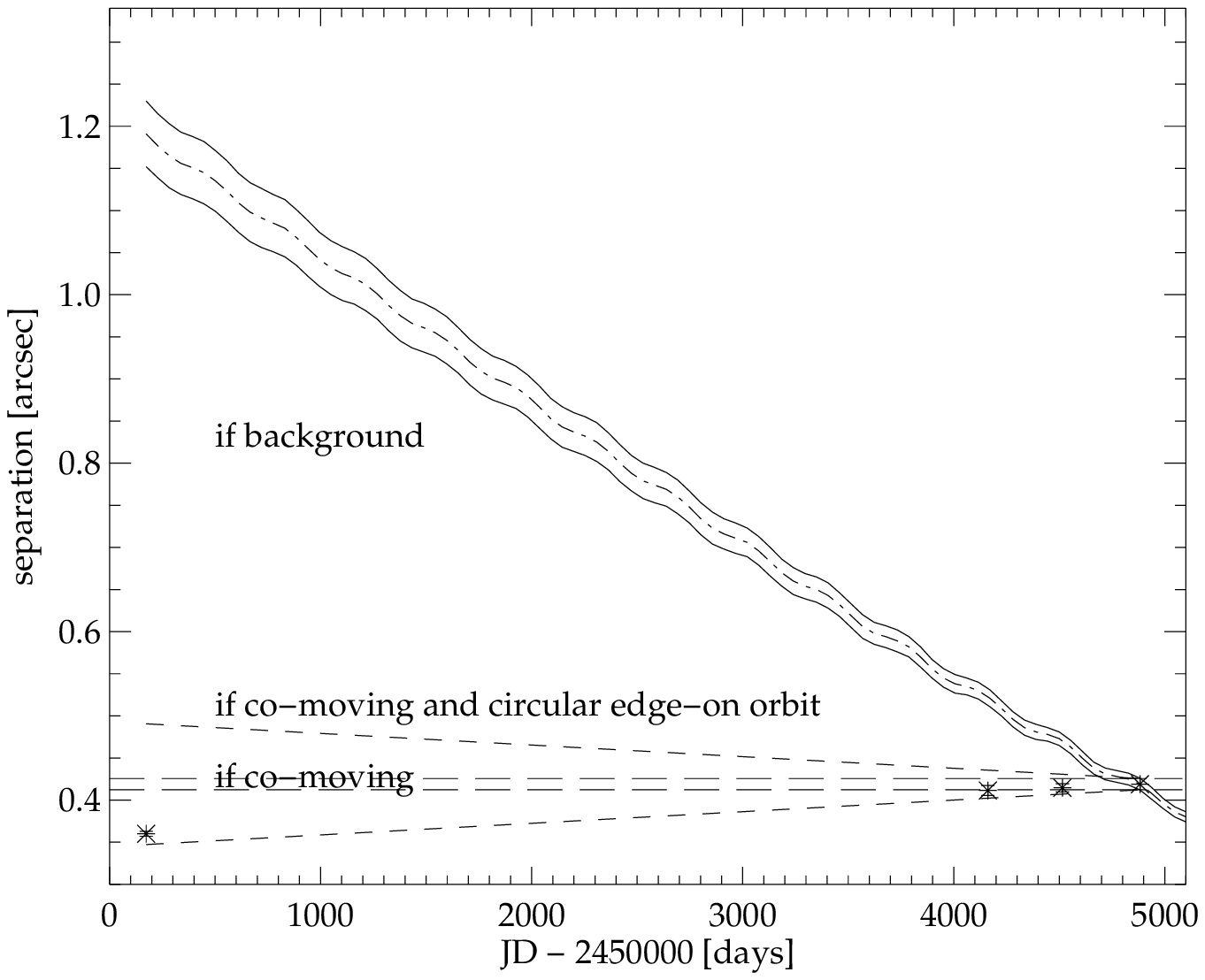}
   \includegraphics[width=0.49\textwidth]{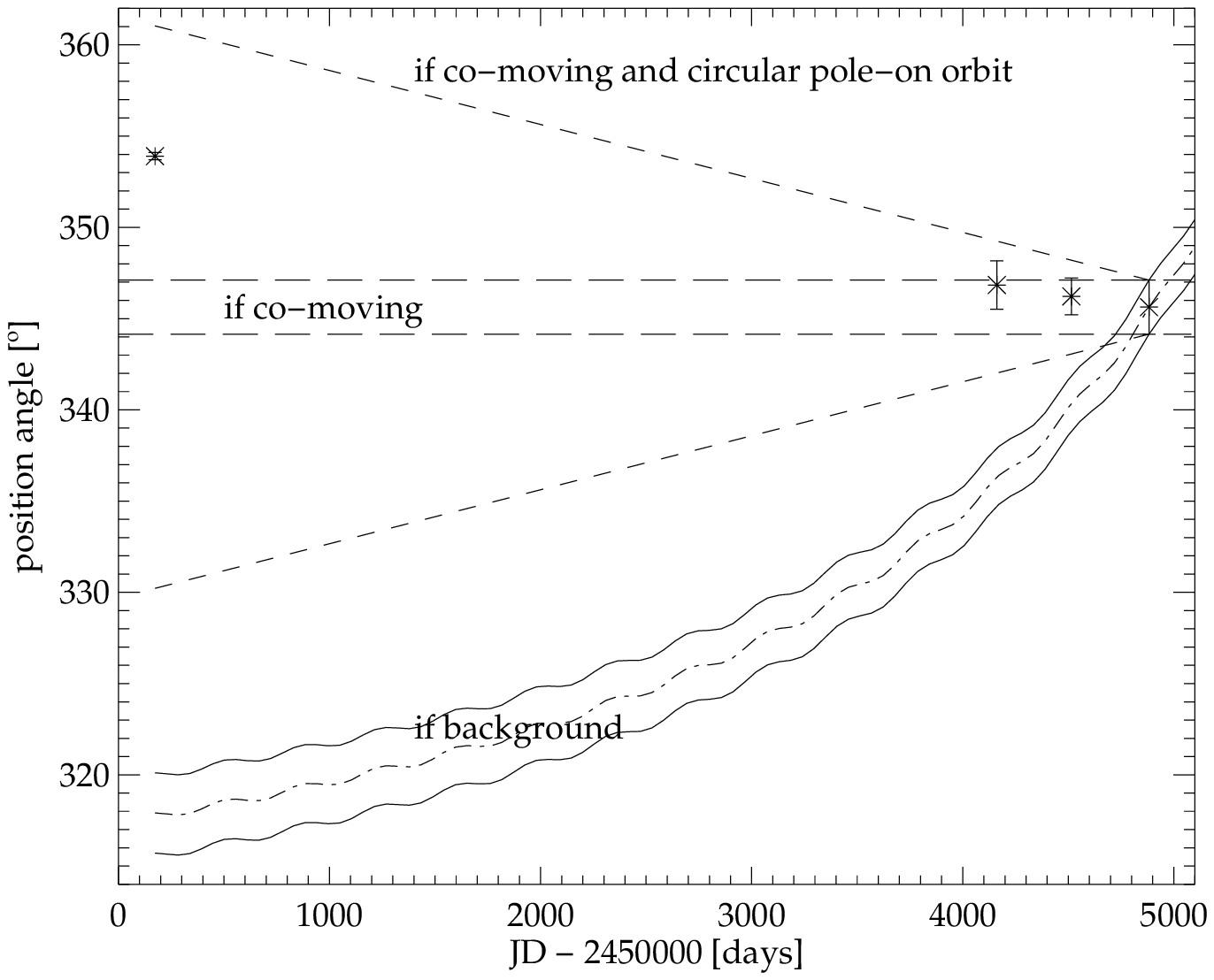}
   \includegraphics[width=0.49\textwidth]{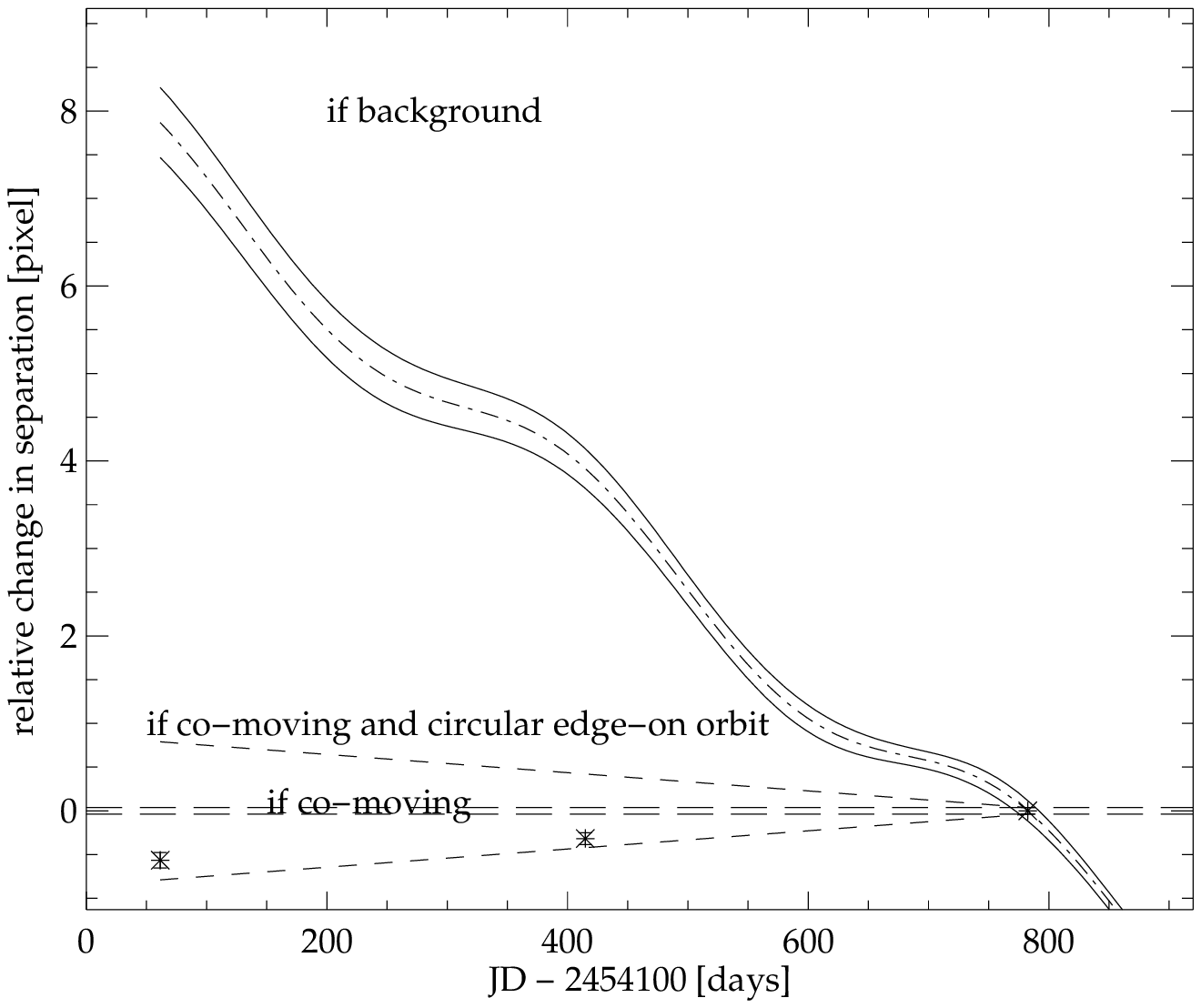}
   \includegraphics[width=0.49\textwidth]{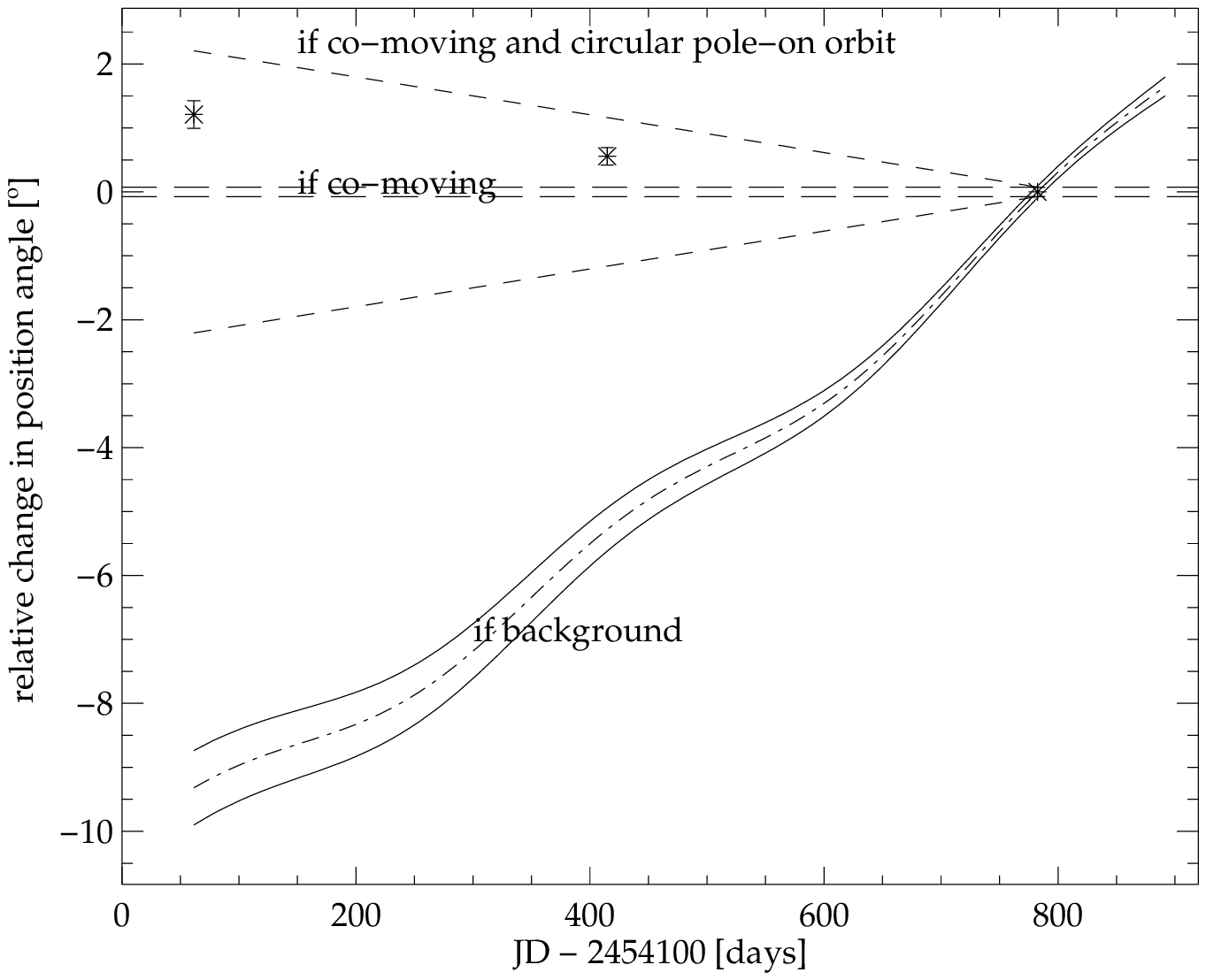}
   \caption{Proper Motion Diagrams (PMD) for separation and position angle change from absolute astrometric measurements (top, left to 
   right) and from relative astrometric measurements (bottom, left to right) in the 
   RX J0935.0-7804 AB system. See text for more information.}
   \label{RX J0935.0-7804}
   \end{figure}
}

\onlfig{4}{
\begin{figure}
   \includegraphics[width=0.49\textwidth]{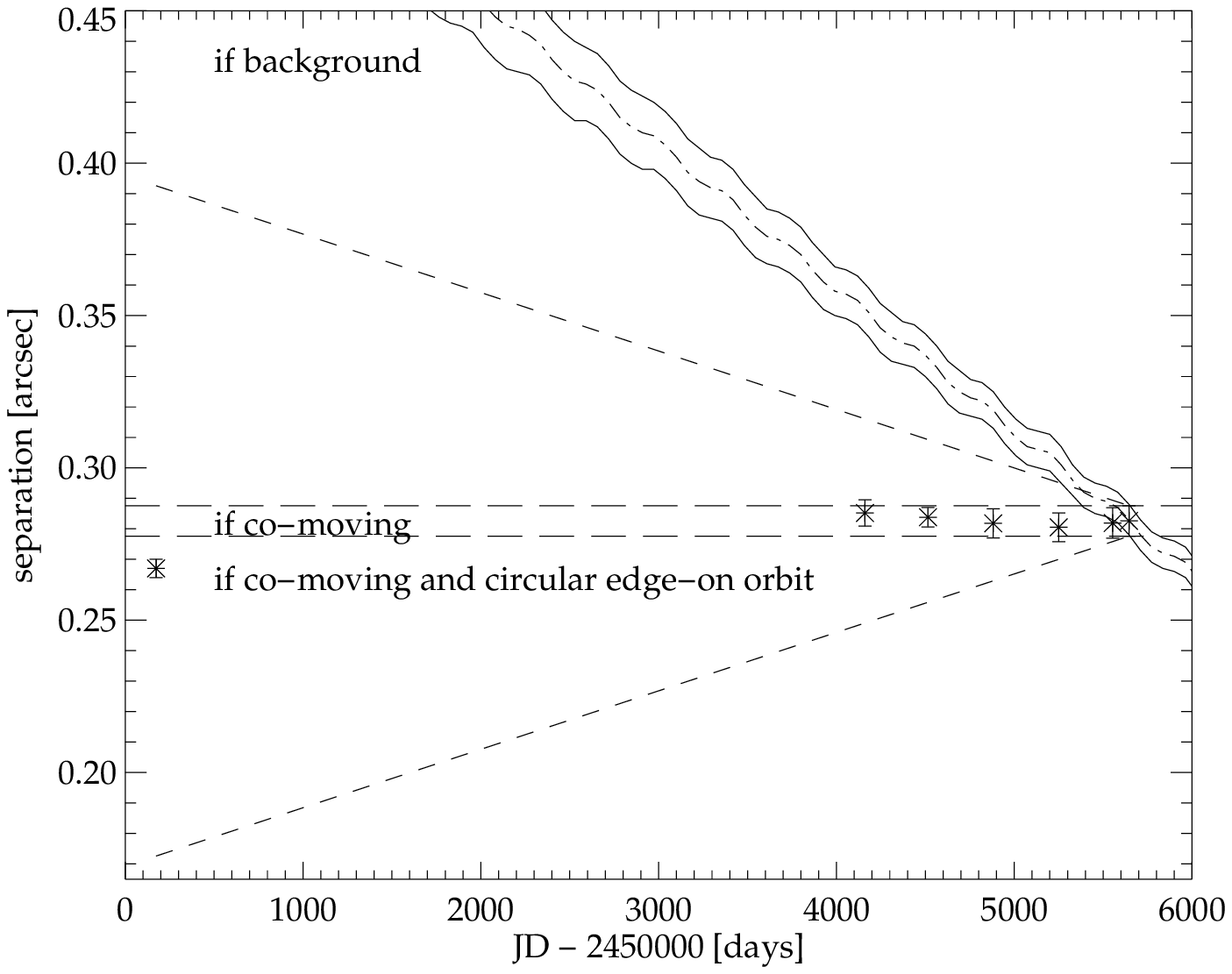}
   \includegraphics[width=0.49\textwidth]{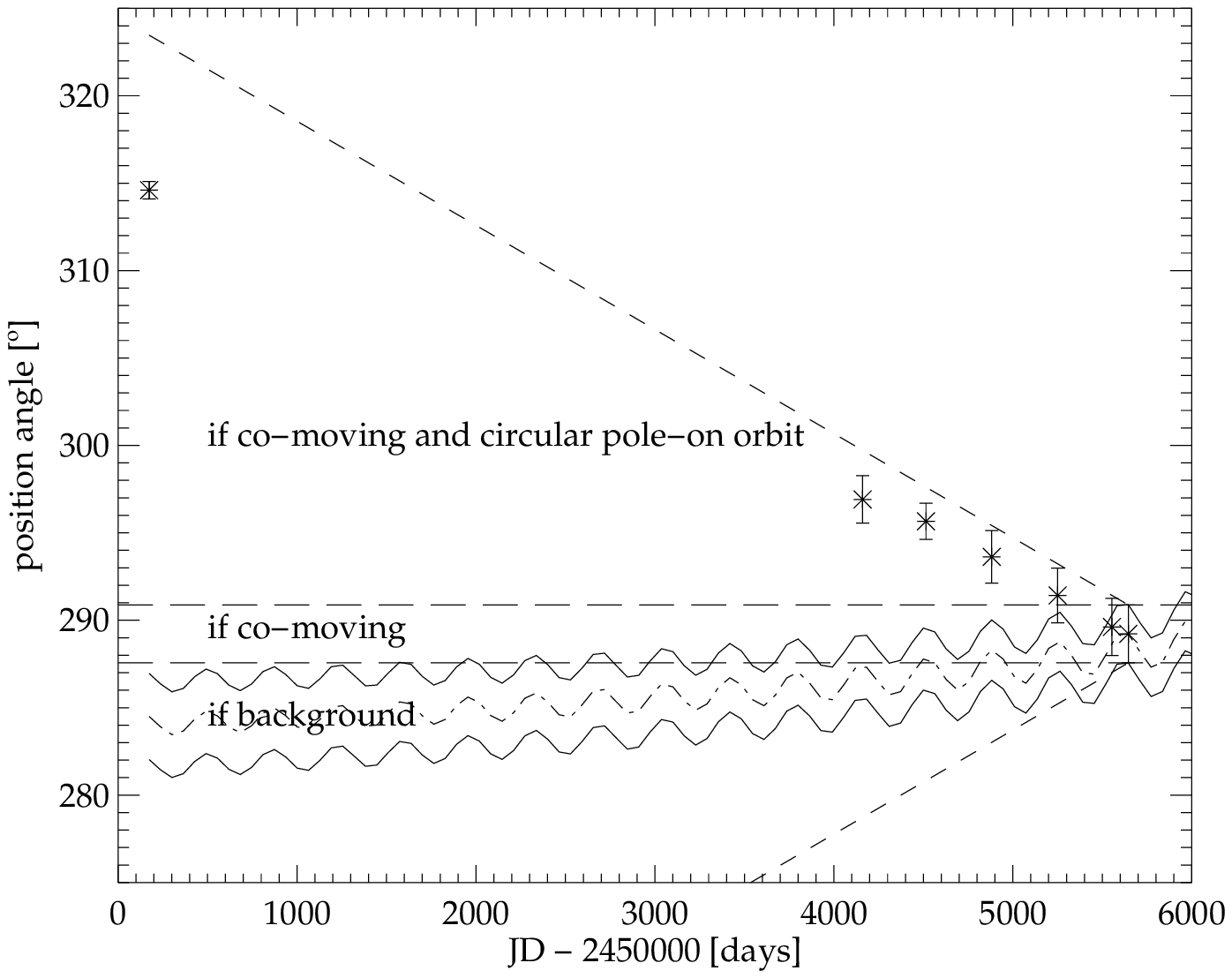}
   \includegraphics[width=0.49\textwidth]{Anhangfigures/RXJ09527_AB_Separation_1107_rel.eps}
   \includegraphics[width=0.49\textwidth]{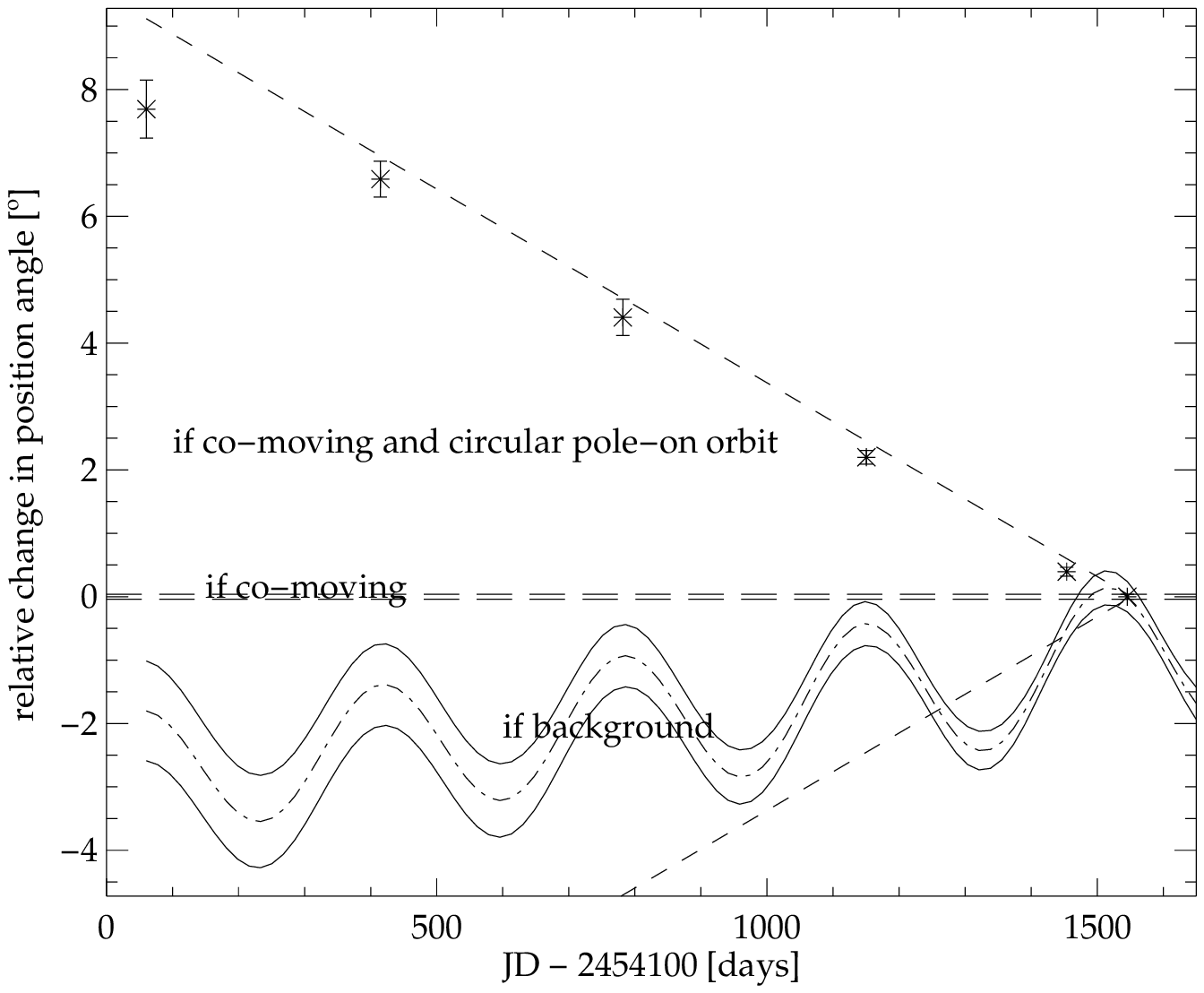}
   \caption{Proper Motion Diagrams (PMD) for separation and position angle change from absolute astrometric measurements (top, left to 
   right) and from relative astrometric measurements (bottom, left to right) in the 
   RX J0952.7-7933 AB system. See text for more information.}
   \label{RX J0952.7-7933}
   \end{figure}
}

\onlfig{5}{
\begin{figure}
   \includegraphics[width=0.49\textwidth]{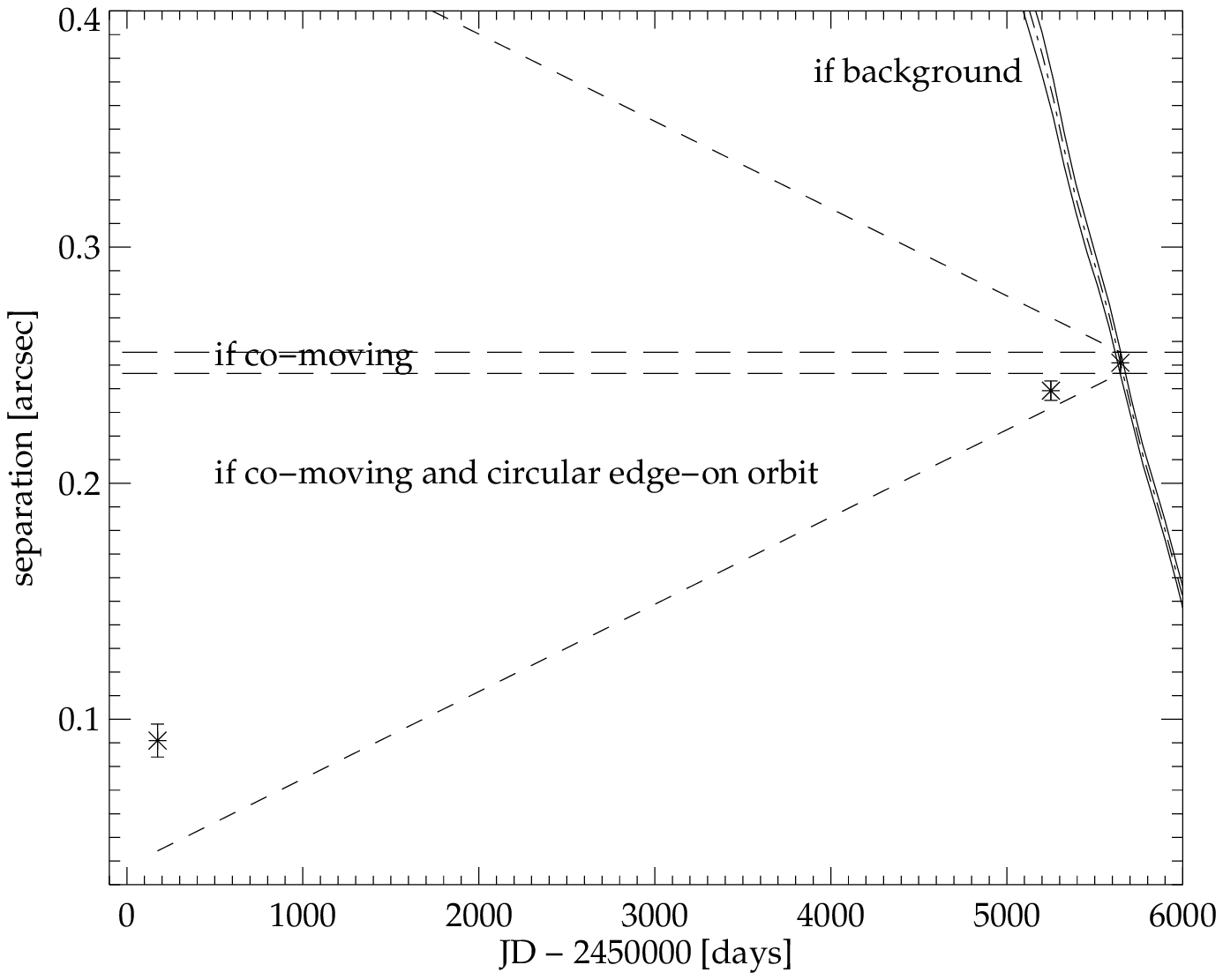}
   \includegraphics[width=0.49\textwidth]{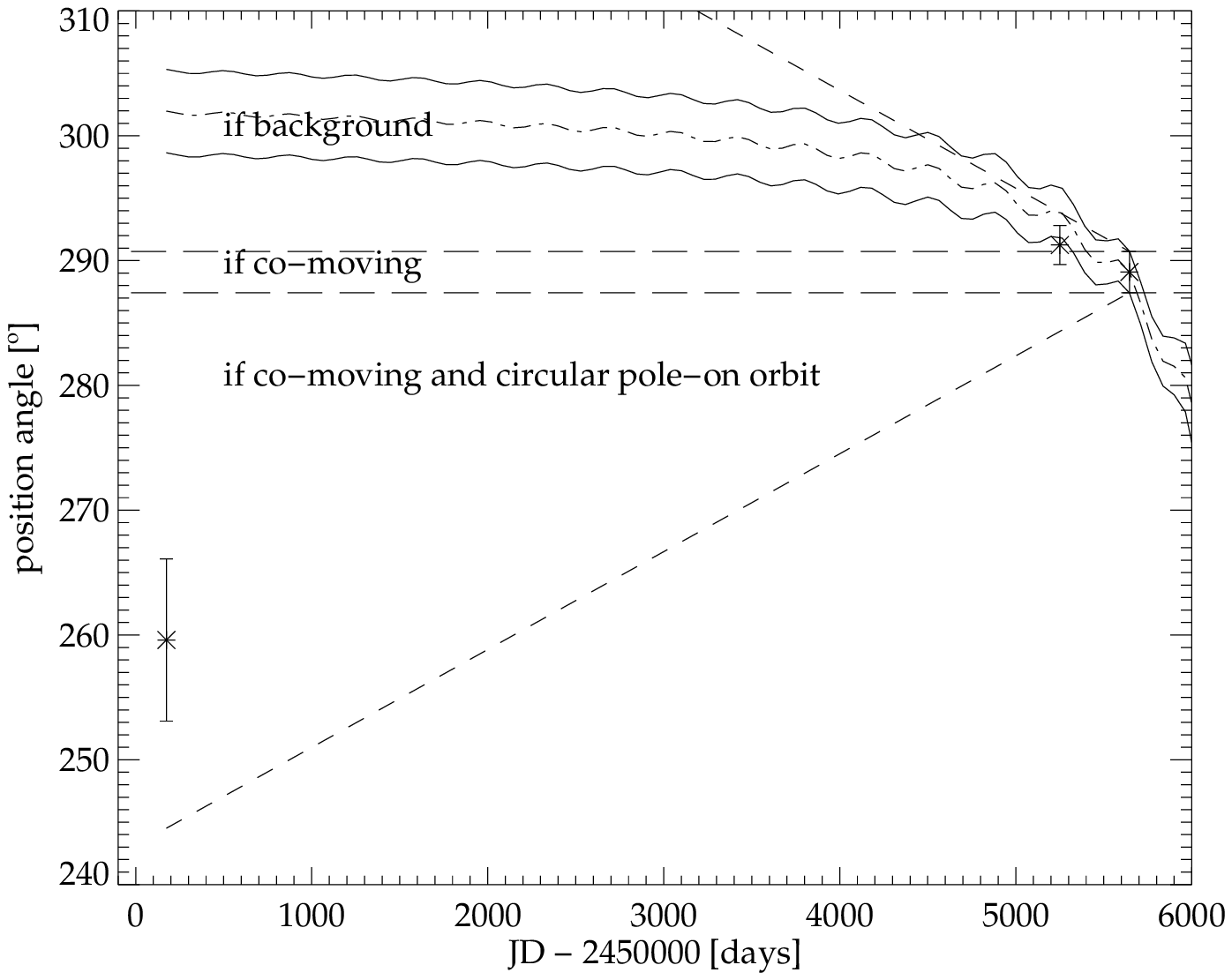}
   \includegraphics[width=0.49\textwidth]{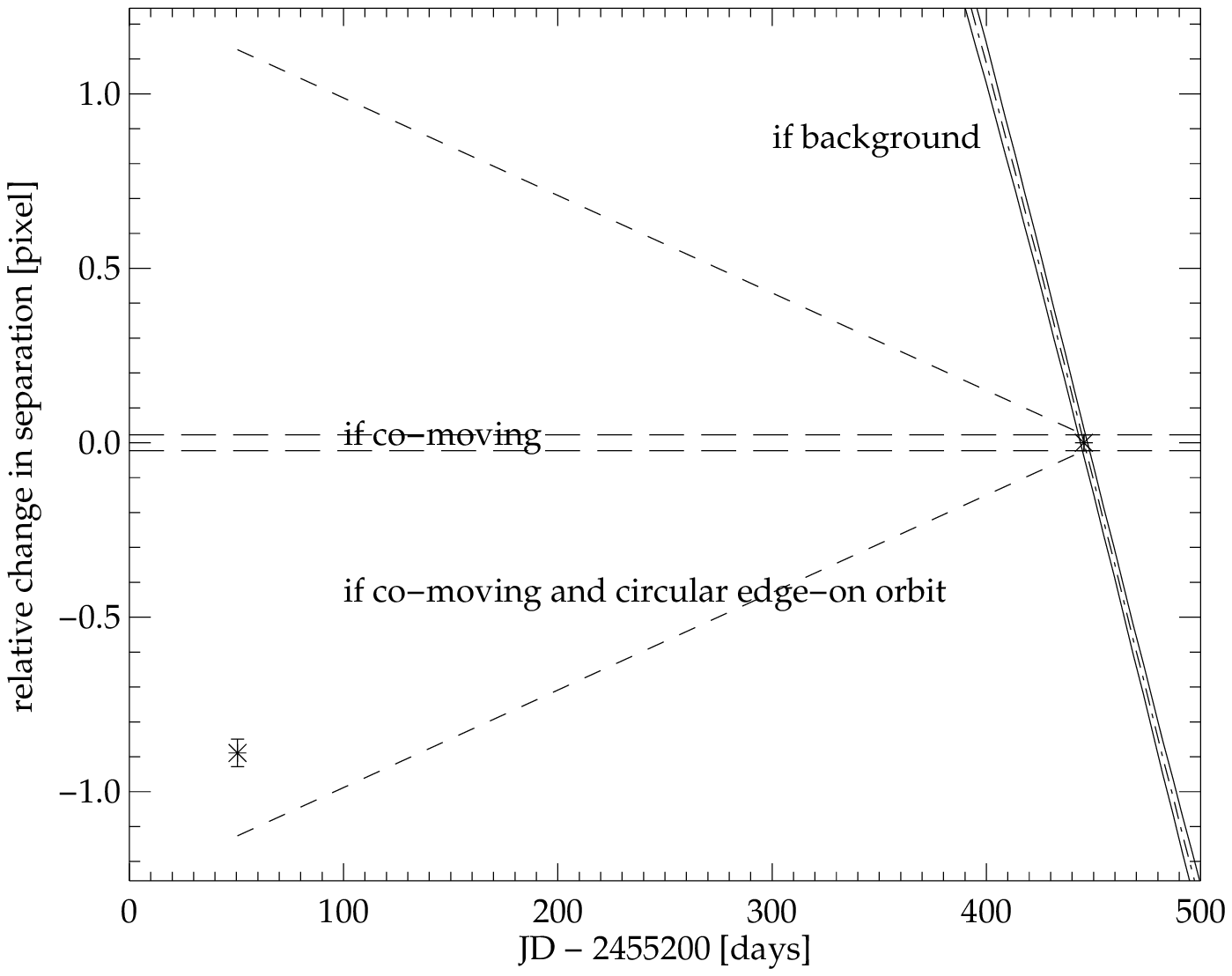}
   \includegraphics[width=0.49\textwidth]{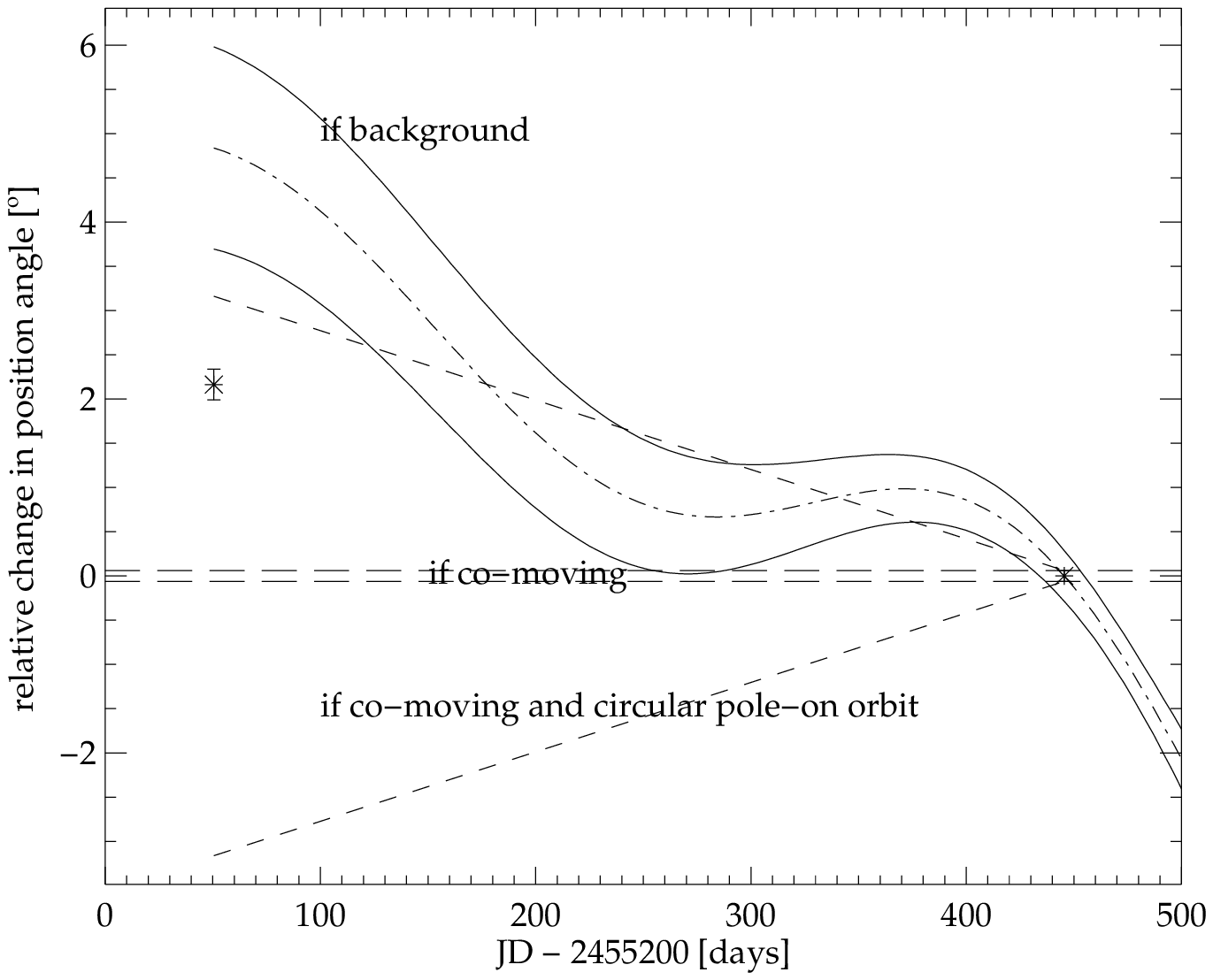}
   \caption{Proper Motion Diagrams (PMD) for separation and position angle change from absolute astrometric measurements (top, left to 
   right) and from relative astrometric measurements (bottom, left to right) in the 
   RX J1014.2-7636 AB system. See text for more information.}
   \label{RX J1014.2-7636}
   \end{figure}
}

\onlfig{6}{
\begin{figure}
   \includegraphics[width=0.49\textwidth]{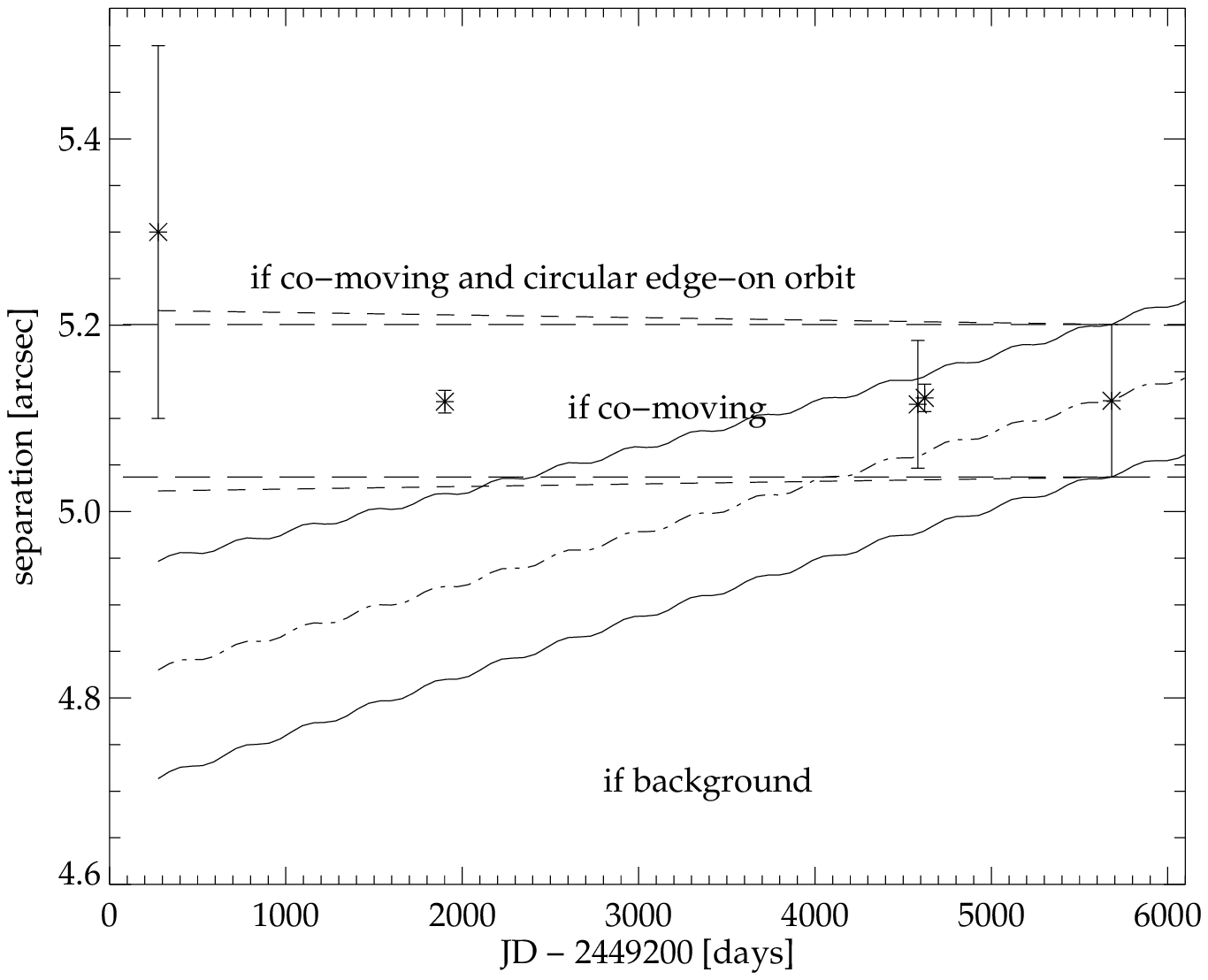}
   \includegraphics[width=0.49\textwidth]{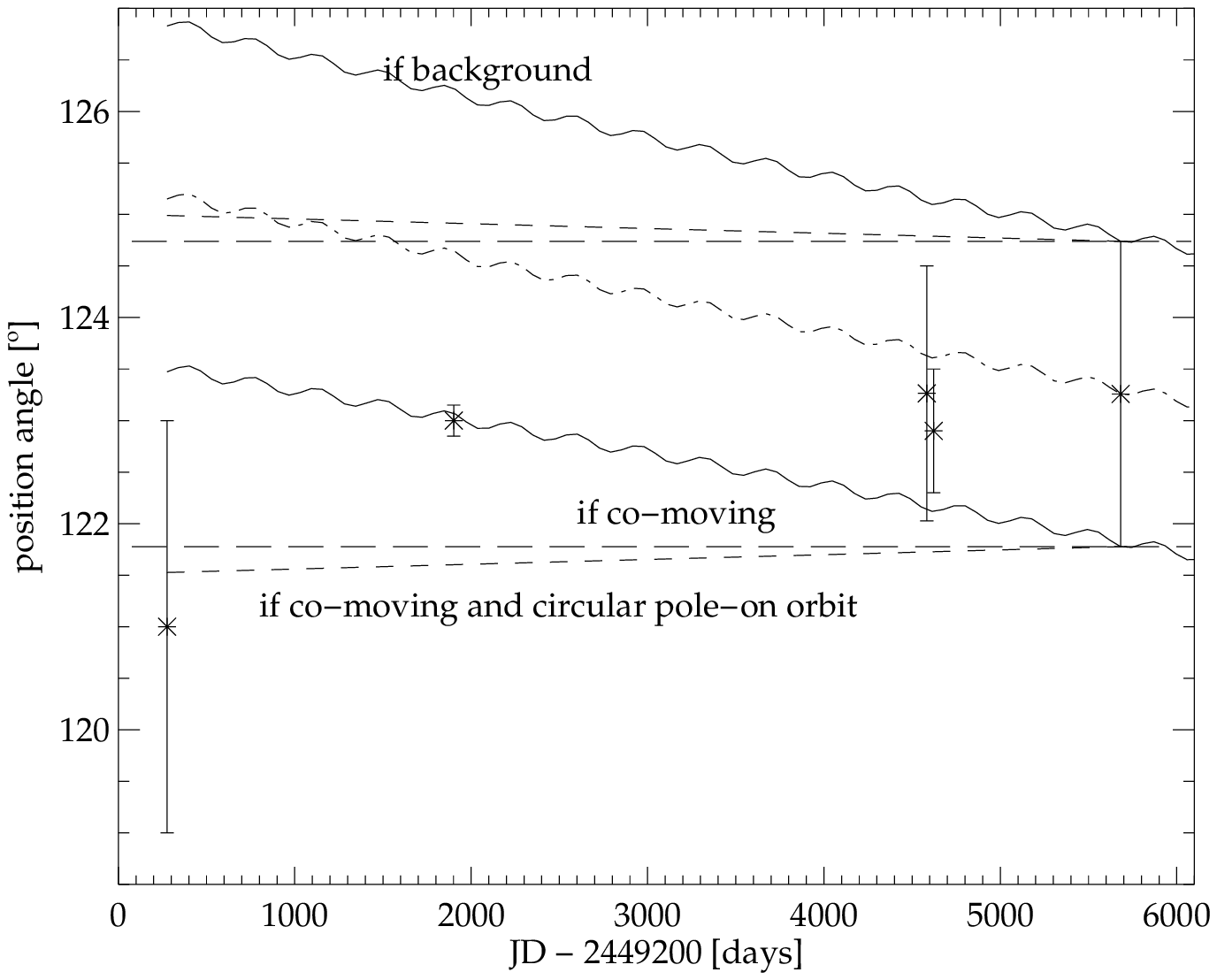}
   \includegraphics[width=0.49\textwidth]{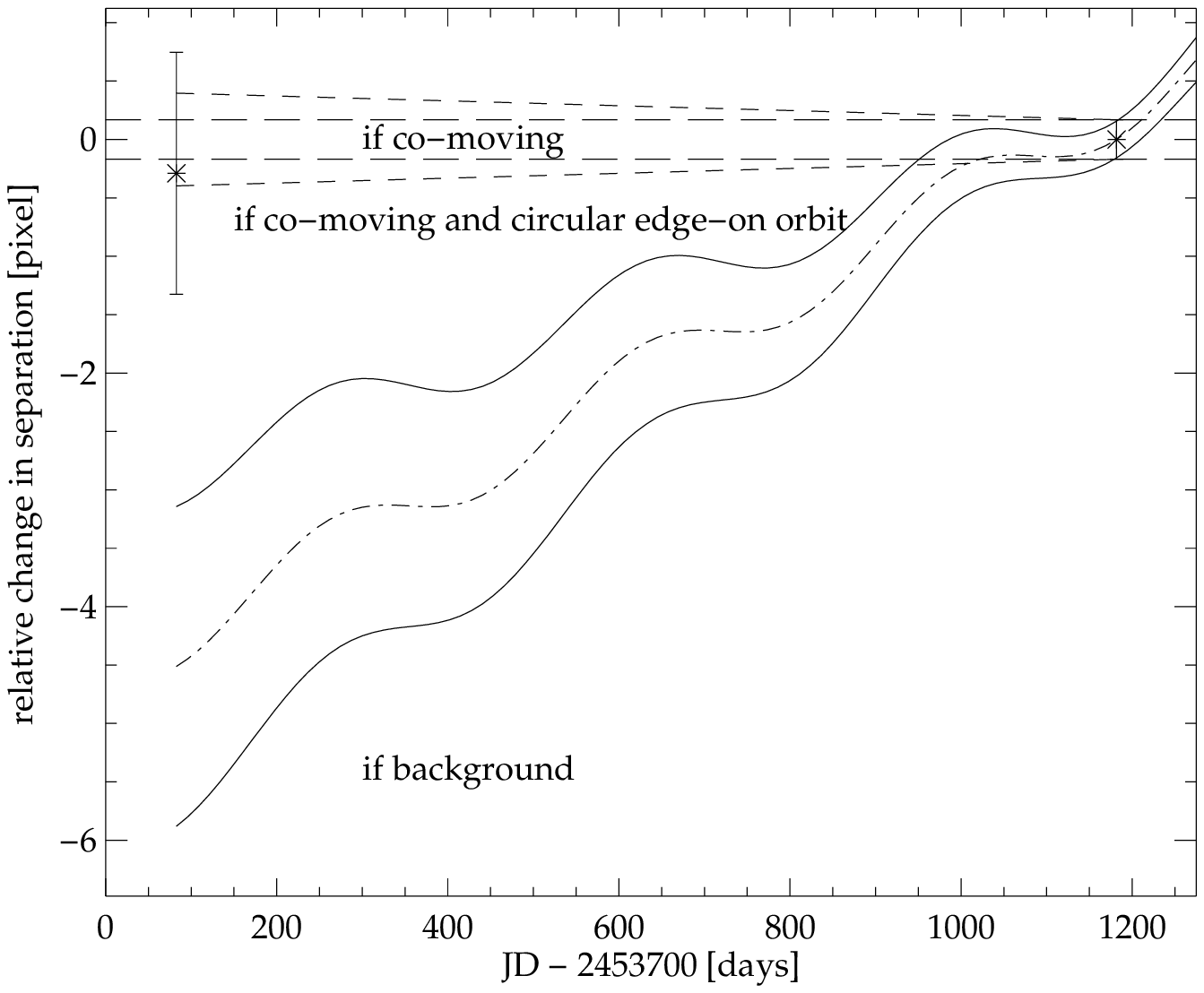}
   \includegraphics[width=0.49\textwidth]{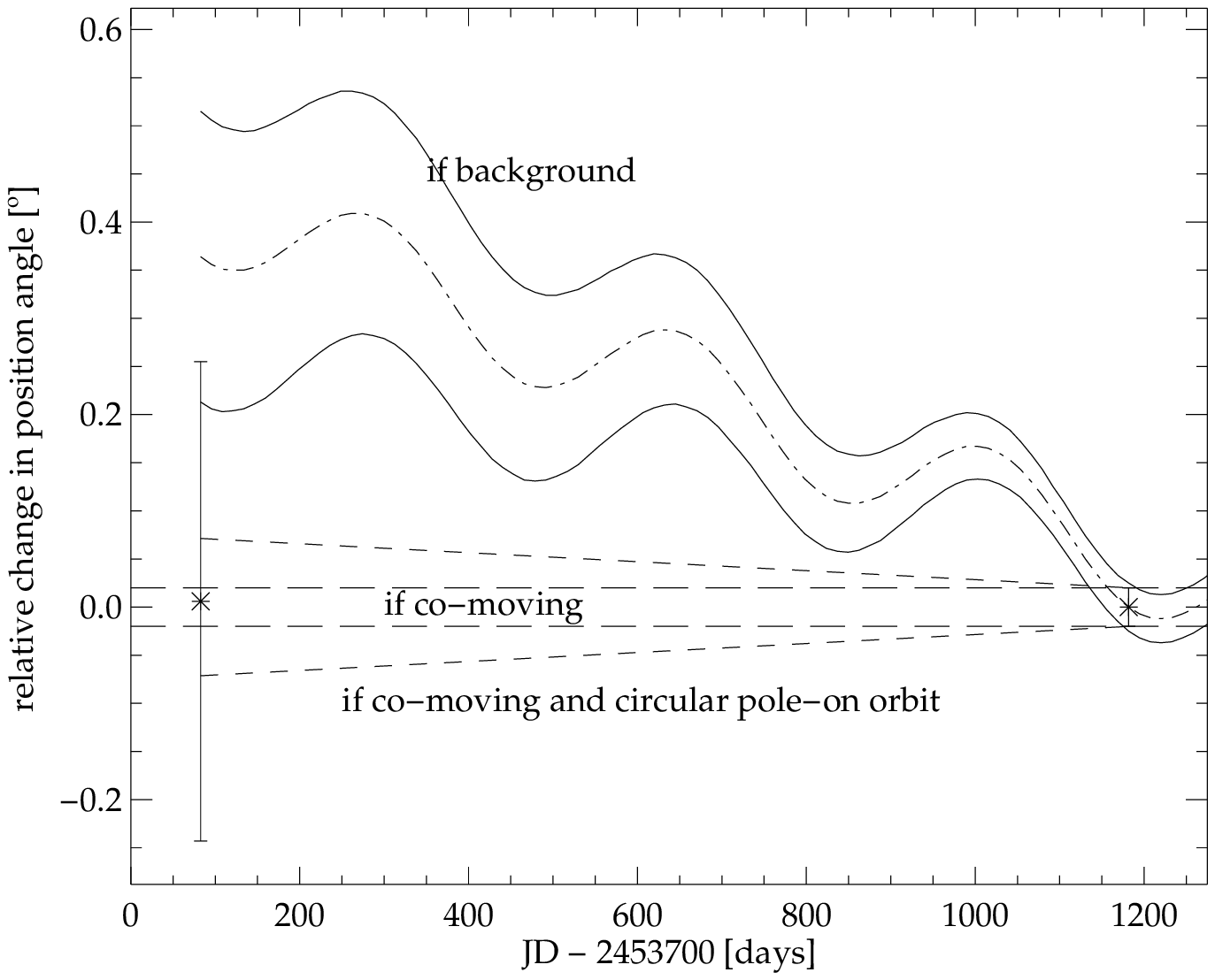}
   \caption{Proper Motion Diagrams (PMD) for separation and position angle change from absolute astrometric measurements (top, left to 
   right) and from relative astrometric measurements (bottom, left to right) in the 
   SZ Cha AB system. See text for more information.}
   \label{SZ Cha}
   \end{figure}
}

\onlfig{7}{
\begin{figure}
   \includegraphics[width=0.49\textwidth]{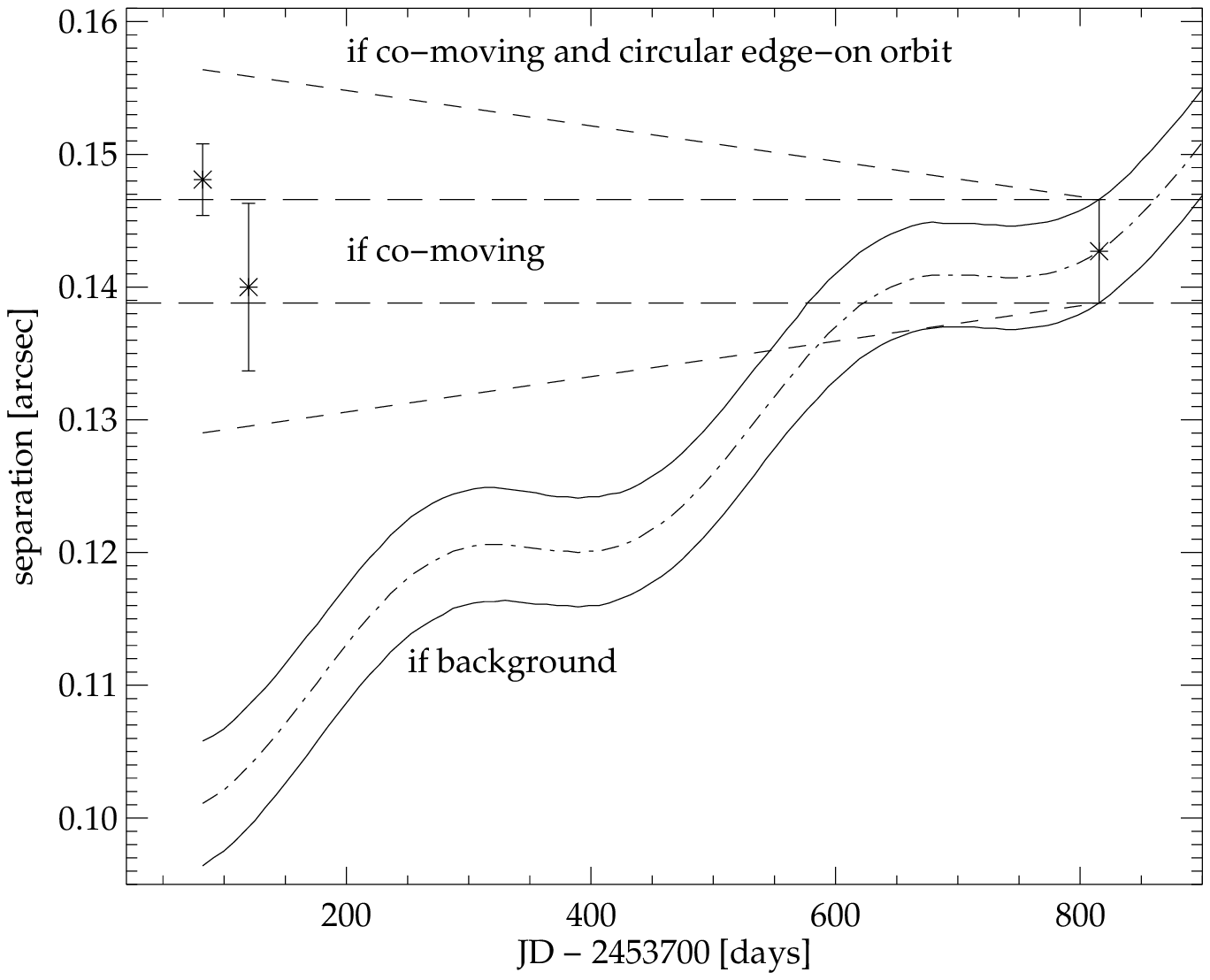}
   \includegraphics[width=0.49\textwidth]{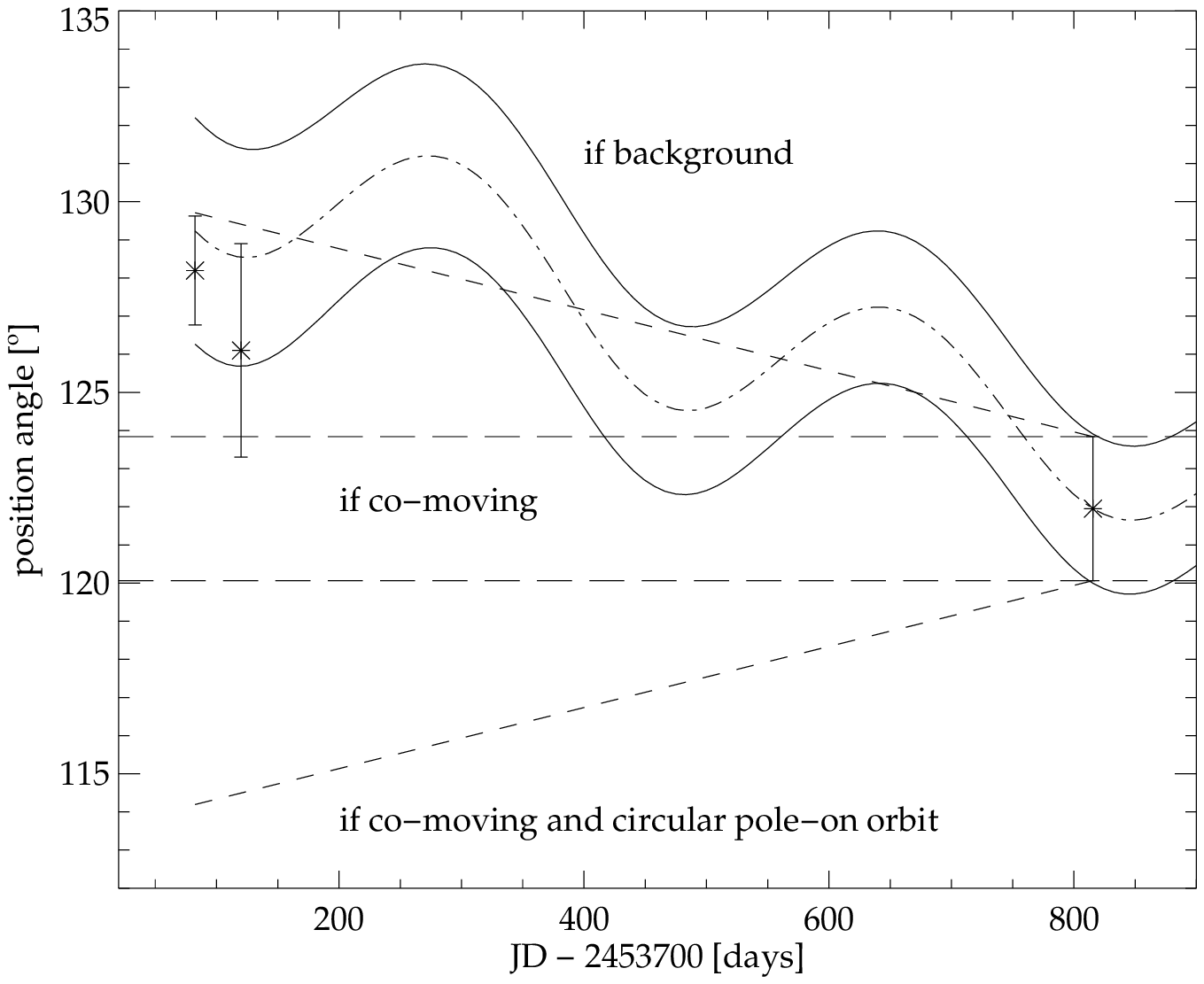}
   \includegraphics[width=0.49\textwidth]{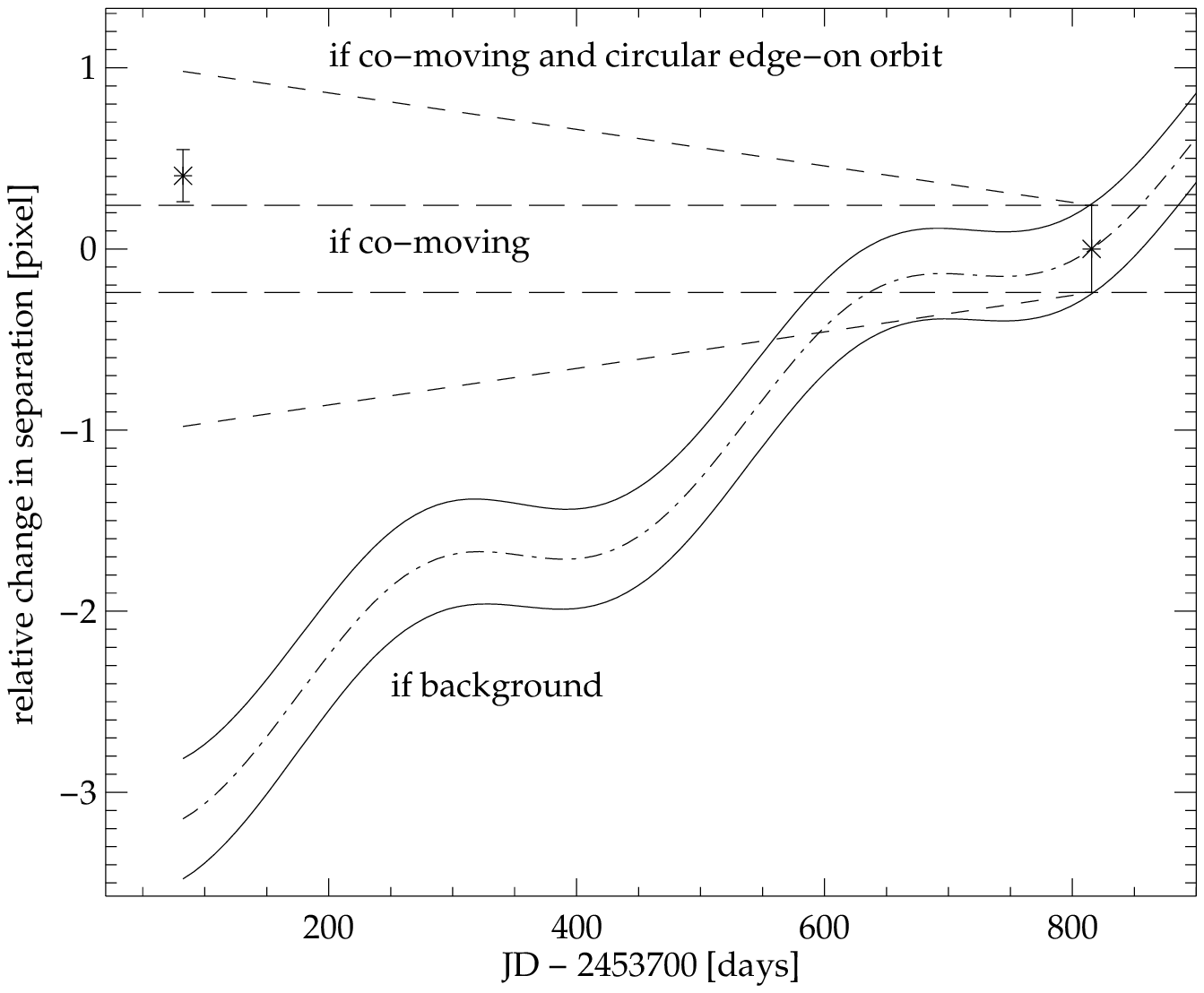}
   \includegraphics[width=0.49\textwidth]{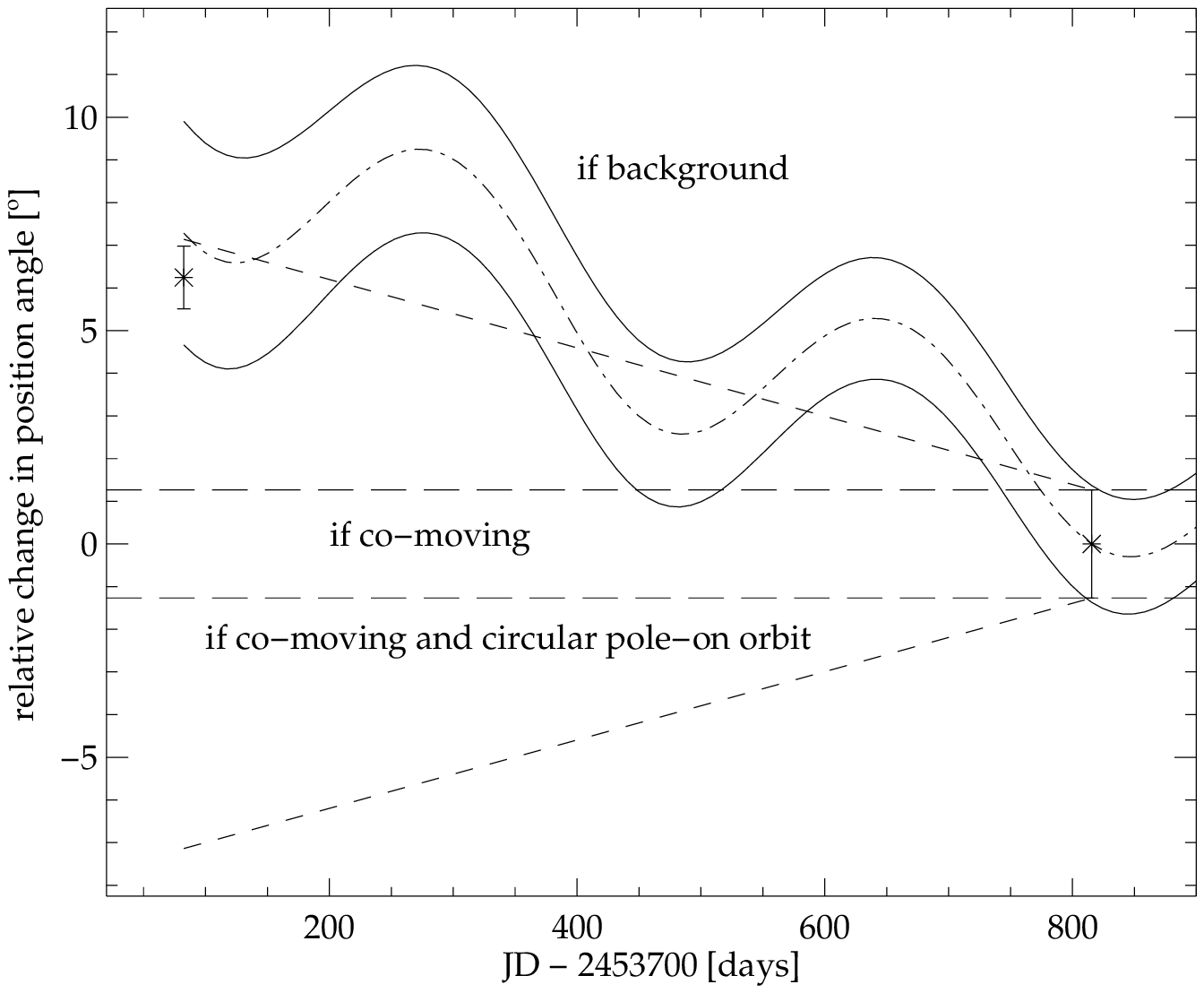}
   \caption{Proper Motion Diagrams (PMD) for separation and position angle change from absolute astrometric measurements (top, left to 
   right) and from relative astrometric measurements (bottom, left to right) in the 
   Ced 110 IRS 2 AB system. See text for more information.}
   \label{Ced 110 IRS 2}
   \end{figure}
}

\onlfig{8}{
\begin{figure}
   \includegraphics[width=0.49\textwidth]{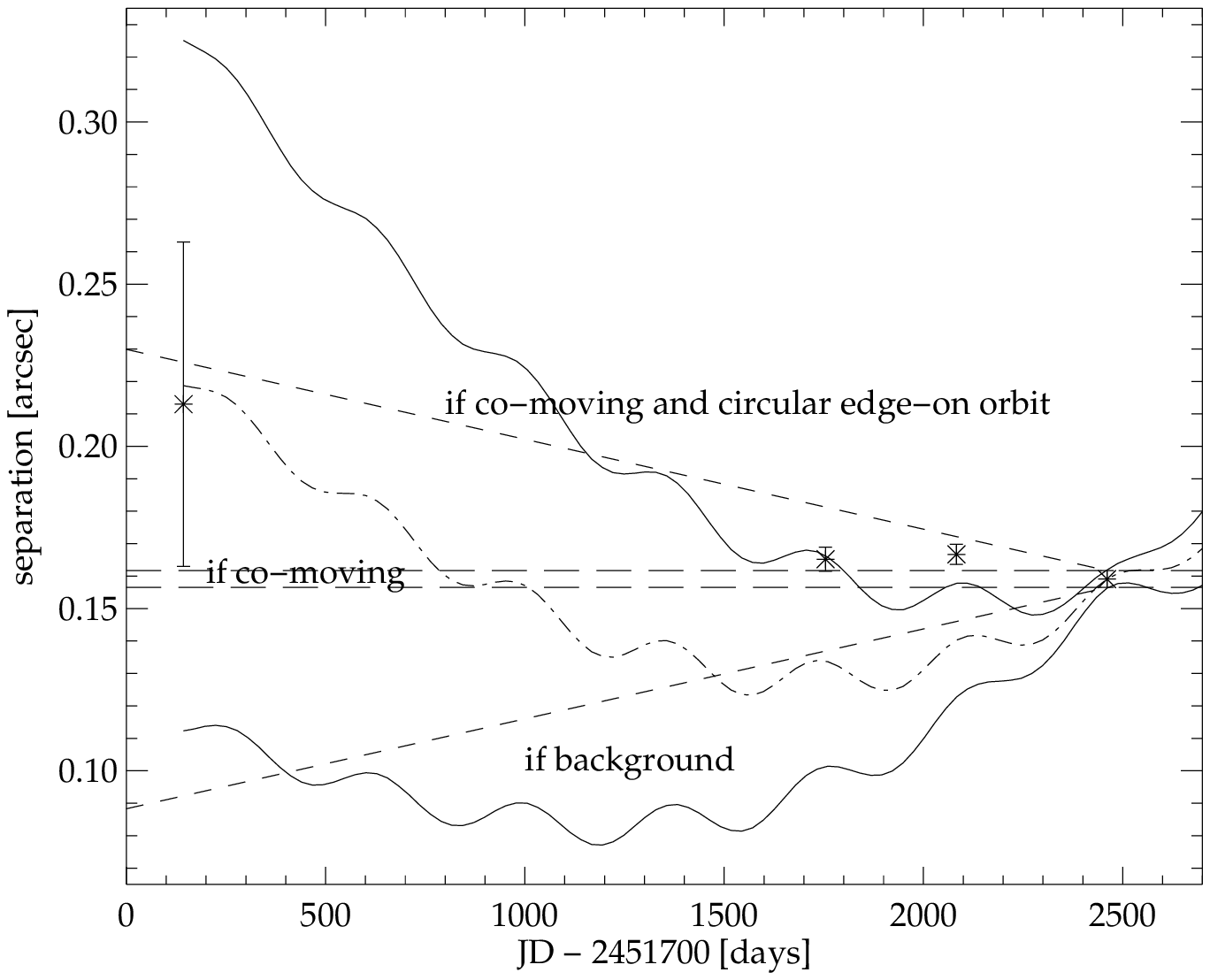}
   \includegraphics[width=0.49\textwidth]{Anhangfigures/ChaHa2_AB_SW_Poswinkel_0700_abs.eps}
   \includegraphics[width=0.49\textwidth]{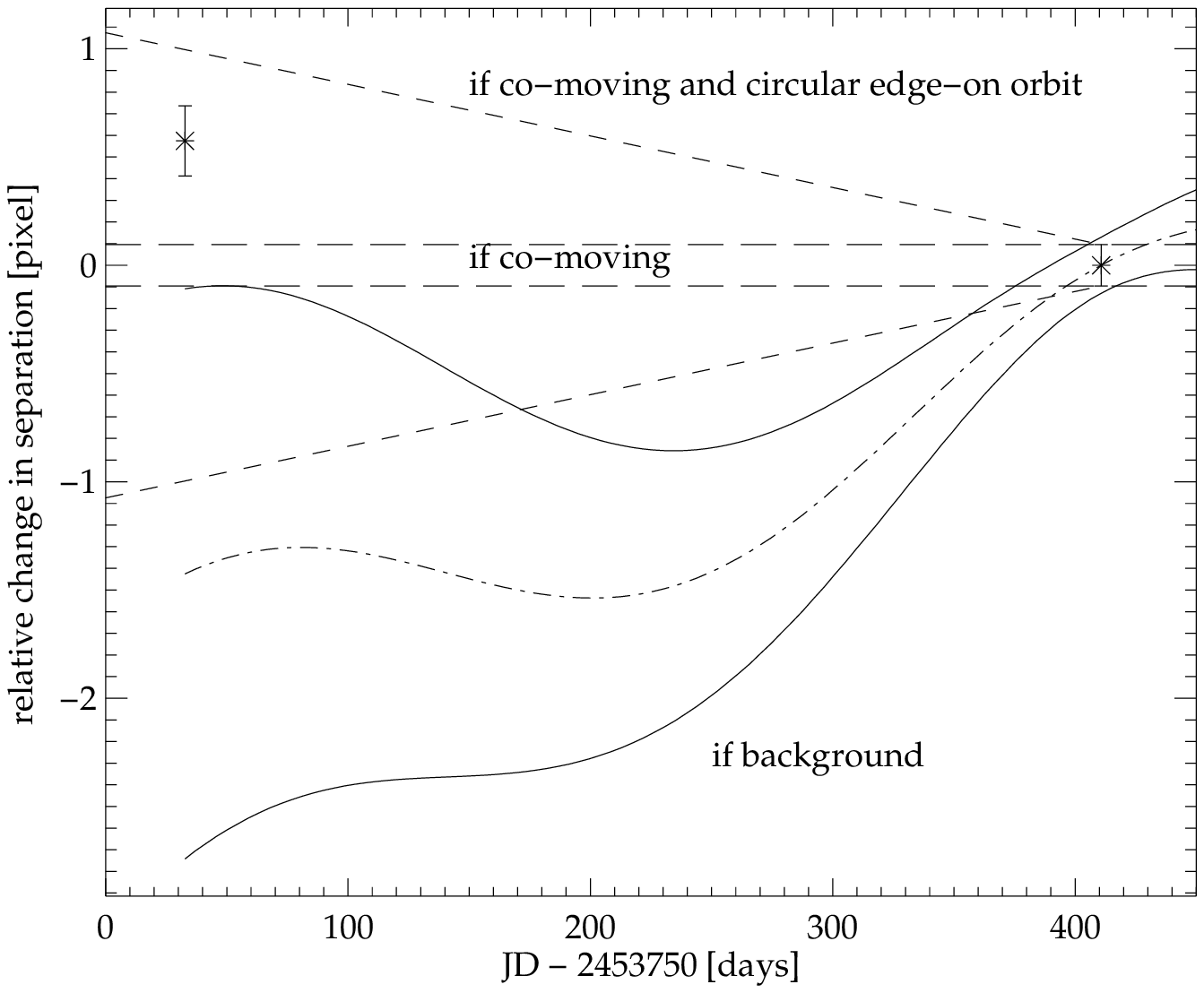}
   \includegraphics[width=0.49\textwidth]{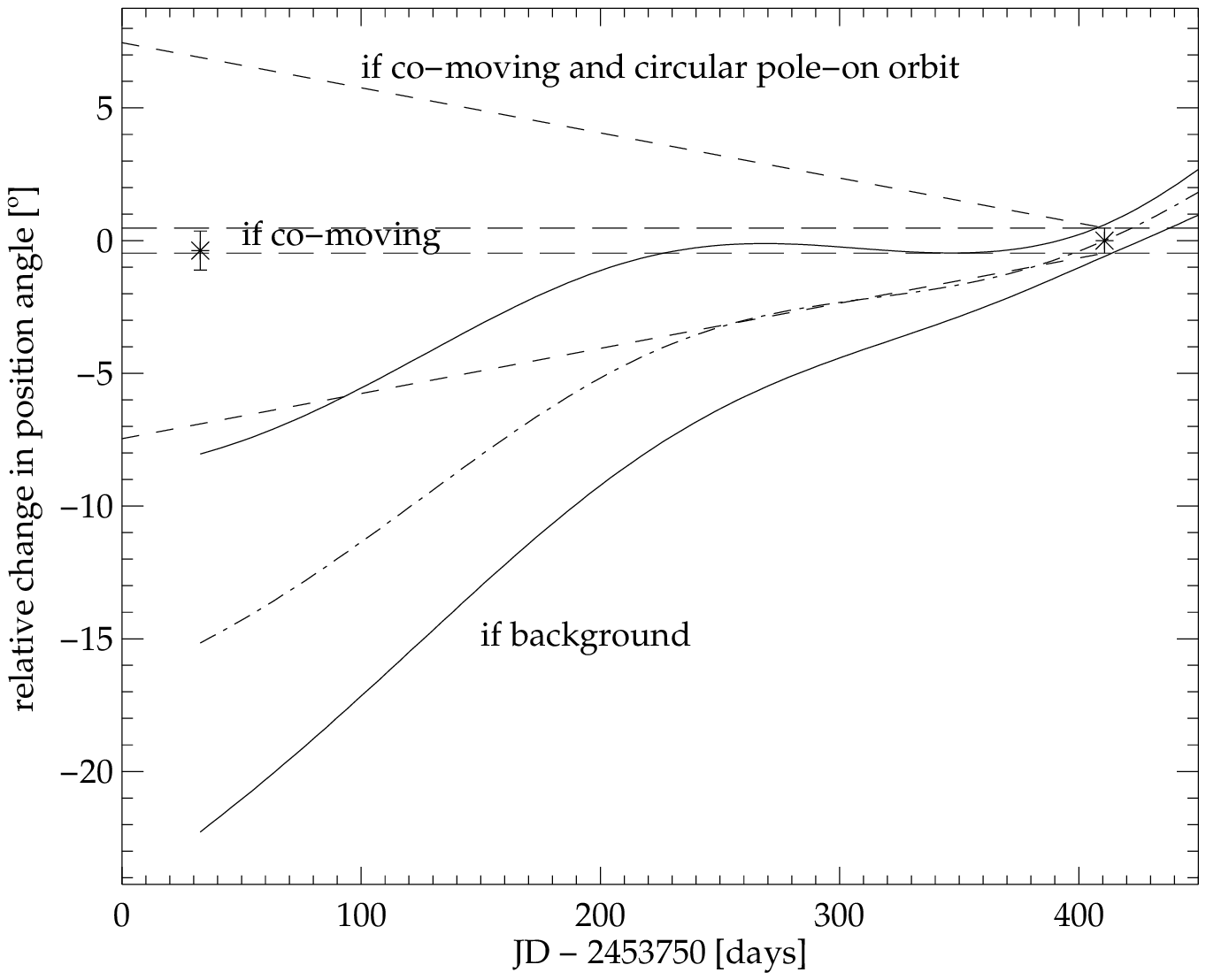}
   \caption{Proper Motion Diagrams (PMD) for separation and position angle change from absolute astrometric measurements (top, left to 
   right) and from relative astrometric measurements (bottom, left to right) in the 
   Cha H$\alpha$ 2 AB system. See text for more information.}
   \label{Cha Ha 2}
   \end{figure}
}

\onlfig{9}{
\begin{figure}
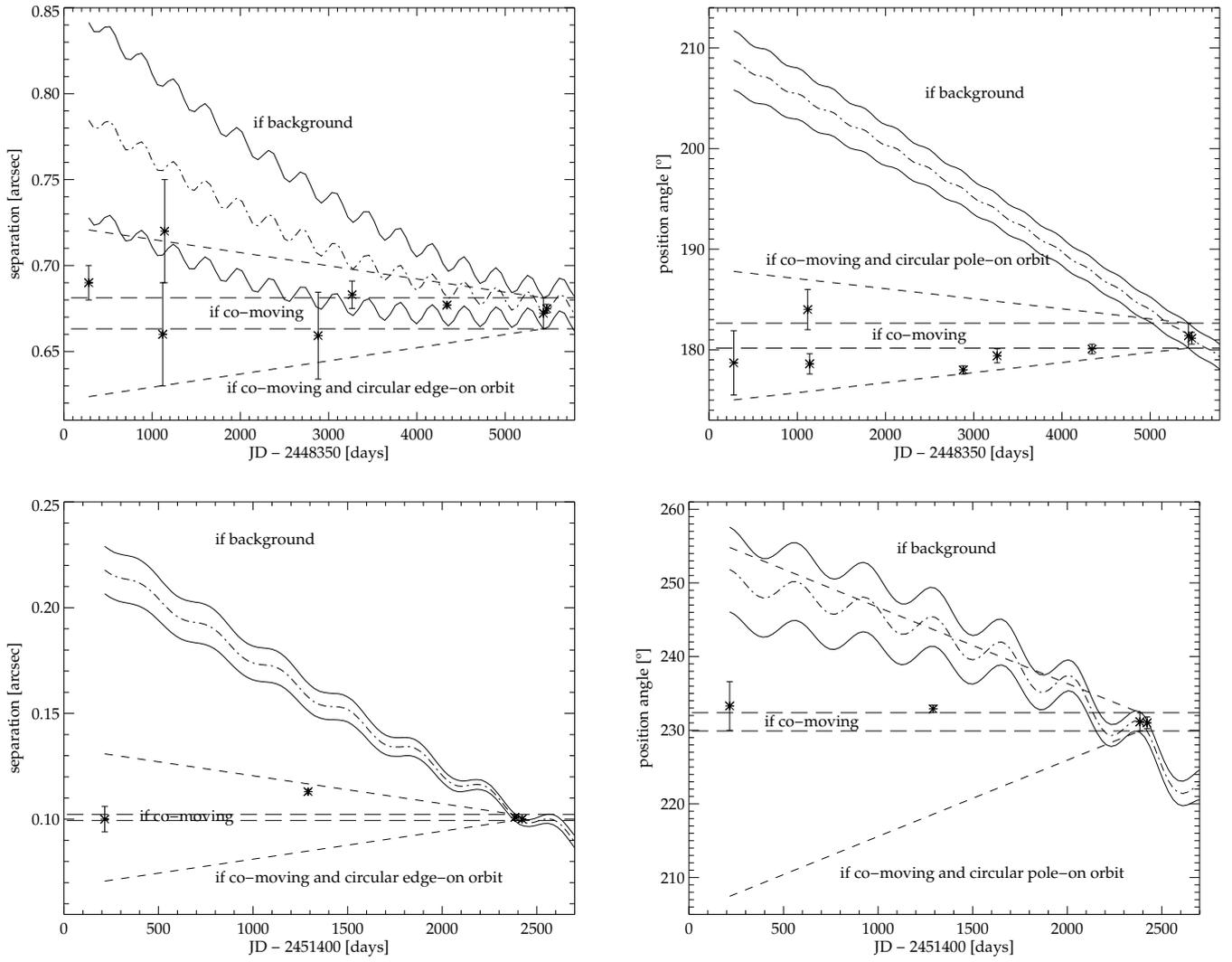

   \includegraphics[width=0.49\textwidth]{Anhangfigures/VWCha_A_BC_Separation_0994_abs.eps}
   \includegraphics[width=0.49\textwidth]{Anhangfigures/VWCha_A_BC_Poswinkel_0994_abs.eps}
   \includegraphics[width=0.49\textwidth]{Anhangfigures/VWCha_BC_Separation_0900_abs.eps}
   \includegraphics[width=0.49\textwidth]{Anhangfigures/VWCha_BC_Poswinkel_0900_abs.eps}
   \caption{Proper Motion Diagrams (PMD) for separation and position angle change from absolute astrometric measurements of VW Cha A 
   relative to the centroid of B \& C (top, left to right) and from absolute astrometric measurements of VW Cha B relative to VW Cha C
   (bottom, left to right). See text for more information.}
   \label{VW Cha}
   \end{figure}
}

\onlfig{10}{
\begin{figure}
   \includegraphics[width=0.49\textwidth]{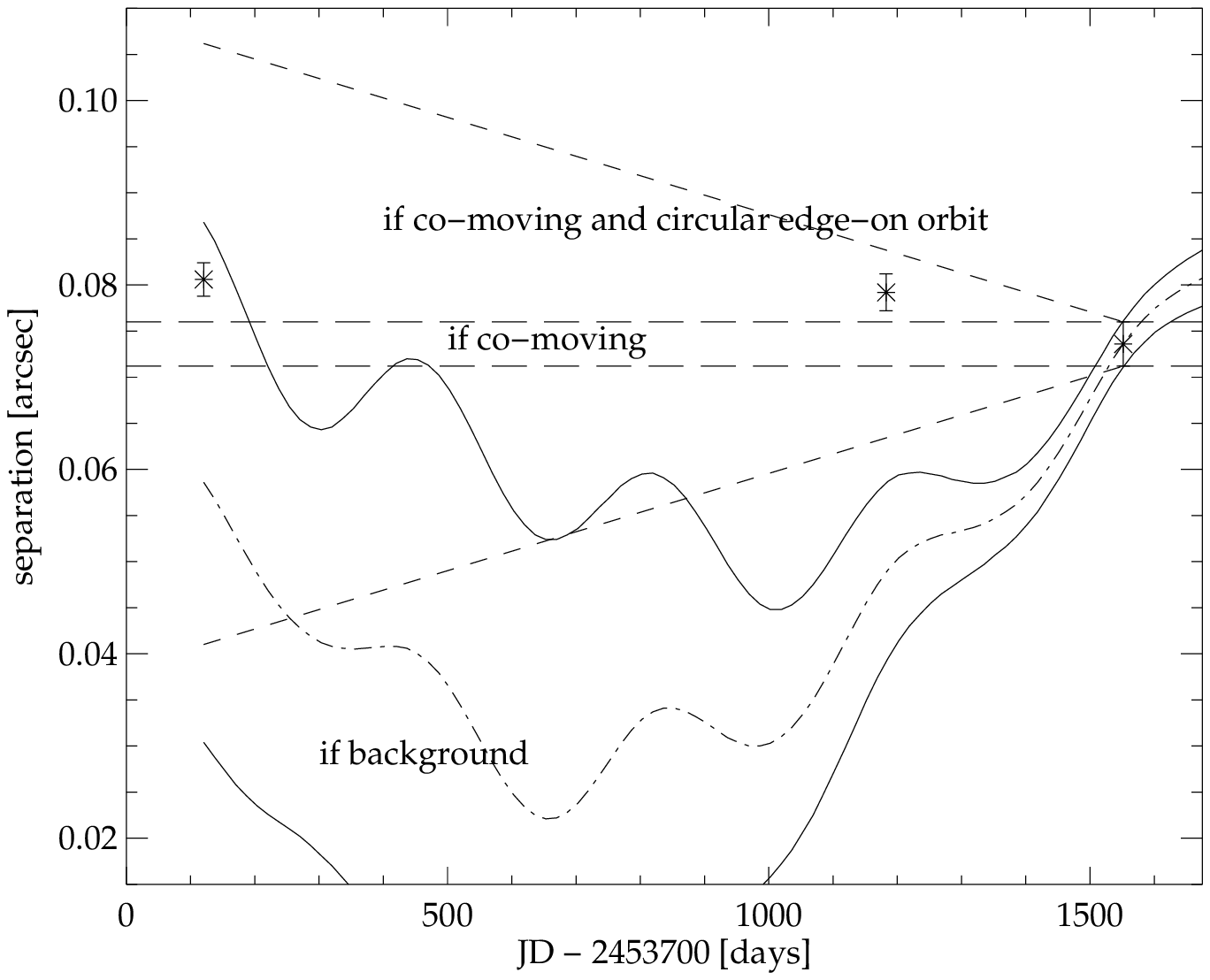}
   \includegraphics[width=0.49\textwidth]{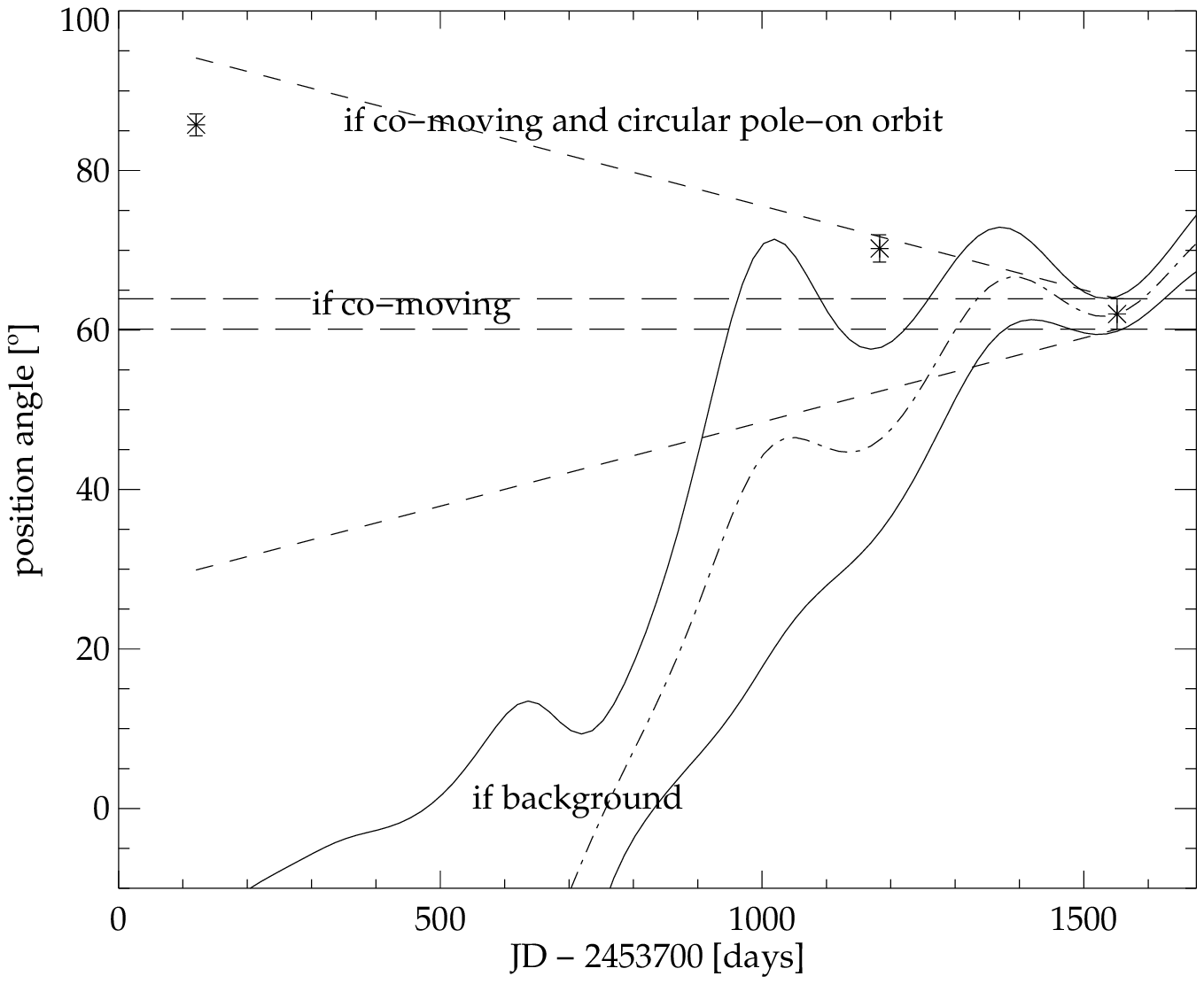}
   \includegraphics[width=0.49\textwidth]{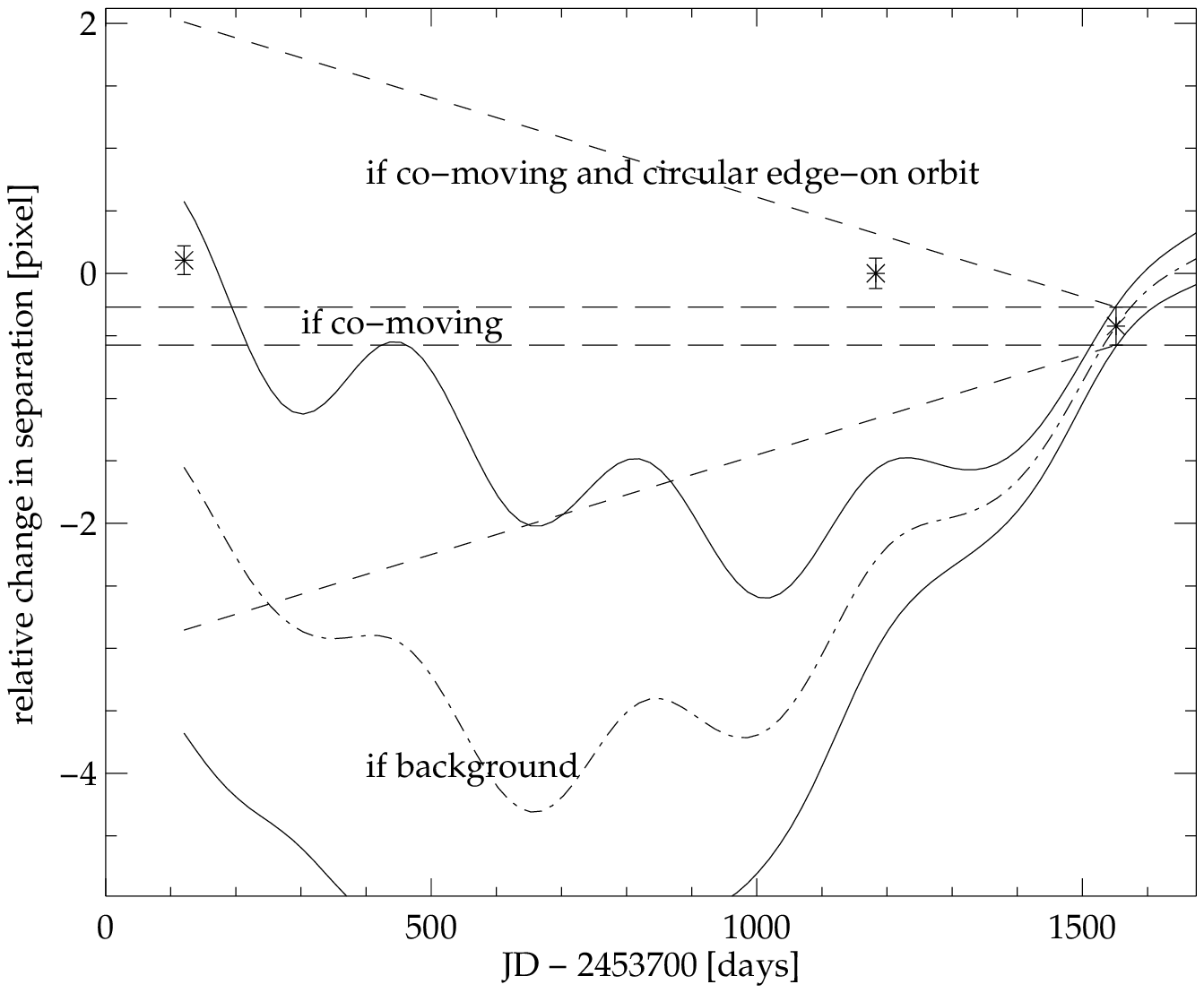}
   \includegraphics[width=0.49\textwidth]{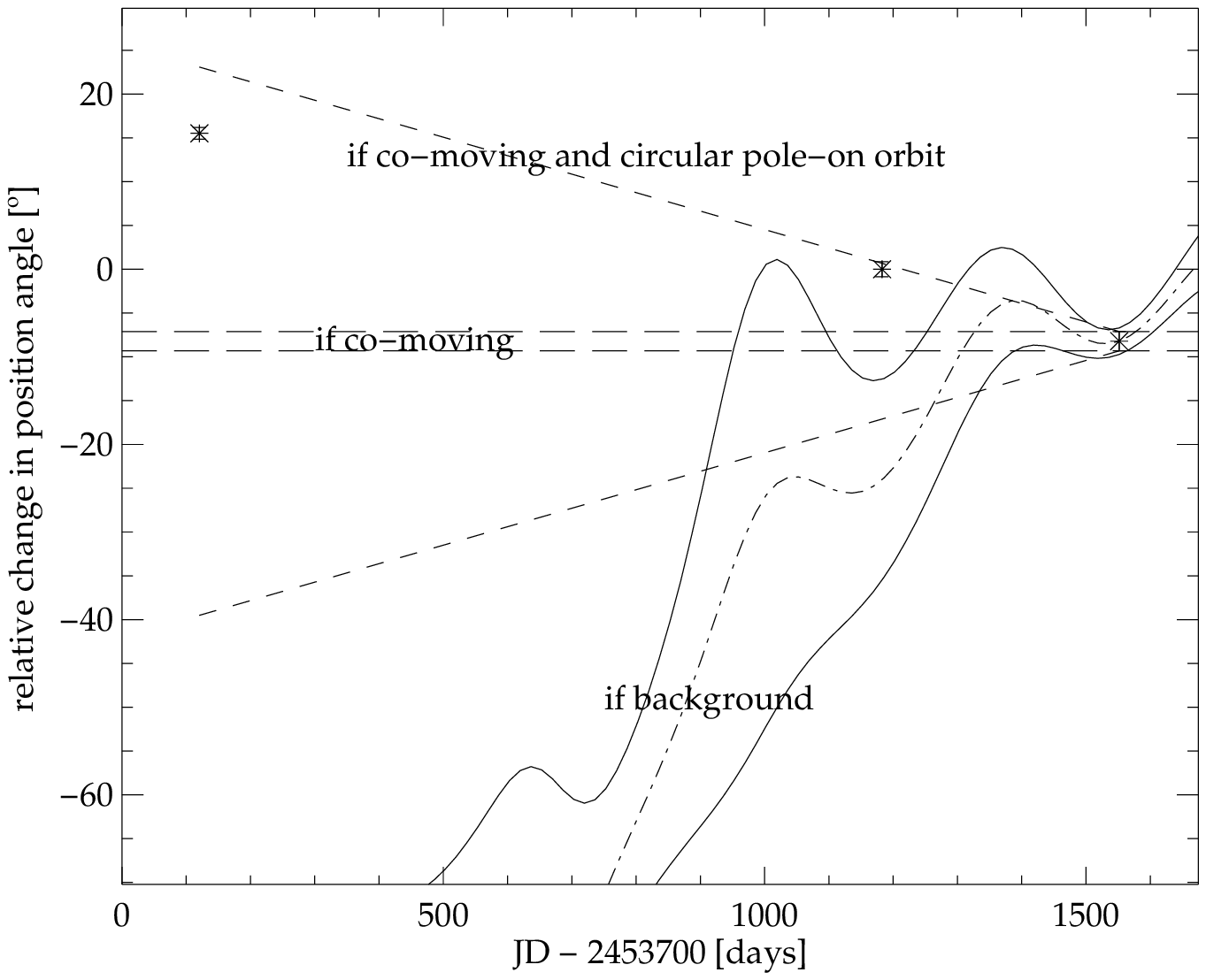}
   \caption{Proper Motion Diagrams (PMD) for separation and position angle change from absolute astrometric measurements (top, left to 
   right) and from relative astrometric measurements (bottom, left to right) in the 
   RX J1109.4-7627 AB system. See text for more information.}
   \label{RX J1109.4-7627}
   \end{figure}
}

\onlfig{11}{
\begin{figure}
   \includegraphics[width=0.49\textwidth]{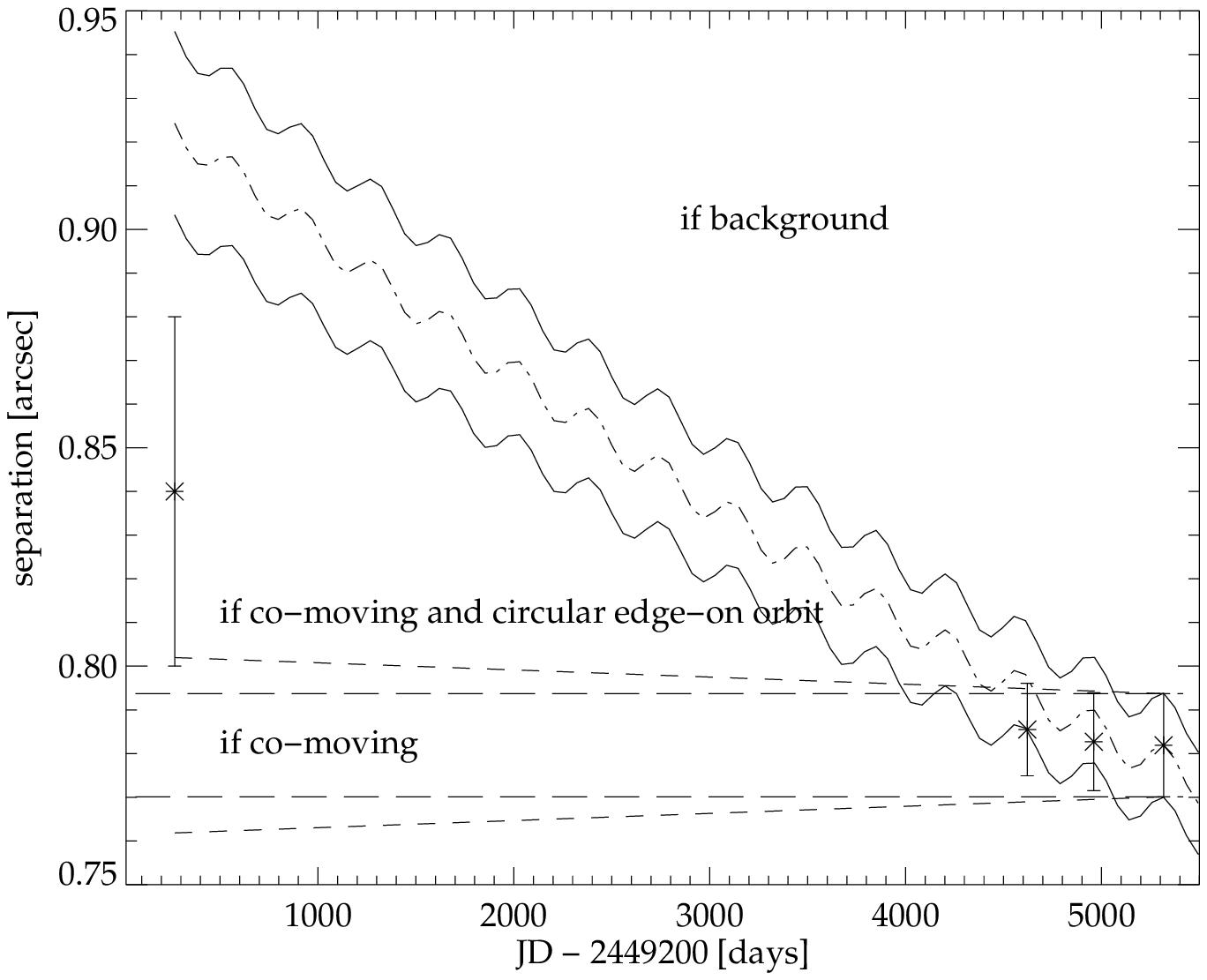}
   \includegraphics[width=0.49\textwidth]{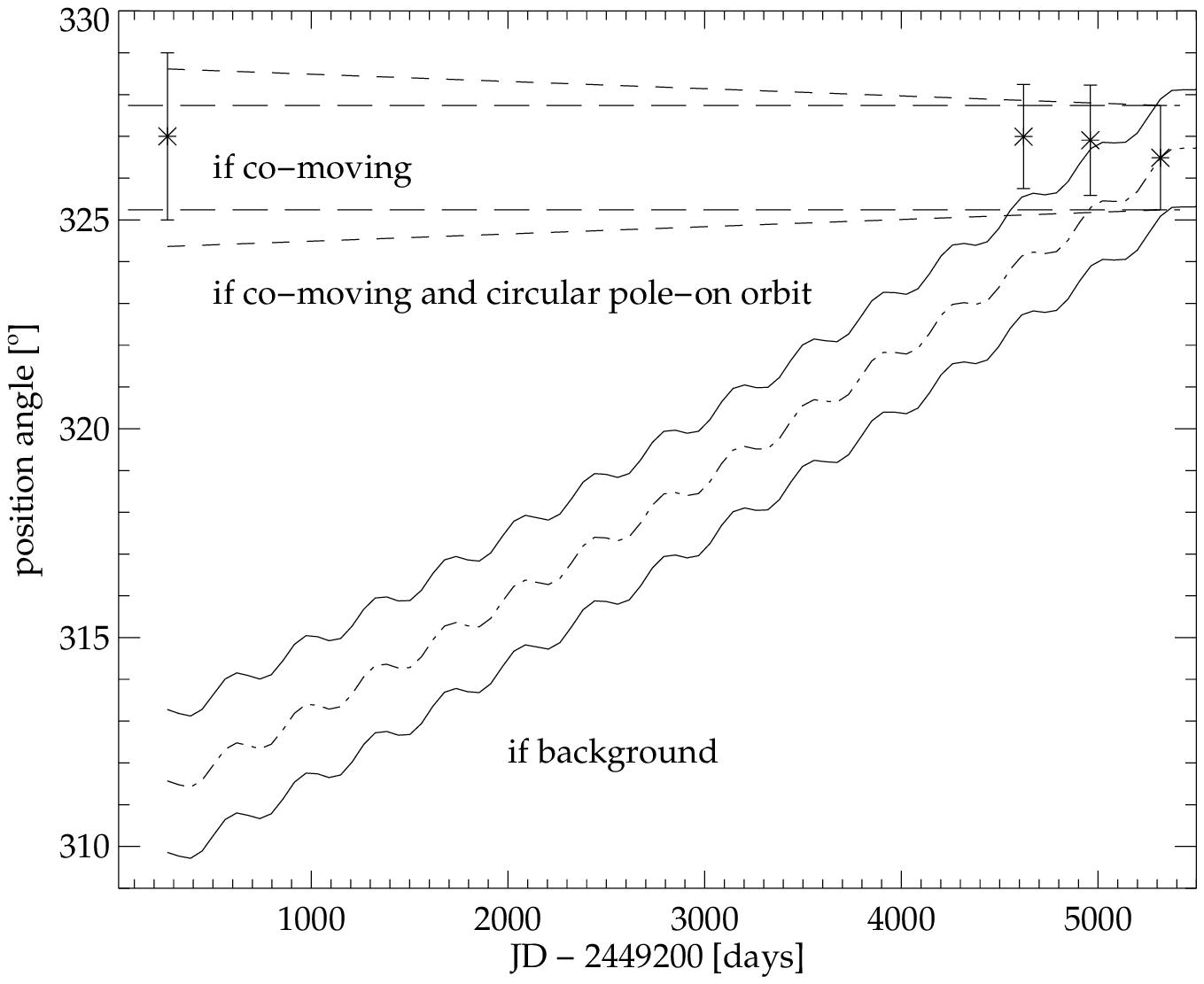}
   \includegraphics[width=0.49\textwidth]{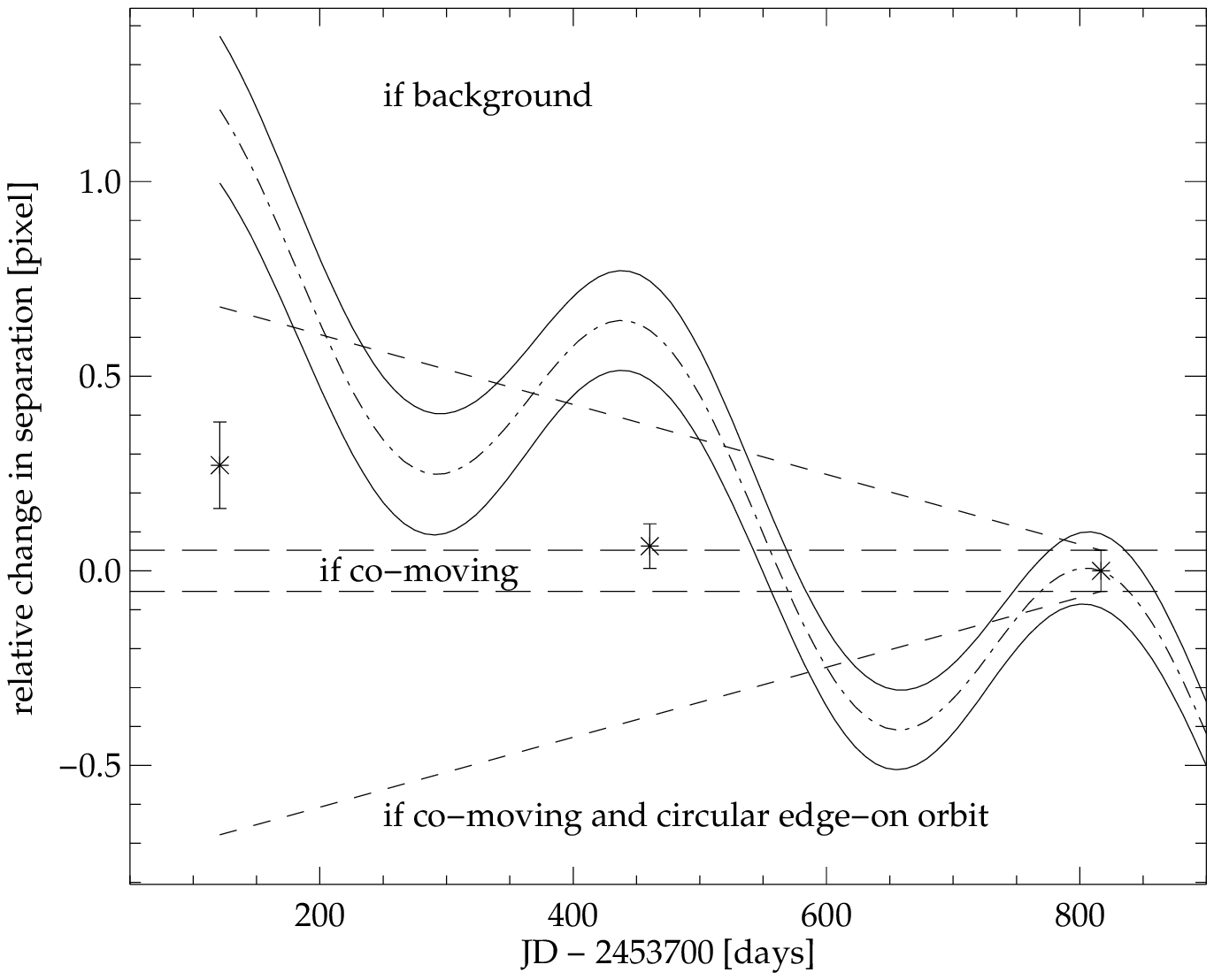}
   \includegraphics[width=0.49\textwidth]{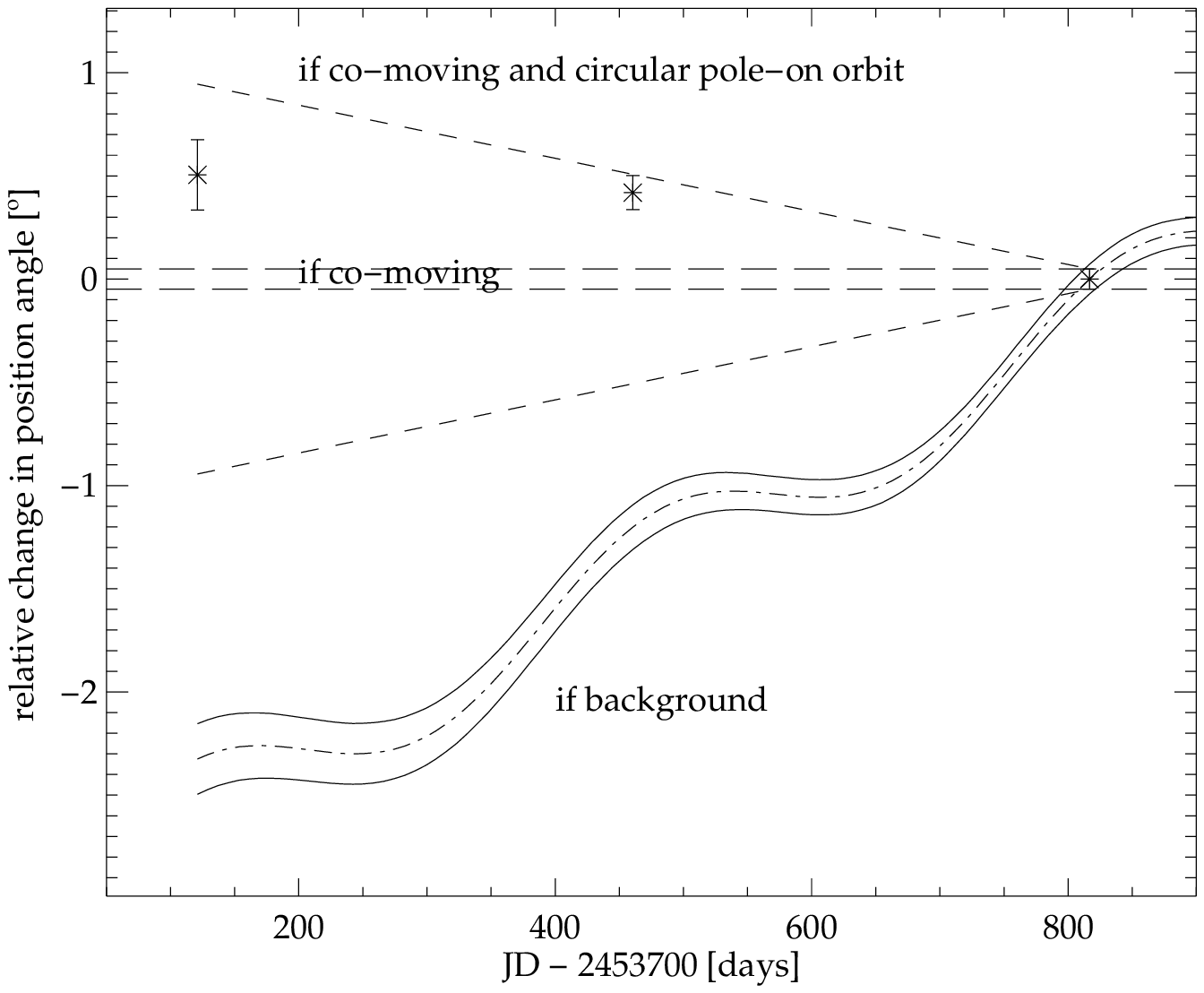}
   \caption{Proper Motion Diagrams (PMD) for separation and position angle change from absolute astrometric measurements (top, left to 
   right) and from relative astrometric measurements (bottom, left to right) in the 
   HD 97300 AB system. See text for more information.}
   \label{HD 97300}
   \end{figure}
}

\onlfig{12}{
\begin{figure}
   \includegraphics[width=0.49\textwidth]{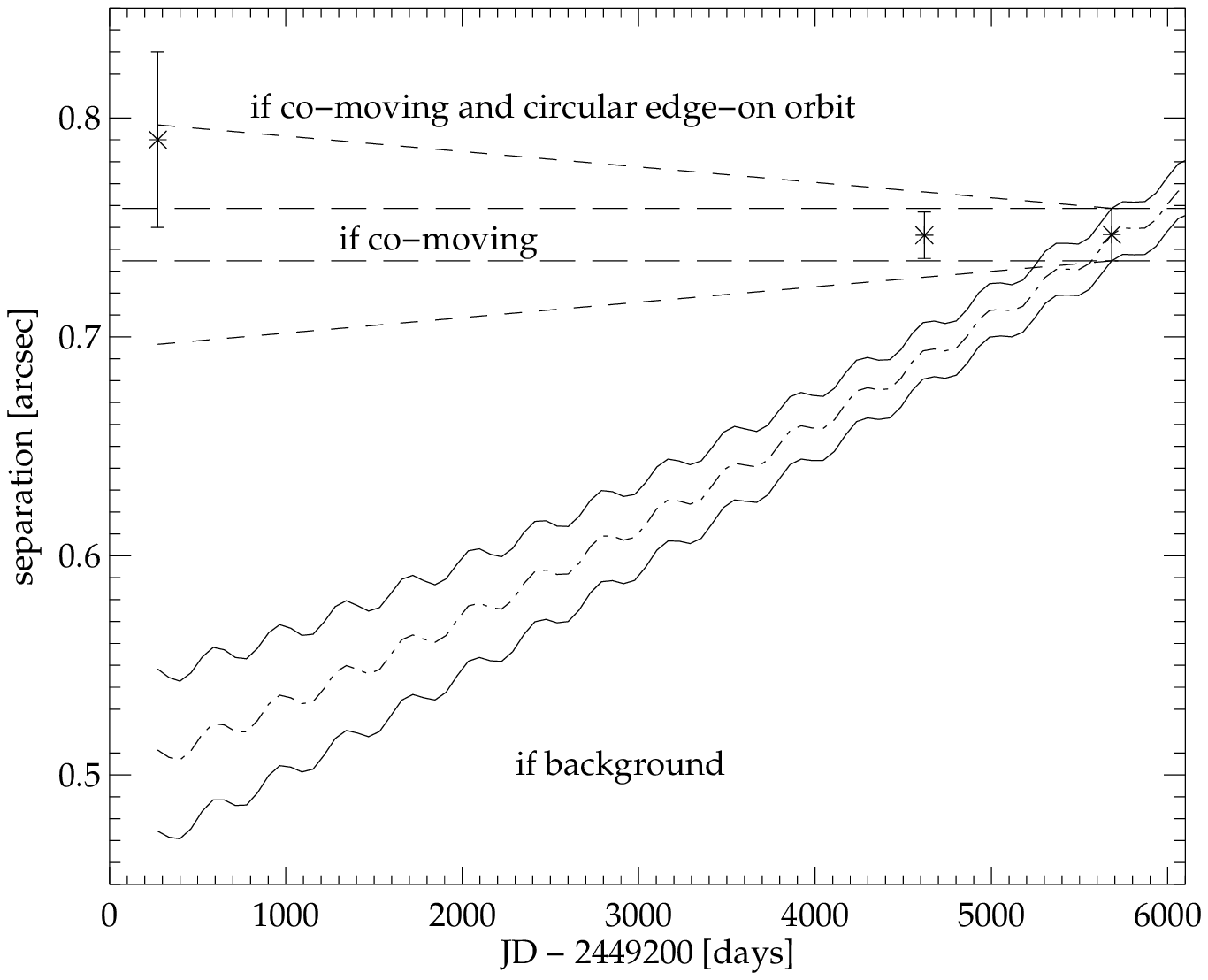}
   \includegraphics[width=0.49\textwidth]{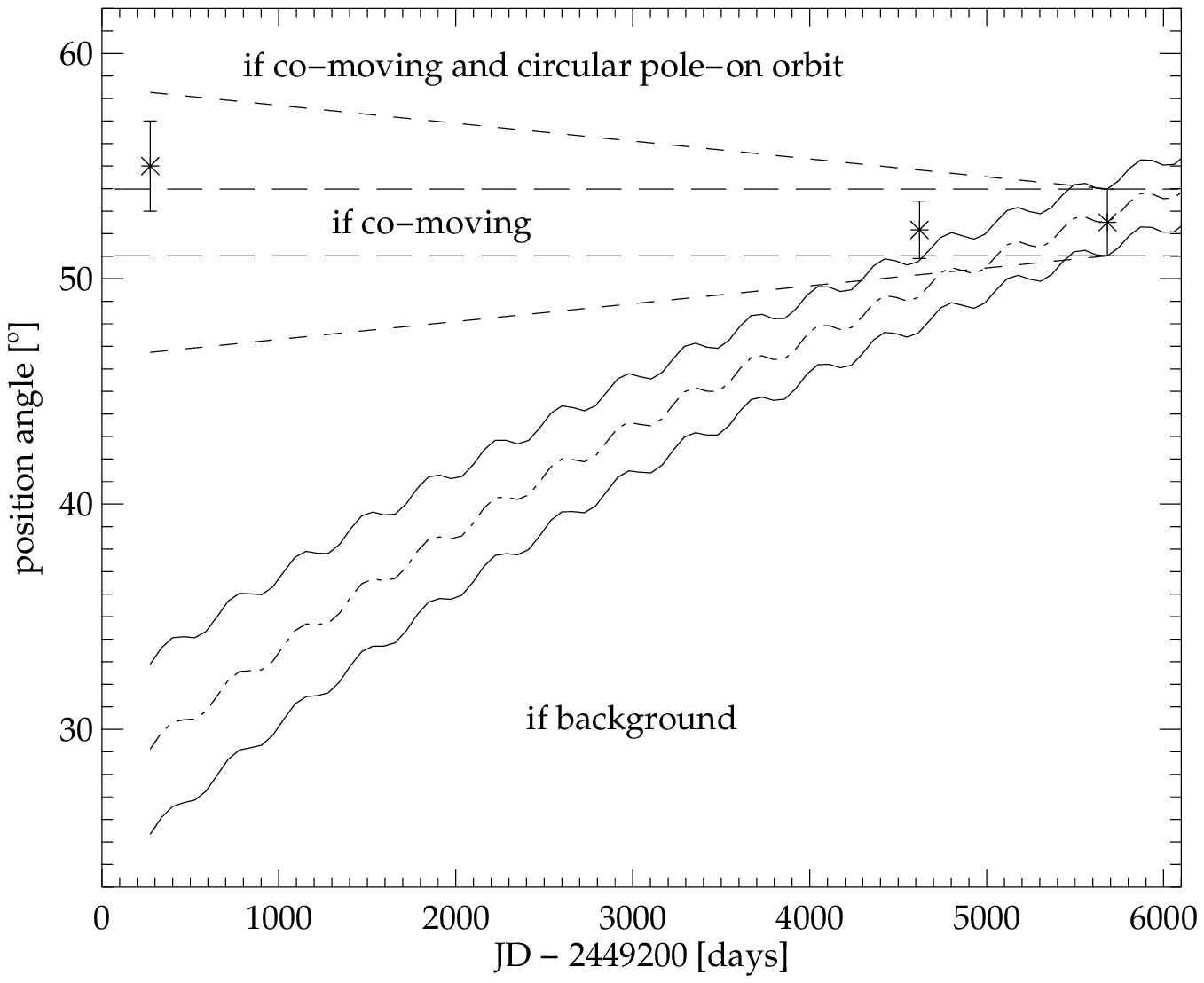}
   \includegraphics[width=0.49\textwidth]{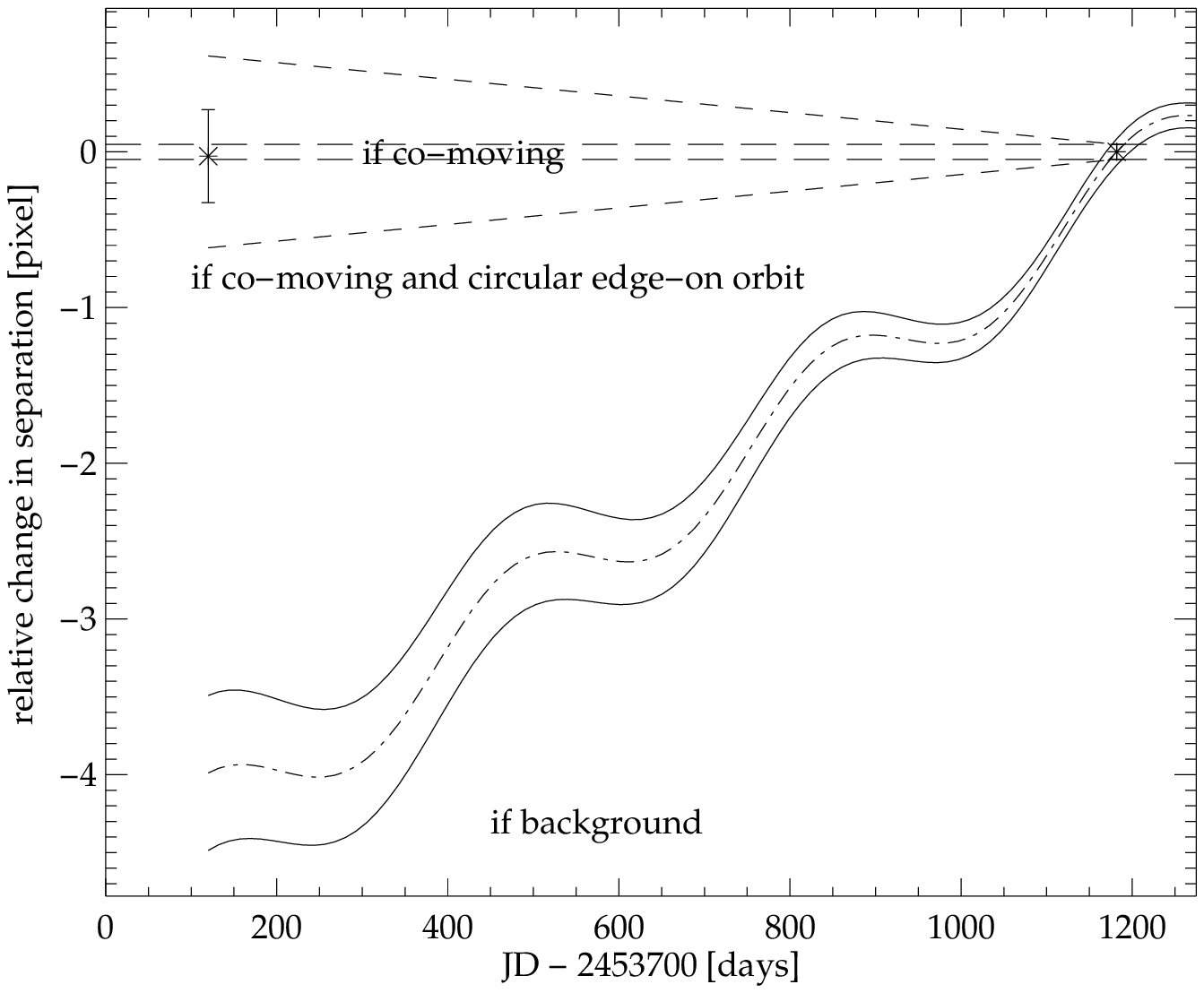}
   \includegraphics[width=0.49\textwidth]{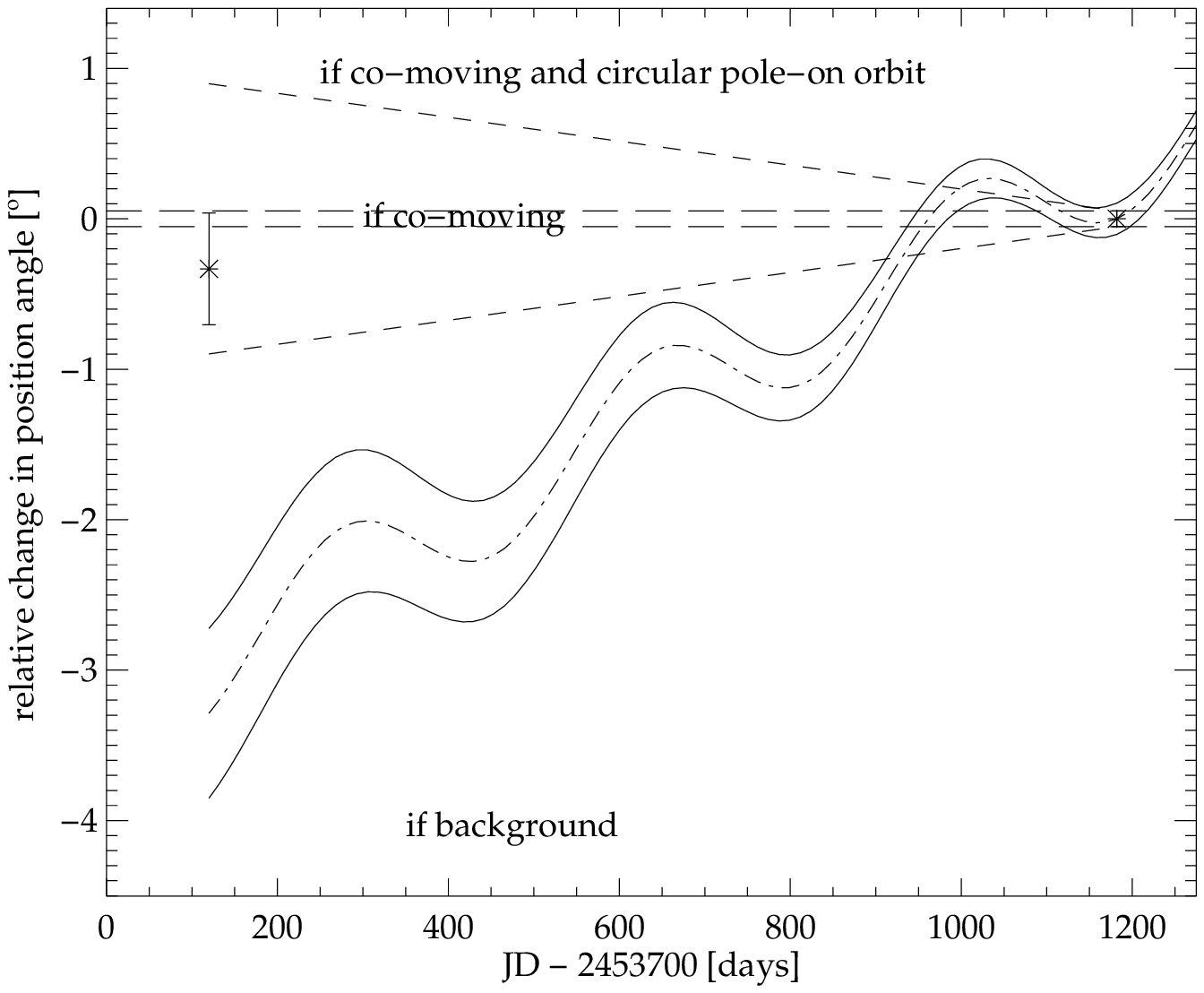}
   \caption{Proper Motion Diagrams (PMD) for separation and position angle change from absolute astrometric measurements (top, left to 
   right) and from relative astrometric measurements (bottom, left to right) in the 
   WX Cha AB system. See text for more information.}
   \label{WX Cha}
   \end{figure}
}

\onlfig{13}{
\begin{figure}
   \includegraphics[width=0.49\textwidth]{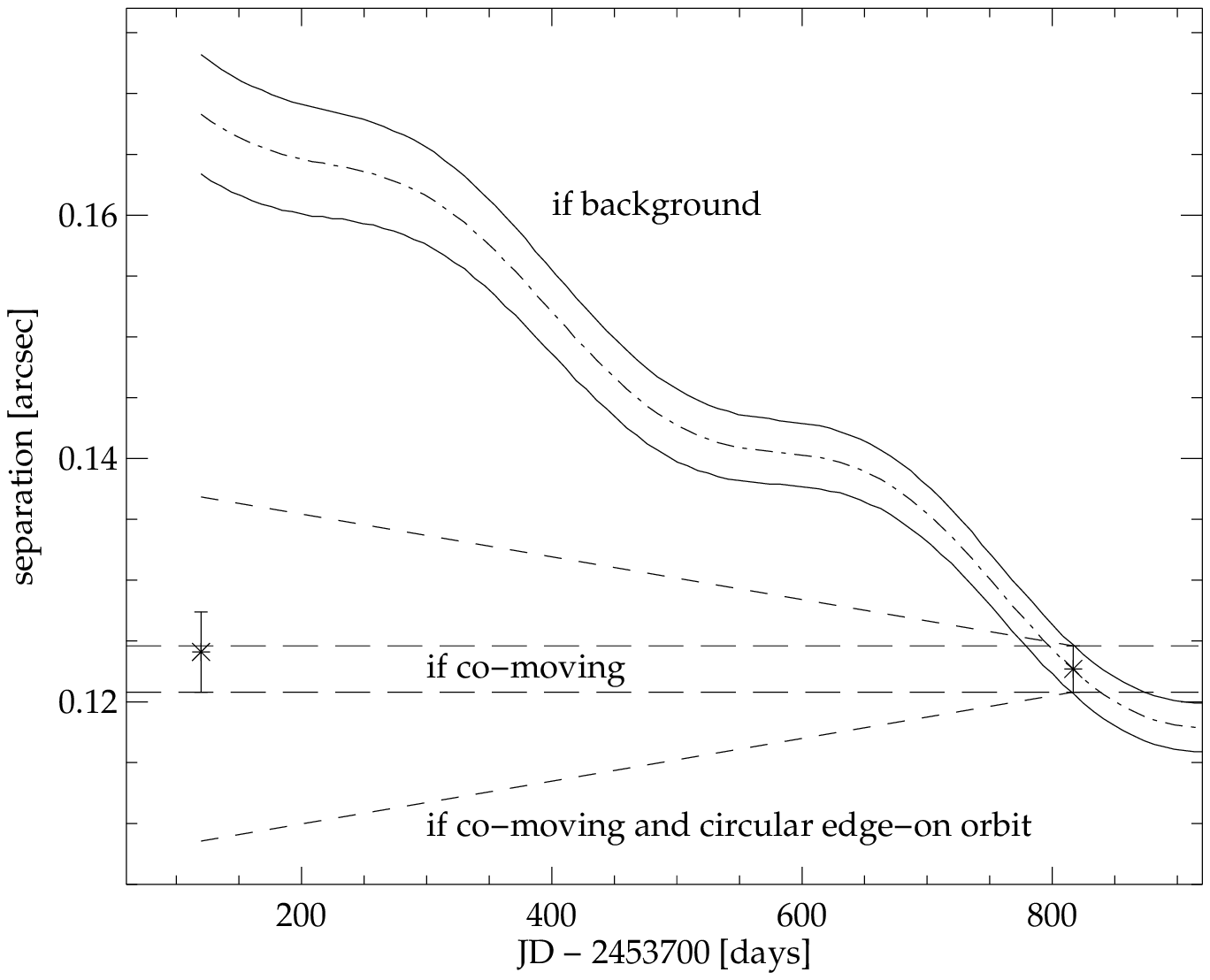}
   \includegraphics[width=0.49\textwidth]{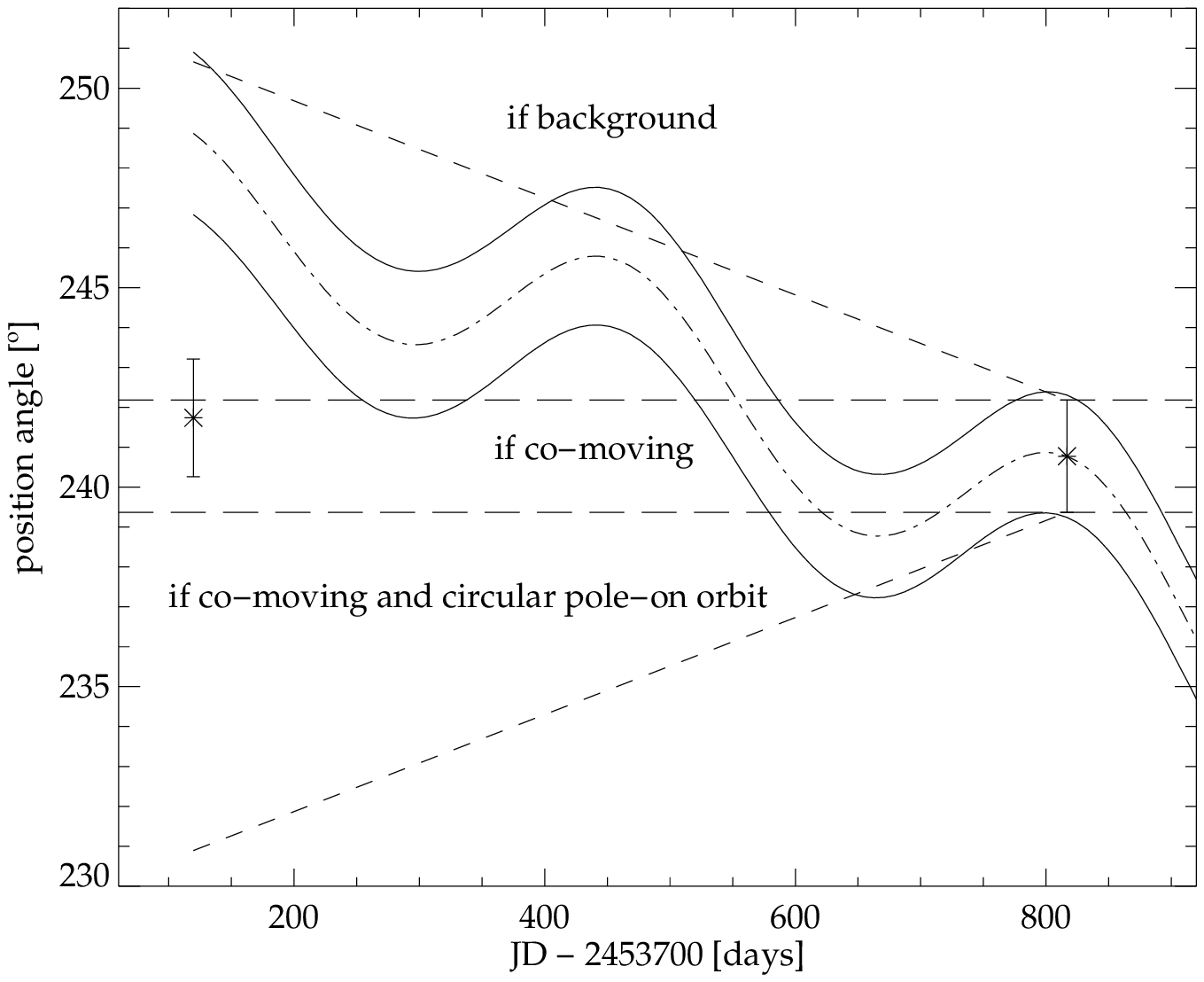}
   \includegraphics[width=0.49\textwidth]{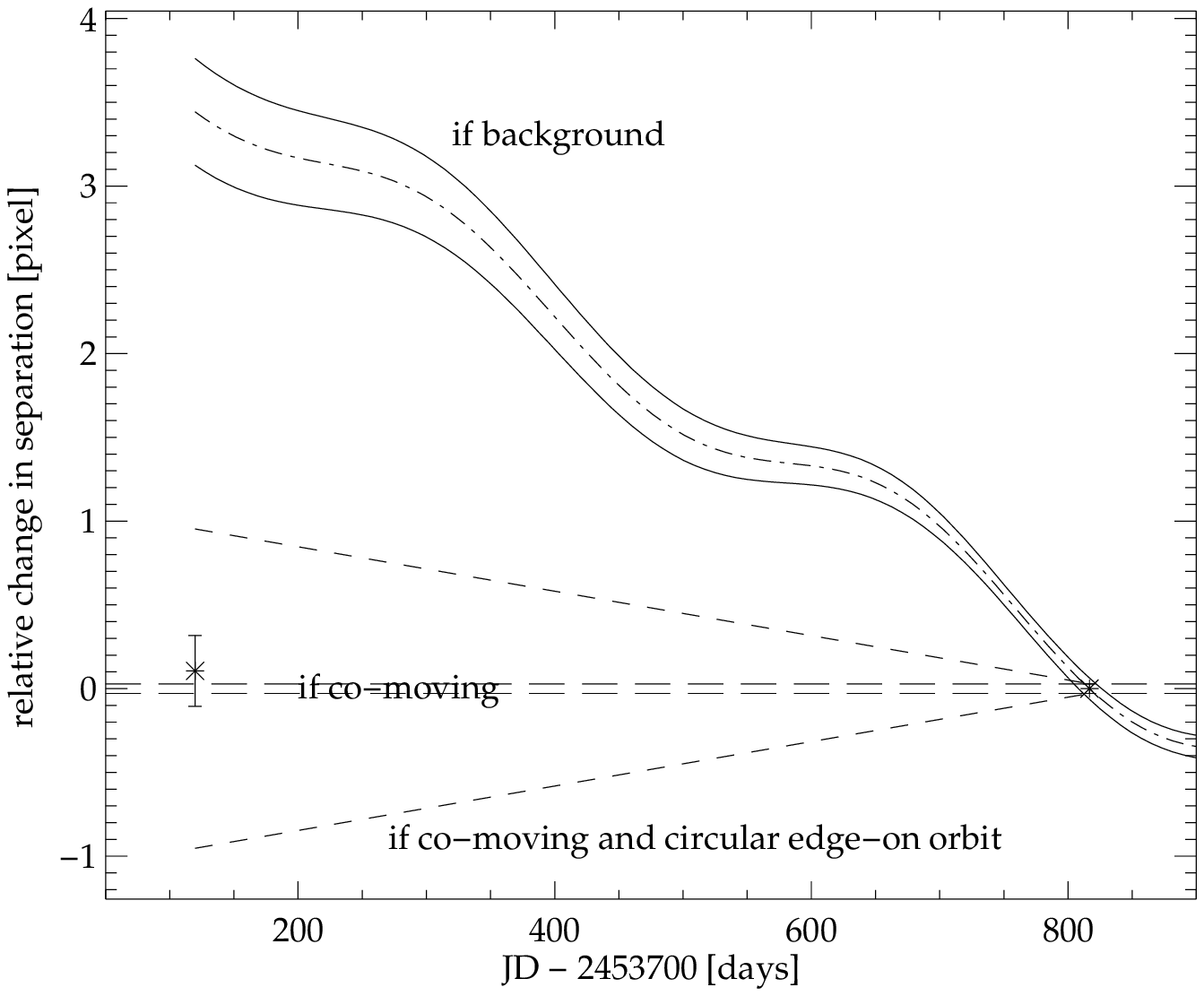}
   \includegraphics[width=0.49\textwidth]{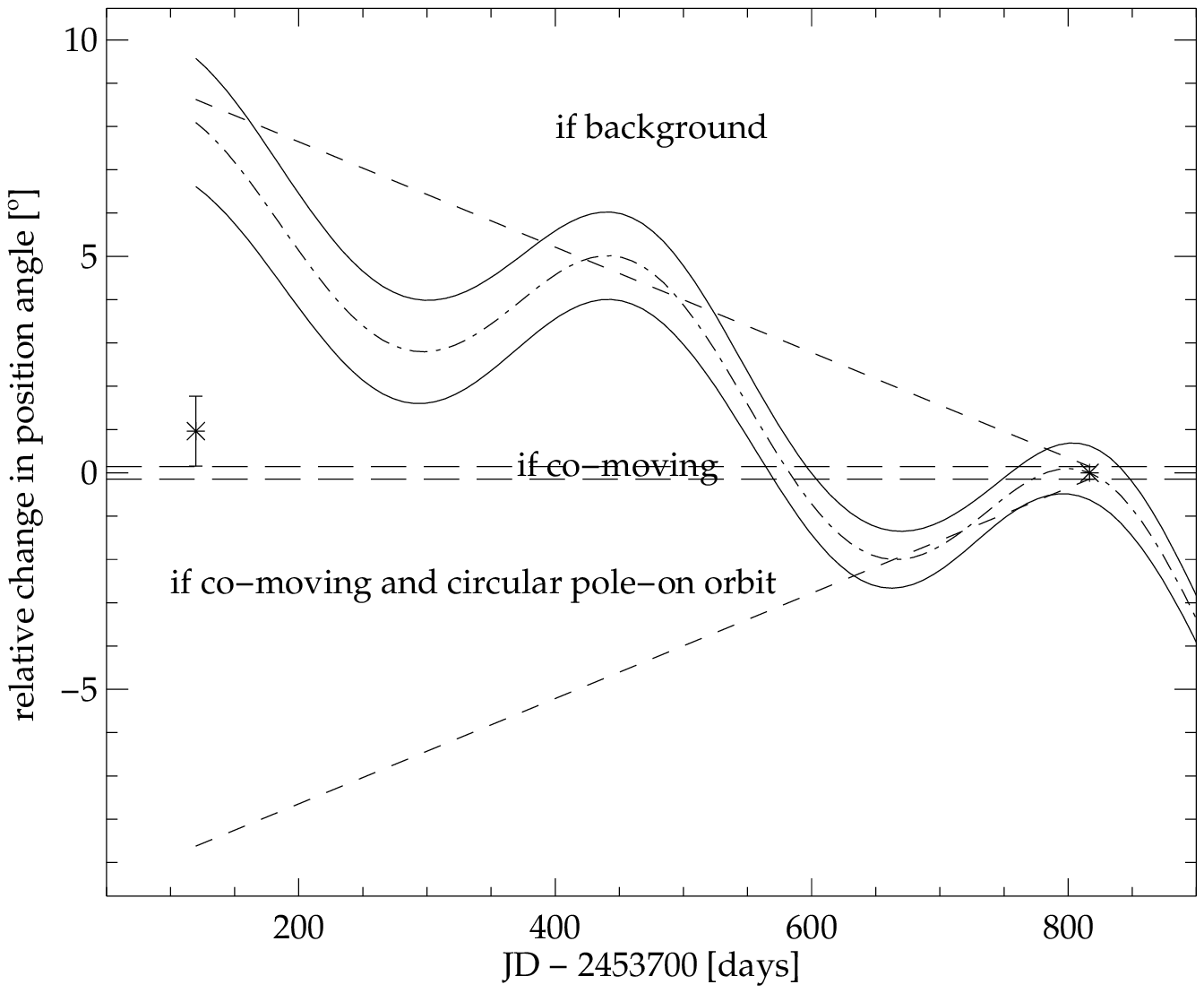}
   \caption{Proper Motion Diagrams (PMD) for separation and position angle change from absolute astrometric measurements (top, left to 
   right) and from relative astrometric measurements (bottom, left to right) in the 
   WY Cha AB system. See text for more information.}
   \label{WY Cha}
   \end{figure}
}        

\onlfig{14}{
\begin{figure}
   \includegraphics[width=0.49\textwidth]{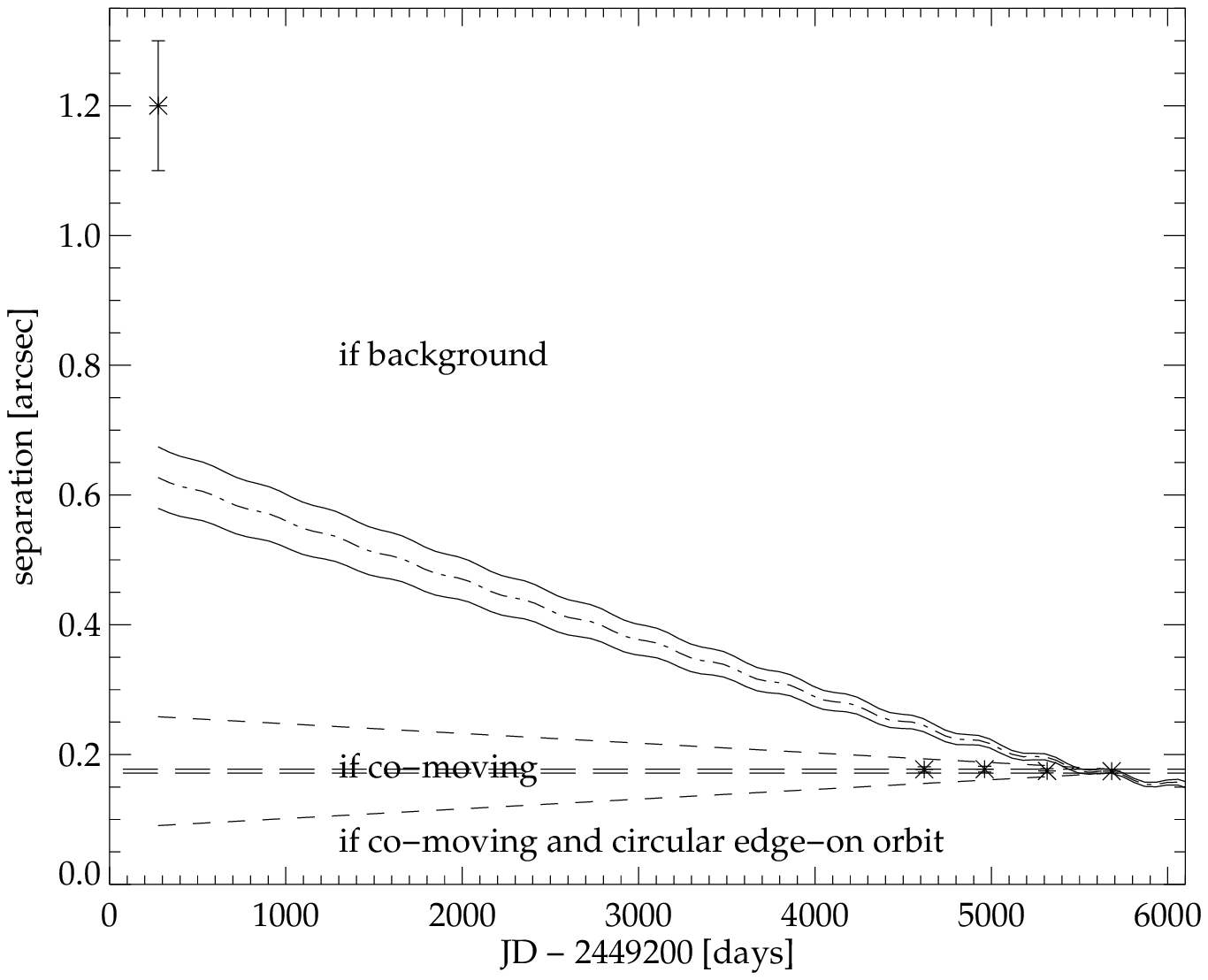}
   \includegraphics[width=0.49\textwidth]{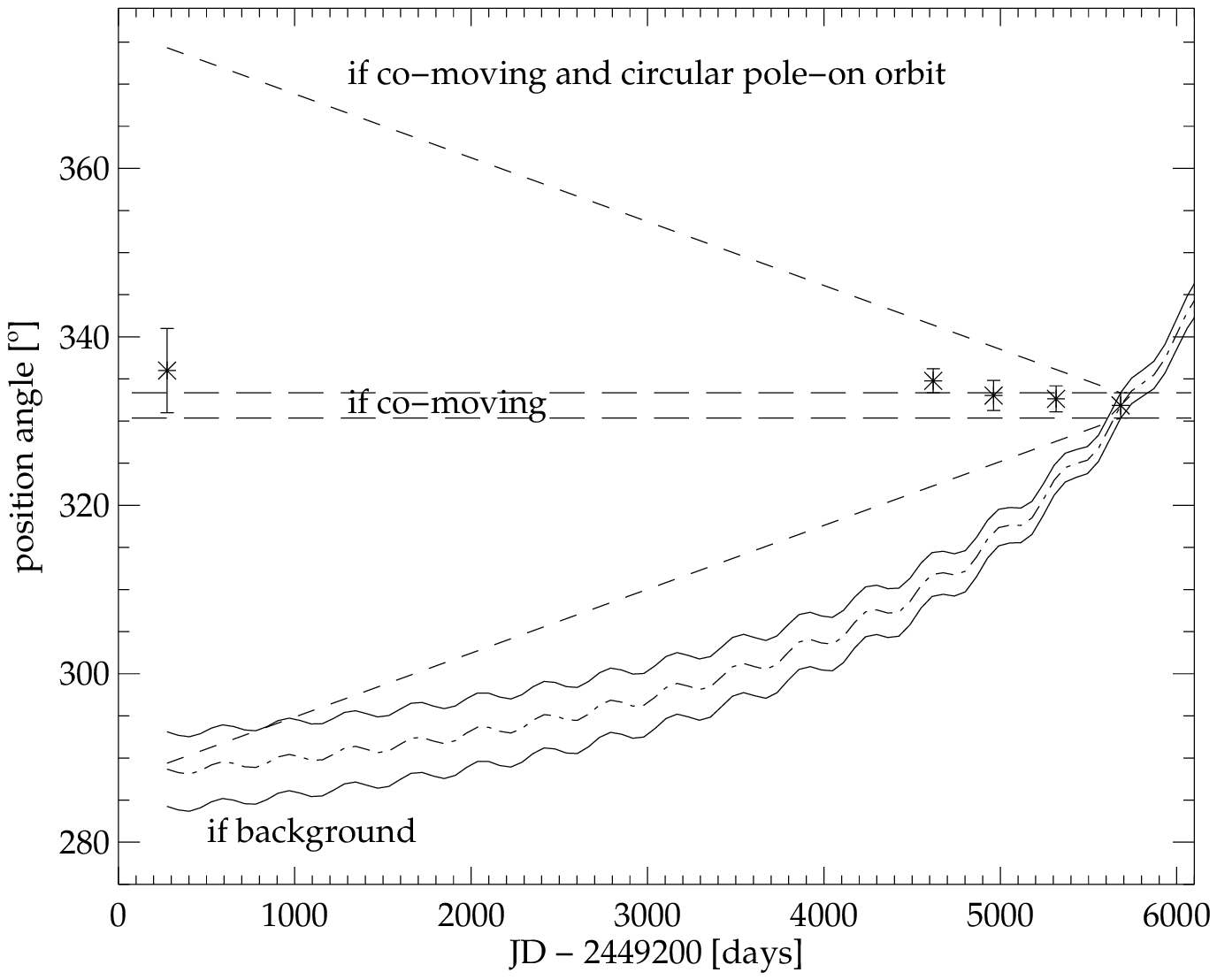}
   \includegraphics[width=0.49\textwidth]{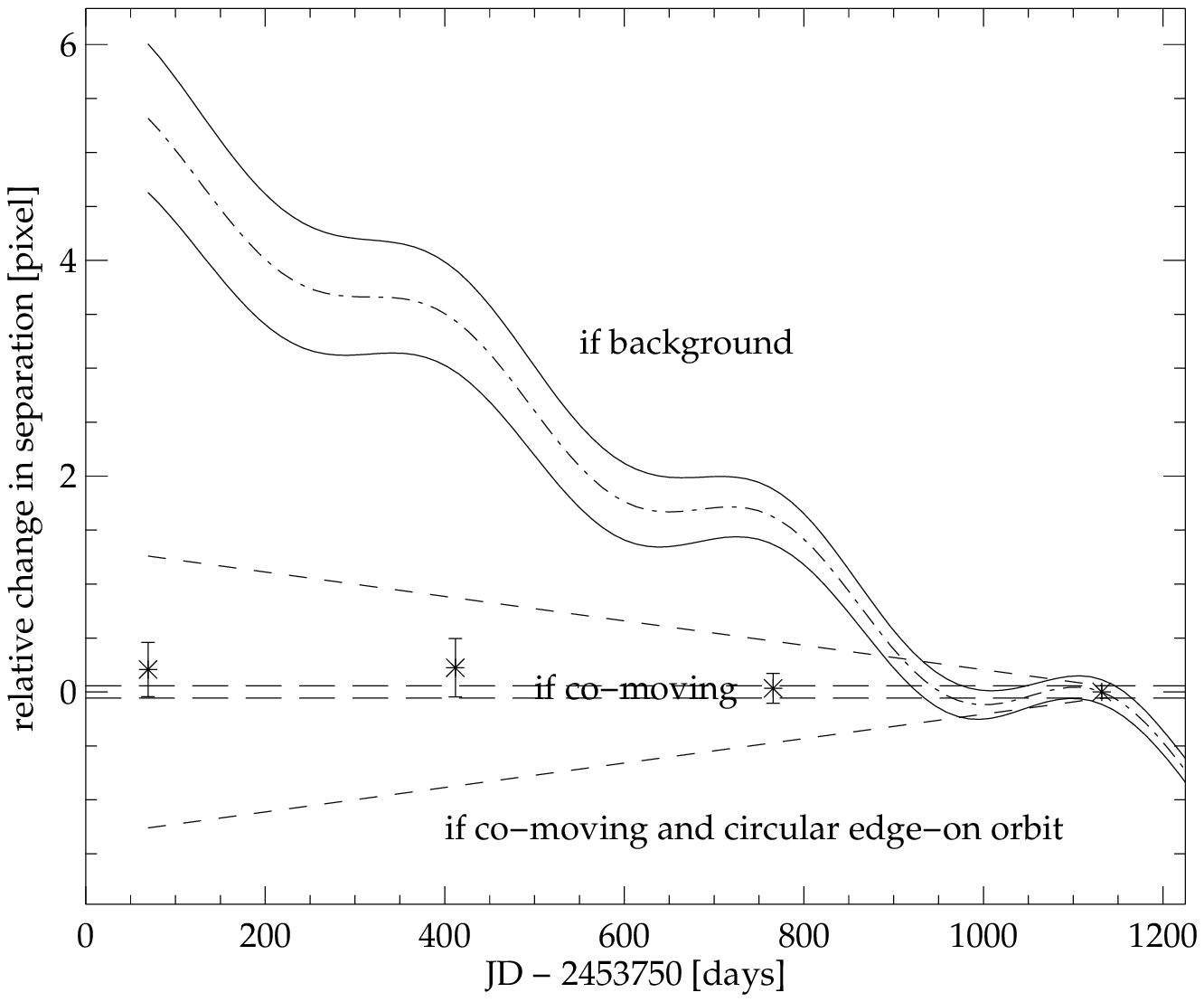}
   \includegraphics[width=0.49\textwidth]{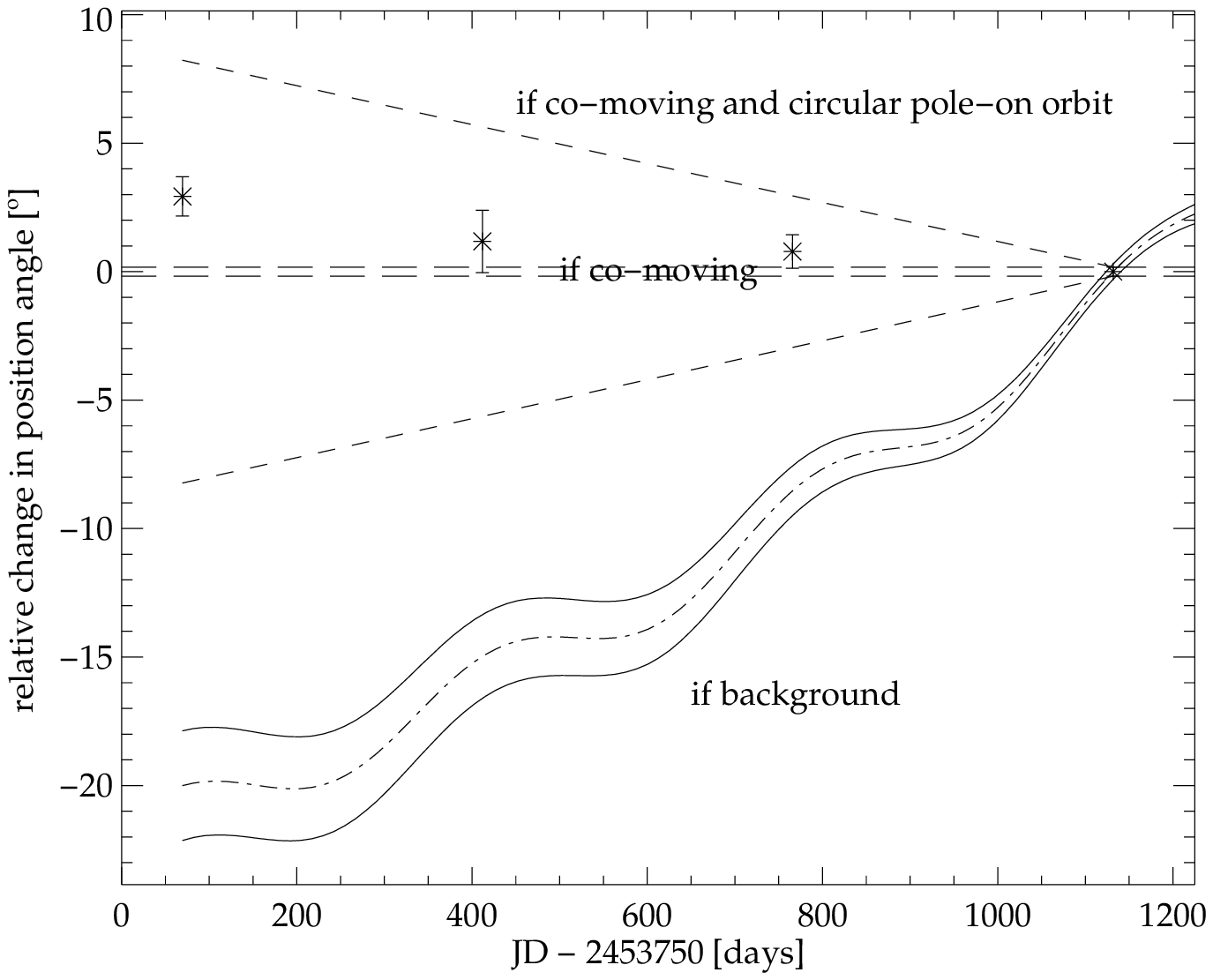}
   \caption{Proper Motion Diagrams (PMD) for separation and position angle change from absolute astrometric measurements (top, left to 
   right) and from relative astrometric measurements (bottom, left to right) in the 
   HJM C 7-11 AB system. See text for more information.}
   \label{HJM C 7-11}
   \end{figure}
}

\onlfig{15}{
\begin{figure}
   \includegraphics[width=0.49\textwidth]{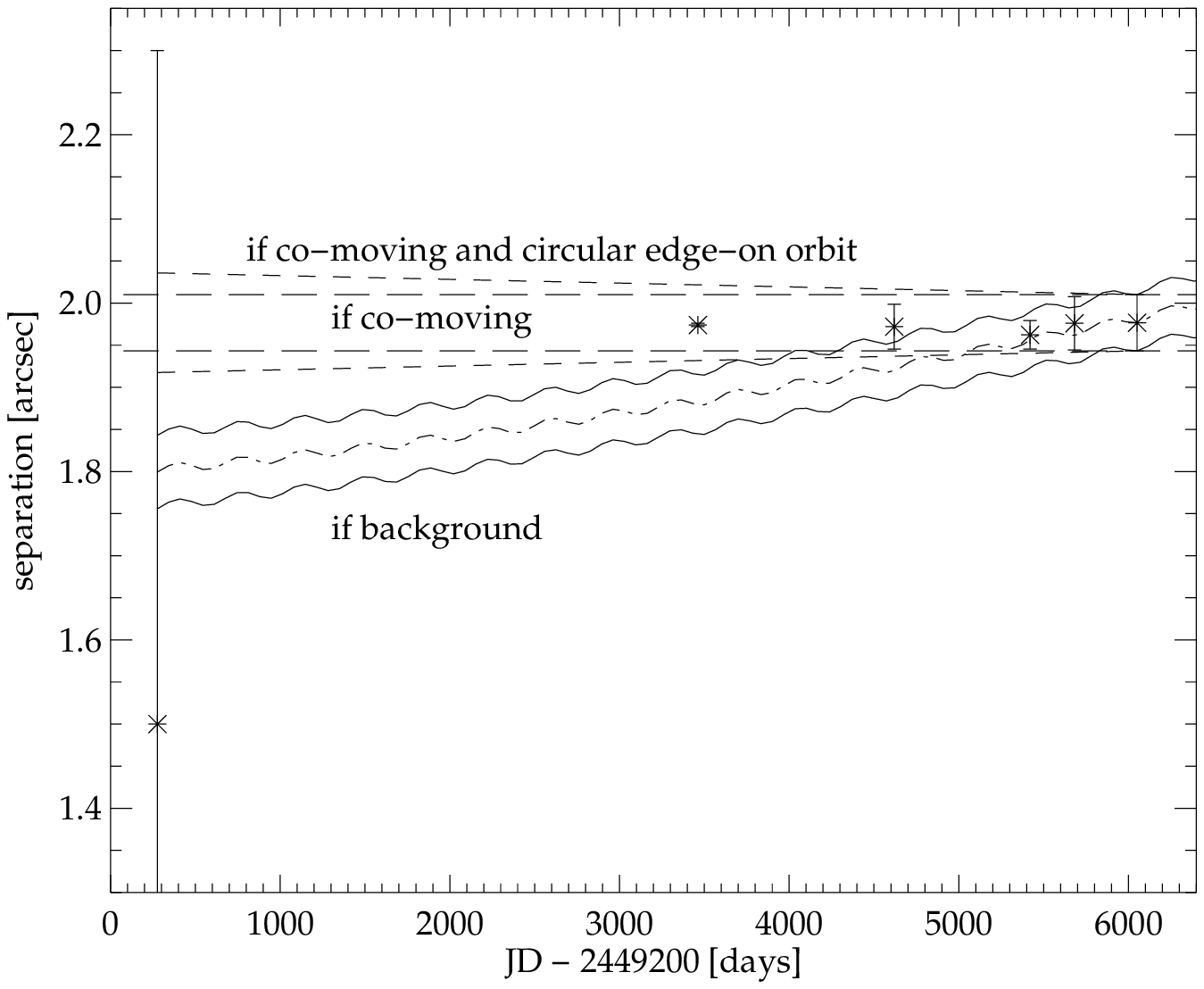}
   \includegraphics[width=0.49\textwidth]{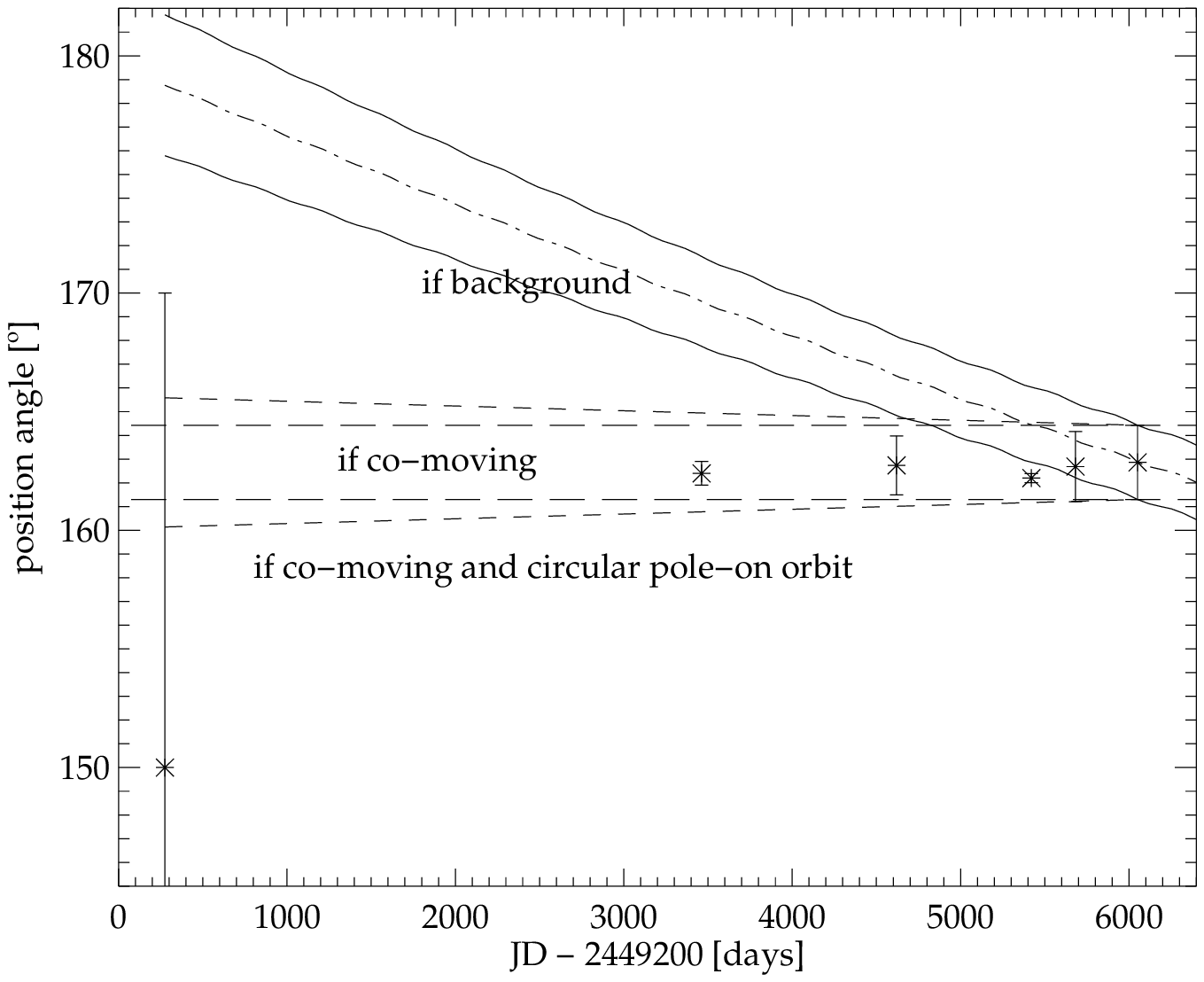}
   \includegraphics[width=0.49\textwidth]{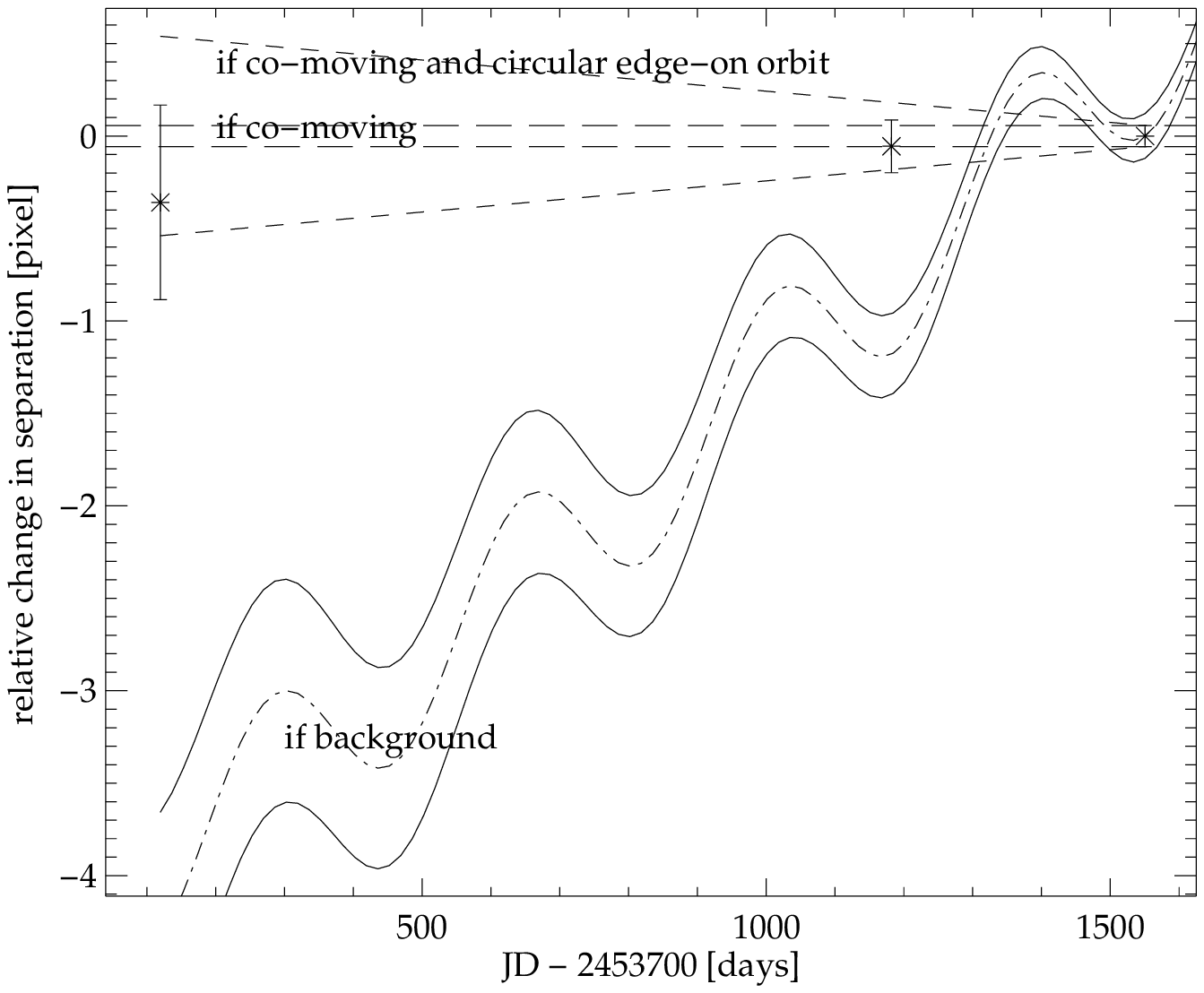}
   \includegraphics[width=0.49\textwidth]{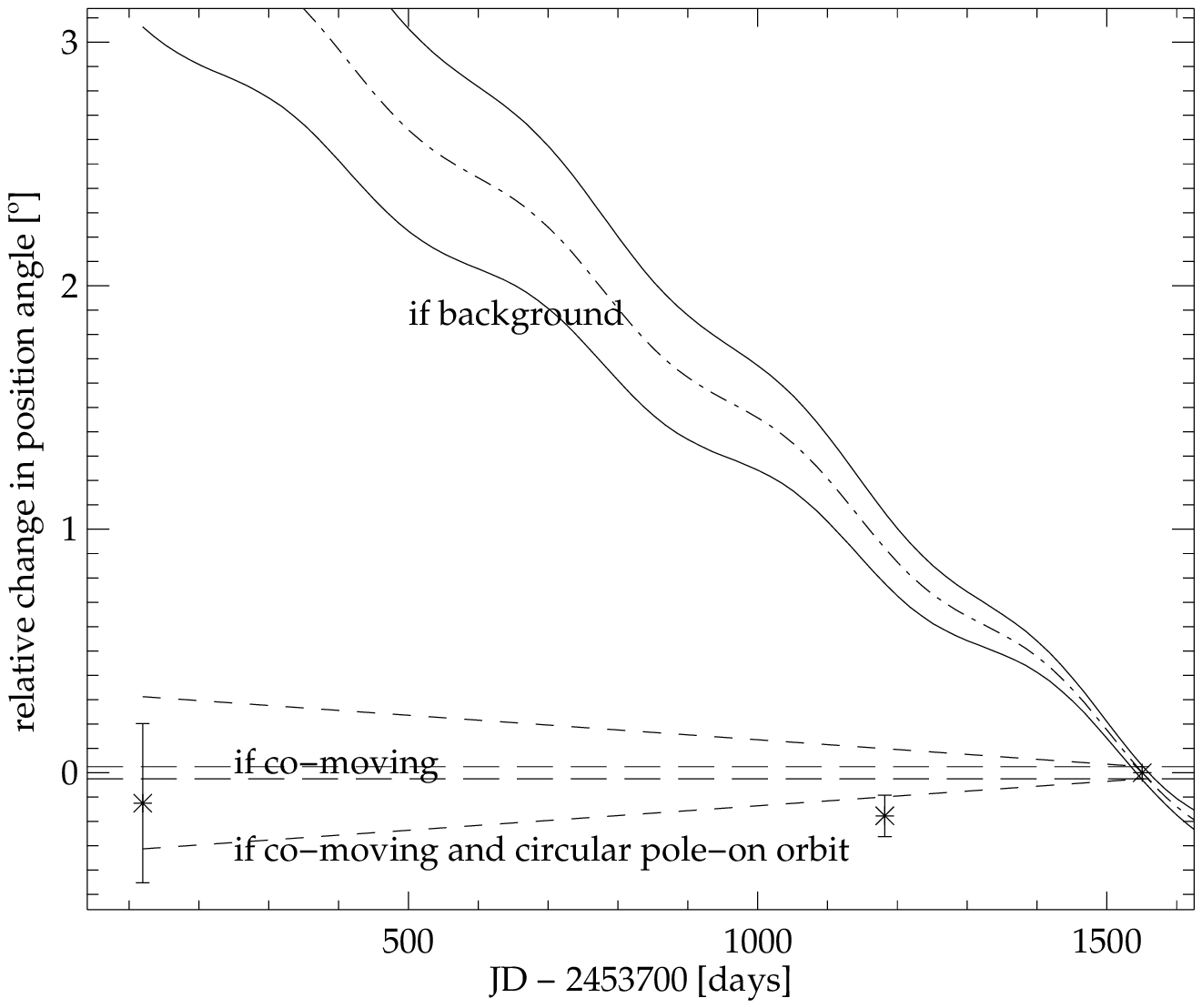}
   \caption{Proper Motion Diagrams (PMD) for separation and position angle change from absolute astrometric measurements (top, left to 
   right) and from relative astrometric measurements (bottom, left to right) in the 
   Sz 41 AB system. See text for more information.}
   \label{Sz 41}
   \end{figure}
}

\onlfig{16}{
\begin{figure}
   \includegraphics[width=0.49\textwidth]{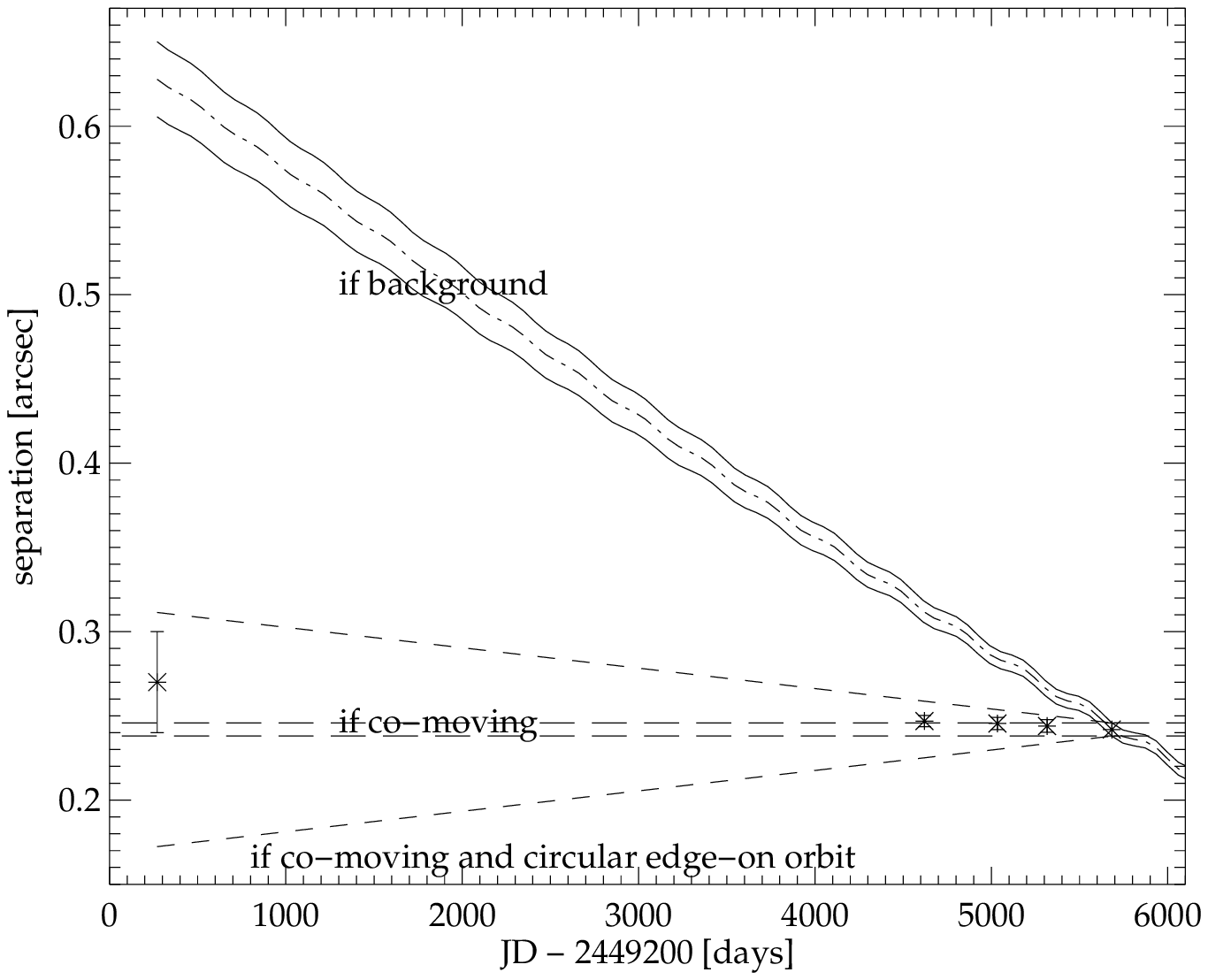}
   \includegraphics[width=0.49\textwidth]{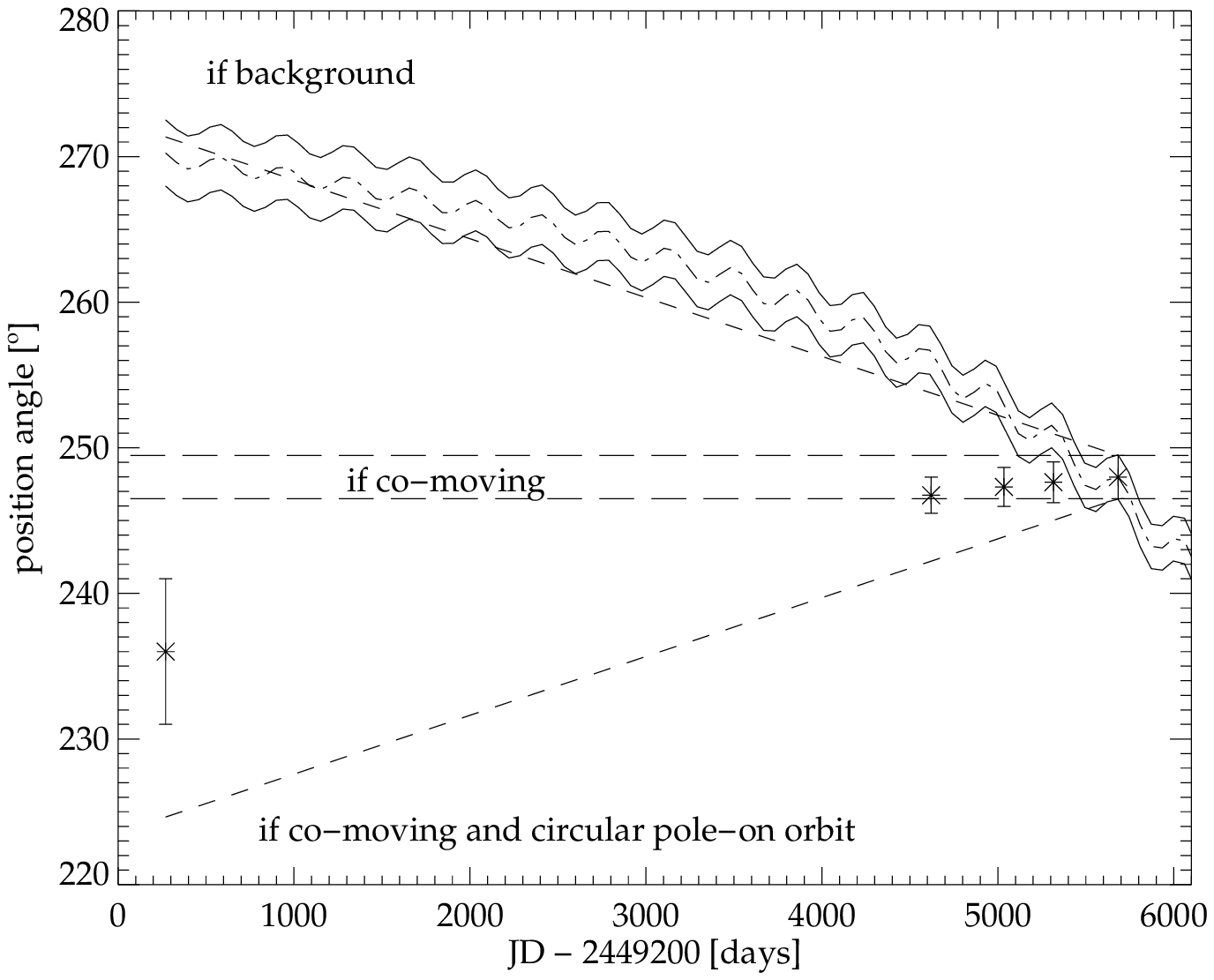}
   \includegraphics[width=0.49\textwidth]{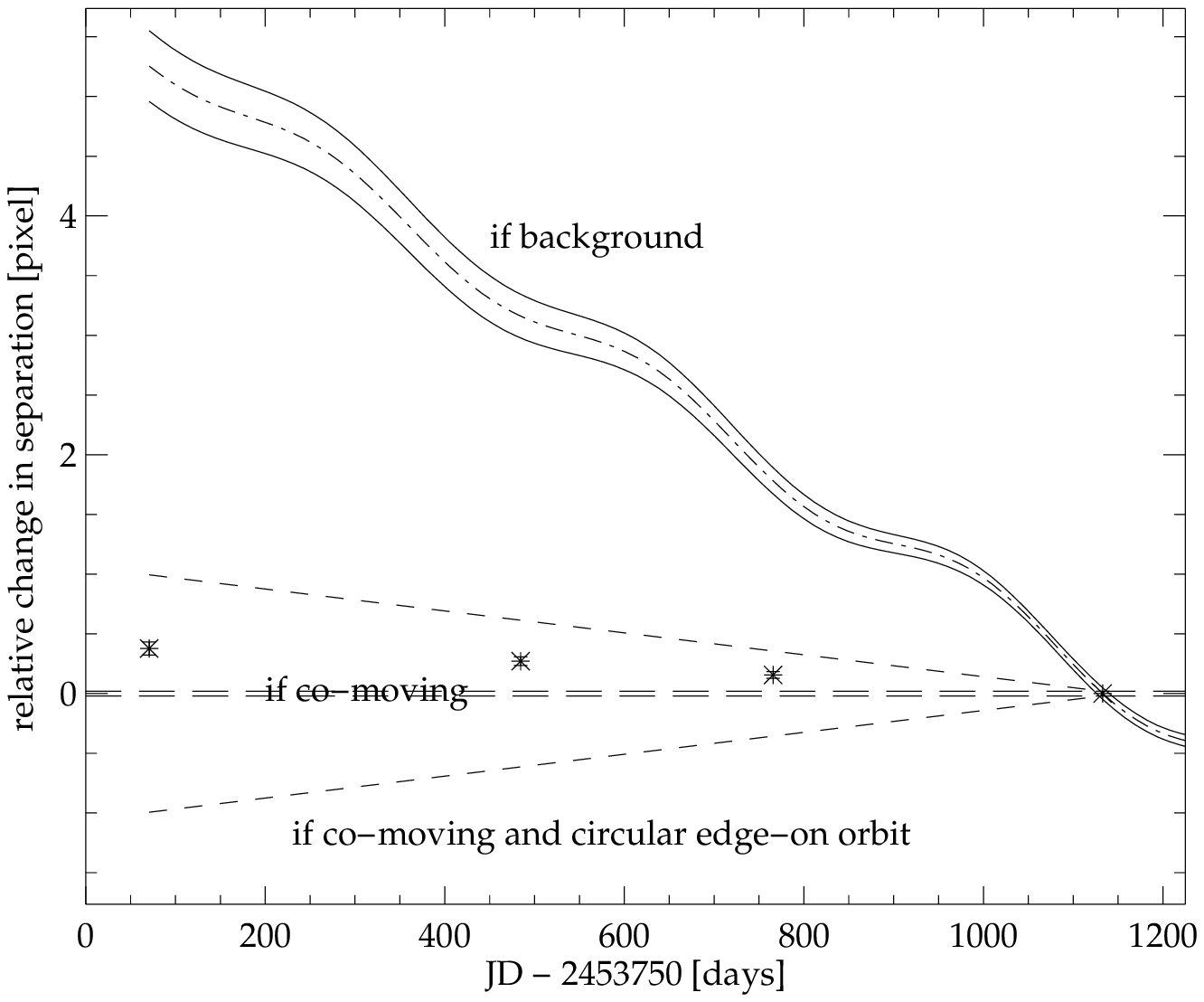}
   \includegraphics[width=0.49\textwidth]{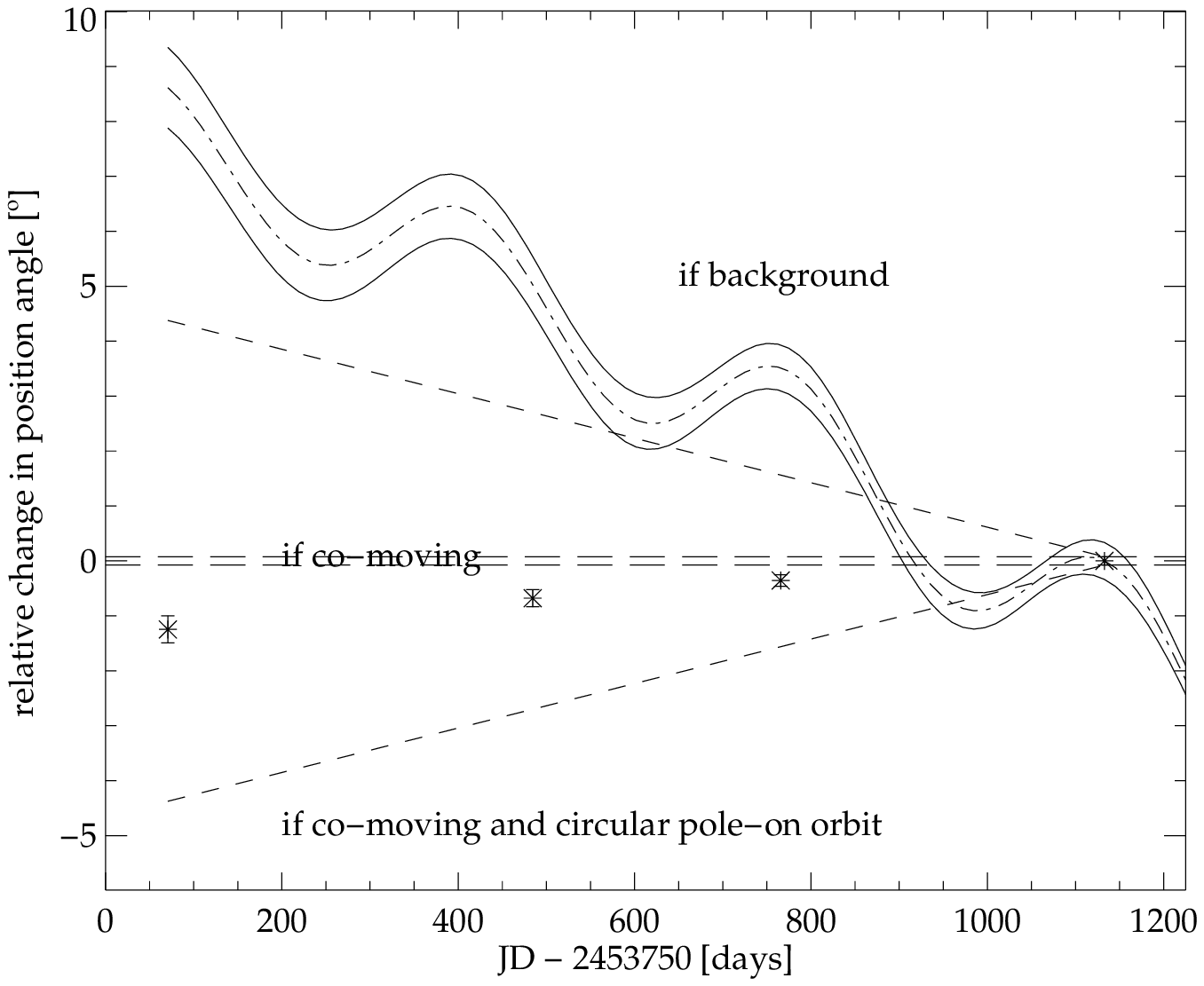}
   \caption{Proper Motion Diagrams (PMD) for separation and position angle change from absolute astrometric measurements (top, left to 
   right) and from relative astrometric measurements (bottom, left to right) in the 
   HM Anon AB system. See text for more information.}
   \label{HM Anon}
   \end{figure}
}

\end{document}